\documentclass{article}

\usepackage{amssymb,amsfonts,amsmath}
\usepackage{cite,enumerate,float}
\usepackage{color}

\usepackage{physics}

\usepackage{mathrsfs}

\usepackage{tikz}
\usetikzlibrary{arrows}
\usetikzlibrary{arrows.meta}
\usetikzlibrary{positioning}
\usetikzlibrary{shapes,snakes}
\usetikzlibrary{fit}
\usetikzlibrary{decorations.pathmorphing,decorations.pathreplacing,decorations.markings}
\usetikzlibrary{calc}
\usepackage{inconsolata}
\usepackage{ytableau}

\usepackage{xcolor}
\definecolor{burgundy}{rgb}{0.5, 0.0, 0.13}
\definecolor{olive}{rgb}{0.50, 0.50, 0.0}
\usepackage[linktocpage=true,colorlinks=true,linkcolor=burgundy,citecolor=black!20!blue,urlcolor=black!40!violet]{hyperref}

\usepackage[most]{tcolorbox}

\usepackage{pdflscape}
\usepackage{array, longtable}
\newcolumntype{C}{>{$}c<{$}}

\def\be{\begin{eqnarray}}
\def\ee{\end{eqnarray}}

\def\p{\partial}

\def\Tr{{\rm Tr}\,}

\definecolor{red}{rgb}{1,0,0}
\definecolor{orange}{rgb}{1,0.5,0}
\definecolor{violet}{rgb}{0.7,0,1}

\def\CA{{\cal A}}

\def\CI {{\cal I}}

\def\CL {{\cal L}}

\def\CN {{\cal N}}

\def\CV {{\cal V}}

\def\CI {{{\cal I}}}

\def\CQ {{\cal Q}}

\def\IC{\mathbb{C}}

\def\IN{\mathbb{N}}

\def\IR{{\mathbb{R}}}

\def\IZ{{\mathbb{Z}}}

\def\ff{\mathfrak{f}}

\def\fg{\mathfrak{g}}

\def\fl{\mathfrak{l}}

\def\fs{\mathfrak{s}}

\def\fs{\mathfrak{s}}

\def\bPsi{{\boldsymbol{\Psi}}}

\def\lm{\limits}

\DeclareSymbolFont{bbsymbol}{U}{bbold}{m}{n}
\DeclareMathSymbol{\bbzero}{\mathbin}{bbsymbol}{"30}
\DeclareMathSymbol{\bbone}{\mathbin}{bbsymbol}{"31}
\DeclareMathSymbol{\bbtwo}{\mathbin}{bbsymbol}{"32}
\DeclareMathSymbol{\bbthree}{\mathbin}{bbsymbol}{"33}
\DeclareMathSymbol{\bbfour}{\mathbin}{bbsymbol}{"34}
\DeclareMathSymbol{\bbfive}{\mathbin}{bbsymbol}{"35}
\DeclareMathSymbol{\bbsix}{\mathbin}{bbsymbol}{"36}
\DeclareMathSymbol{\bbseven}{\mathbin}{bbsymbol}{"37}
\DeclareMathSymbol{\bbeight}{\mathbin}{bbsymbol}{"38}
\DeclareMathSymbol{\bbnine}{\mathbin}{bbsymbol}{"39}

\newcommand\sqbox[1]{{
		\setbox0=\hbox{\mbox{$\Box$}}
		\setbox1=\hbox{\mbox{\raisebox{0.35ex}{\small #1}}}
		\mbox{\raisebox{-0.2ex}{\rlap{\hbox to \wd0{\hss{\box1}\hss}}\box0}}
}}
\newcommand\ssqbox[1]{{
		\setbox0=\hbox{\mbox{$\scriptstyle\Box$}}
		\setbox1=\hbox{\mbox{\raisebox{0.35ex}{\tiny #1}}}
		\mbox{\raisebox{-0.2ex}{\rlap{\hbox to \wd0{\hss{\box1}\hss}}\box0}}
}}

\def\myblue{white!40!blue}

\tcbset{boxrule=0.6pt,colback=white!97!blue}

\newcommand{\Eul}{\text{Eul}}
\renewcommand{\d}{\partial}
\renewcommand{\bar}{\overline}

\usepackage{tabularray} 

\hoffset= - 1.0in         

\numberwithin{equation}{section}
\begin{document}

\hfill MIPT/TH-14/24

\hfill ITEP/TH-19/24

\hfill IITP/TH-16/24

\vskip 1.5in
\begin{center}
	
	{\bf\Large Algorithms for representations of quiver Yangian algebras}
	
	\vskip 0.2in
	\renewcommand{\thefootnote}{\alph{footnote}}
	{Dmitry Galakhov$^{2,3,4,}$\footnote[1]{e-mail: galakhov@itep.ru},
		Alexei Gavshin$^{1,2,4,}$\footnote[2]{e-mail: gavshin.an@phystech.edu},	
		Alexei Morozov$^{1,2,3,4,}$\footnote[3]{e-mail: morozov@itep.ru} and Nikita Tselousov$^{1,2,4,}$\footnote[4]{e-mail: tselousov.ns@phystech.edu}}
	\vskip 0.2in
	\renewcommand{\thefootnote}{\roman{footnote}}
	{\small{
			\textit{$^1$MIPT, 141701, Dolgoprudny, Russia}
			\vskip 0 cm
			\textit{$^2$NRC ``Kurchatov Institute'', 123182, Moscow, Russia}
			\vskip 0 cm
			\textit{$^3$IITP RAS, 127051, Moscow, Russia}
			\vskip 0 cm
			\textit{$^4$ITEP, Moscow, Russia}
	}}
\end{center}

\vskip 0.2in
\baselineskip 16pt

\begin{abstract}
	In this note, we aim to review algorithms for constructing crystal representations of quiver Yangians in detail.
	Quiver Yangians are believed to describe an action of the BPS algebra on BPS states in systems of D-branes wrapping toric Calabi-Yau three-folds.
	Crystal modules of these algebras originate from molten crystal models for Donaldson-Thomas invariants of respective three-folds.
	Despite the fact that this subject was originally at the crossroads of algebraic geometry with effective supersymmetric field theories, equivariant toric action simplifies applied calculations drastically.
	So the sole pre-requisite for this algorithm's implementation is linear algebra.
	It can be easily taught to a machine with the help of any symbolic calculation system.
	Moreover, these algorithms may be generalized to toroidal and elliptic algebras and exploited in various numerical experiments with those algebras.
	We illustrate the work of the algorithms in applications to simple cases of $\mathsf{Y}(\mathfrak{sl}_2)$, $\mathsf{Y}(\widehat{\mathfrak{gl}}_{1})$ and $\mathsf{Y}(\widehat{\mathfrak{gl}}_{1|1})$.
\end{abstract}

\ytableausetup{boxsize = 0.5em}
\tableofcontents

\section{Introduction}\label{sec: Introduction}

One of the first definitions of the Yangian algebra was presented by Drinfeld \cite{zbMATH03966417}.
Since then, this mysterious algebra has appeared in various fields of mathematics \cite{zbMATH04091684}, \cite{molev2002yangians}, and physics \cite{Polyakov:1976fu}.
In practice, one could consider Yangians as a deformation/refinement of ordinary Lie algebras.
Yangians originate from symmetries of integrable spin-chains \cite{Faddeev:1996iy, du2021yangian}, and due to celebrated gauge/Bethe correspondence \cite{Nekrasov:2009uh,Nekrasov:2009ui} may represent symmetries of supersymmetric quantum field theories \cite{Beisert:2010jq, Ferro:2011ph}.

In this paper we are interested in quiver Yangians, which appeared in \cite{Li:2020rij} -- 
affine and non-affine \cite{Bao:2023ece}.
These algebras play the role of BPS algebras introduced in \cite{Harvey:1995fq, Harvey:1996gc, Kontsevich:2010px} (see also a recent review in \cite{Harrison:2021gnp}) for BPS states in systems of D-branes wrapping toric Calabi-Yau three-folds.
These quiver Yangians may be classified by quivers describing effective field theories on D-brane volumes.
Quiver classification covers and goes \emph{beyond} canonical affine Dynkin diagram classification.
This fact offers hope for \emph{new} integrable spin-chain models in addition to those based on Lie algebras (even despite certain pessimistic issue observations \cite{Galakhov:2022uyu}).
Alternative description is in the language of Calogero systems \cite{calogero1971solution,MOSER1975197,olshanetsky1976completely} and WLZZ models
\cite{Wang:2022fxr,Mironov:2023pnd}
and involves infinitely many integrable systems, associated with
``rays and cones'' \cite{Mironov:2023zwi, Mironov:2020pcd, Mironov:2023wga}.
It can be further raised \cite{Liu:2023trf,Mironov:2024sbc} to the DIM \cite{1996q.alg.....8002D,miki2007q,Awata:2016riz} level,
where all rays become related by Miki automorphisms \cite{miki1999toroidal}, not so transparent
at Yangian level.

The most interesting issue is representation theory of  quiver Yangians.
It is still not fully/exhaustively developed, and there are different approaches to it.
In this paper we focus on one of them, which leads to representations in terms
of Young-like diagrams and their generalizations, called crystals.
Classification even of this type of representation is still not complete.
String theory allows one to bootstrap \cite{Prochazka:2015deb, Li:2020rij} crystal representations of quiver Yangians.
This construction is based on the molten crystal model \cite{Ooguri:2009ijd,Aganagic:2010qr,Yamazaki:2010fz,Sulkowski:2009rw} for Donaldson-Thomas invariants of toric Calabi-Yau three-folds enumerating BPS D-brane states, so that vectors of a selected basis in the quiver Yangian module are labeled by crystals.
Then a requirement for the Yangian generators to modify these vectors transforming crystals into crystals allows one to bootstrap respective matrix elements in the form of meromorphic functions of Yangian complex parameters $\epsilon_{1,2}$.
We could call this representation a \emph{square-root} quiver Yangian representation.

Another approach to the representation construction \cite{Rapcak:2018nsl,Rapcak:2020ueh,Galakhov:2020vyb} based on application of equivariant  integration over quiver moduli spaces allows one to represent explicit matrix elements as \emph{rational} functions of $\epsilon_{1,2}$.
For systematic computations and numerical experiments the second approach is more preferable as there is no need to keep track of which square-root branch is used in each operation.
On the other hand Duistermaat-Heckman integration formulae \cite{Pestun:2016qko,guillemin2013supersymmetry,LocReview,DTandIM} allows one to shrink the integration process to a neighborhood of fixed points where solely elementary differential geometry and linear algebra methods are required.
All in all the procedure of enumerating quiver Yangian module vectors and calculating generator matrix elements between them becomes exceptionally simple and may be automatized with the help of any symbolic or numeric calculation system.

In this paper we provide a detailed description of these algorithms and illustrate its explicit work with few examples of the following Yangian algebras: $\mathsf{Y}(\fs\fl_2)$, $\mathsf{Y}(\widehat{\fg\fl}_1)$ and $\mathsf{Y}(\widehat{\fg\fl}_{1|1})$.
Respectively, one can extend the algorithm easily to trigonometric (quantum toroidal) and elliptic algebras \cite{Noshita:2021ldl, Galakhov:2021vbo}.
During illustrations we also discuss connections of the Yangian algebra with families of orthogonal polynomials.
The generator action is encoded in the coefficients of raising and lowering operators of the algebra that are rational functions in Fock-like representations.
This provides us with an alternative way to construct families of orthogonal polynomials like Schur, Jack \cite{1996alg.geom.10021N}, Uglov \cite{Uglov:1997ia, Galakhov:2024mbz, Mishnyakov:2024cgl}, Macdonald \cite{macdonald1998symmetric, Mironov:2019exq} and super-Schur \cite{Galakhov:2023mak} polynomials.

This paper is organized as follows. 
In Sec.~\ref{sec: Effective quiver theories for CY}, we review the definition of quiver Yangian algebras and its representations from quivers that correspond to toric Calabi-Yau varieties. 
Sec.~\ref{sec:Algo_Cry} is devoted to a construction algorithm for crystal representations, and in Sec.~\ref{sec:Algo_ME} we describe an algorithm for calculating respective matrix coefficients in those representations. 
As illustrations of the algorithm applications, we discuss $\mathsf{Y}(\fs\fl_{2})$ algebra in Sec.~\ref{sec: Y(sl(2))}, some comments on $\mathsf{Y}(\fs\fl_{n})$ algebras in Sec.~\ref{sec: sl(n)}, and algebras $\mathsf{Y}(\widehat{\fg\fl}_{1})$, $\mathsf{Y}(\widehat{\fg\fl}_{1|1})$ in Sec.~\ref{sec: Y(gl(1))}, and Sec.~\ref{sec: Y(gl(1|1))}, respectively.

\section{Effective quiver gauge theories for toric Calabi-Yau three-folds}\label{sec: Effective quiver theories for CY}

\subsection{Quiver gauge theories}\label{sec: quiver gauge theories}

\begin{figure}[ht!]
	\centering
	\begin{tikzpicture}[scale=0.8, every path/.style={>={Stealth[scale=0.6]}}]
		\begin{scope}[shift={(1,0)}]
			\begin{scope}[rotate=30]
				\draw[postaction={decorate},decoration={markings, 
					mark= at position 0.6 with {\arrow{>}}}] (0,0) to (2.3094,0);
				\draw[postaction={decorate},decoration={markings, 
					mark= at position 0.6 with {\arrow{>}}}] (0,0) to[out=10,in=170] (2.3094,0);
				\draw[postaction={decorate},decoration={markings, 
					mark= at position 0.6 with {\arrow{>}}}] (0,0) to[out=-10,in=-170] (2.3094,0);
			\end{scope}
		\end{scope}
		\begin{scope}[shift={(1,0)}]
			\begin{scope}[rotate=-30]
				\draw[postaction={decorate},decoration={markings, 
					mark= at position 0.6 with {\arrow{>}}}] (0,0) to (2.3094,0);
				\draw[postaction={decorate},decoration={markings, 
					mark= at position 0.6 with {\arrow{>}}}] (0,0) to[out=10,in=170] (2.3094,0);
				\draw[postaction={decorate},decoration={markings, 
					mark= at position 0.6 with {\arrow{>}}}] (0,0) to[out=-10,in=-170] (2.3094,0);
			\end{scope}
		\end{scope}
		\begin{scope}[shift={(5,0)}]
			\begin{scope}[rotate=150]
				\draw[postaction={decorate},decoration={markings, 
					mark= at position 0.6 with {\arrow{>}}}] (0,0) to (2.3094,0);
				\draw[postaction={decorate},decoration={markings, 
					mark= at position 0.6 with {\arrow{>}}}] (0,0) to[out=10,in=170] (2.3094,0);
				\draw[postaction={decorate},decoration={markings, 
					mark= at position 0.6 with {\arrow{>}}}] (0,0) to[out=-10,in=-170] (2.3094,0);
			\end{scope}
		\end{scope}
		\begin{scope}[shift={(5,0)}]
			\begin{scope}[rotate=-150]
				\draw[postaction={decorate},decoration={markings, 
					mark= at position 0.6 with {\arrow{>}}}] (0,0) to (2.3094,0);
				\draw[postaction={decorate},decoration={markings, 
					mark= at position 0.6 with {\arrow{>}}}] (0,0) to[out=10,in=170] (2.3094,0);
				\draw[postaction={decorate},decoration={markings, 
					mark= at position 0.6 with {\arrow{>}}}] (0,0) to[out=-10,in=-170] (2.3094,0);
			\end{scope}
		\end{scope}
		\draw[postaction={decorate},decoration={markings, 
			mark= at position 0.6 with {\arrow{>}}}] (3,1.1547) to[out=-85,in=85] (3,-1.1547);
		\draw[postaction={decorate},decoration={markings, 
			mark= at position 0.6 with {\arrow{>}}}] (3,1.1547) to[out=-75,in=75] (3,-1.1547);
		\draw[postaction={decorate},decoration={markings, 
			mark= at position 0.6 with {\arrow{>}}}] (3,-1.1547) to[out=95,in=-95] (3,1.1547);
		\draw[postaction={decorate},decoration={markings, 
			mark= at position 0.6 with {\arrow{>}}}] (3,-1.1547) to[out=105,in=-105] (3,1.1547);
		\draw[fill=\myblue] (3,1.1547) circle (0.1);
		\draw[fill=\myblue] (3,-1.1547) circle (0.1);
		\foreach \x in {0}
		{ 
			\begin{scope}[rotate = \x * 36]
				\draw[postaction={decorate},decoration={markings, 
					mark= at position 0.6 with {\arrow{>}}}] (1.05,0) -- (0.849468, 0.617175);
				\draw[postaction={decorate},decoration={markings, 
					mark= at position 0.6 with {\arrow{>}}}] (0.768566, 0.558396) -- (0.95,0);
			\end{scope}
		}
		\foreach \x in {1,...,9}
		{ 
			\begin{scope}[rotate = \x * 36]
				\draw[postaction={decorate},decoration={markings, 
					mark= at position 0.6 with {\arrow{>}}}] (1.05,0) -- (0.849468, 0.617175);
				\draw[postaction={decorate},decoration={markings, 
					mark= at position 0.6 with {\arrow{>}}}] (0.768566, 0.558396) -- (0.95,0);
				\draw[postaction={decorate},decoration={markings, 
					mark= at position 0.8 with {\arrow{>}}}] (1.05,0) to [out=315,in=270] (1.5,0) to[out=90,in=45] (1.05,0);
			\end{scope}
		}
		\foreach \x in {0,...,9}
		{ 
			\begin{scope}[rotate = \x * 36]
				\draw[postaction={decorate},decoration={markings, 
					mark= at position 0.6 with {\arrow{>}}}] (0,0) -- (1,0);
				\draw[fill=\myblue] (1,0) circle (0.1);
			\end{scope}
		}
		\draw[fill=burgundy] (-0.1,-0.1) -- (-0.1,0.1) -- (0.1,0.1) -- (0.1,-0.1) -- cycle;
		\begin{scope}[shift={(6,0)}]
			\foreach \x in {0,1,2,3,4,6,7,8,9}
			{ 
				\begin{scope}[rotate = \x * 36]
					\draw[postaction={decorate},decoration={markings, 
						mark= at position 0.6 with {\arrow{>}}}] (1.05,0) -- (0.849468, 0.617175);
					\draw[postaction={decorate},decoration={markings, 
						mark= at position 0.6 with {\arrow{>}}}] (0.768566, 0.558396) -- (0.95,0);
					\draw[postaction={decorate},decoration={markings, 
						mark= at position 0.8 with {\arrow{>}}}] (1.05,0) to [out=315,in=270] (1.5,0) to[out=90,in=45] (1.05,0);
				\end{scope}
			}
			\foreach \x in {5}
			{ 
				\begin{scope}[rotate = \x * 36]
					\draw[postaction={decorate},decoration={markings, 
						mark= at position 0.6 with {\arrow{>}}}] (1.05,0) -- (0.849468, 0.617175);
					\draw[postaction={decorate},decoration={markings, 
						mark= at position 0.6 with {\arrow{>}}}] (0.768566, 0.558396) -- (0.95,0);
				\end{scope}
			}
			\foreach \x in {0,...,9}
			{ 
				\begin{scope}[rotate = \x * 36]
					\draw[postaction={decorate},decoration={markings, 
						mark= at position 0.6 with {\arrow{>}}}] (0,0) -- (1,0);
					\draw[fill=\myblue] (1,0) circle (0.1);
				\end{scope}
			}
			\draw[fill=burgundy] (-0.1,-0.1) -- (-0.1,0.1) -- (0.1,0.1) -- (0.1,-0.1) -- cycle;
		\end{scope}
	\end{tikzpicture}
	\caption{Example of a quiver}\label{fig:Quiv_exmpl}
\end{figure}
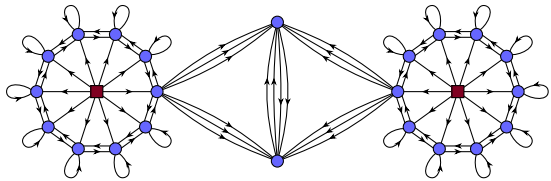

To describe D-branes on Calabi-Yau three-folds, we exploit one of the effective descriptions \cite{Douglas:1996sw} for this system in terms of the effective 4d $\CN=1$ SYM theory.
For simplicity, this theory is further compactified to $\CN=4$ matrix supersymmetric quantum mechanics (SQM).
To fix an effective Lagrangian theory, it suffices to describe the gauge-matter content of this theory.
A simple way to do so is to use a pictorial diagram of a \emph{quiver} -- a graph with all edges oriented.

In general, the theory is uniquely defined by what we call \emph{quiver data}: a quiver diagram $Q$, a dimension vector $\vec d$, a vector of stability parameters $\vec \zeta$, a holomorphic superpotential function $W$, complex equivariant (weight) parameters $h_{I}$, and R-charges.
Let us describe all these elements in detail.

We start with a quiver $Q$, and denote the set of quiver nodes as $Q_0$ and the set of arrows as $Q_1$.
Also, we introduce some helpful notations for the following quantities:
\begin{itemize}
	\item $\{a\to b\}$ -- a set of arrows flowing from node $a$ to node $b$.
	\item $|a\to b|$ -- a number of arrows flowing from node $a$ to node $b$.
\end{itemize}
If we needed to specify the head and the tail of some arrow $I$ we would denote it as $I:a\to b$.

Quiver nodes may be divided into gauge nodes and framing nodes.
We distinguish them pictorially by their shapes: the gauge nodes are denoted as round nodes, whereas framing nodes are denoted as square nodes (see Fig.~\ref{fig:Quiv_exmpl}).
To these two types of nodes, we associate a gauge or a flavor group $U(d_a)$; group ranks $d_a\geq 0$ form the vector of quiver dimensions $\vec d$.
On the level of the gauge theory, these are dynamical or static 4d $\CN=1$ vector multiplets dimensionally reduced to 1d $\CN=4$, where we extract a complex scalar $\phi$:
\begin{equation}
	(A_{\mu},\lambda_{\alpha},D)\rightsquigarrow (A_0,A_3,\Phi=A_1+i A_2,\lambda_{\alpha},D)\,.
\end{equation}
To introduce the flavor action for the framing nodes, we simply assume that all these fields are non-dynamical and are equal to zero except $\Phi$ parameterizing flavor fugacity. 
For gauge nodes, one could introduce Fayet-Illiopolous coupling $\zeta_a\Tr D_a$, $a\in Q_0$ where $\zeta_a$ are real valued parameters also called stability parameters.
Together, they form a vector of stability parameters $\vec\zeta$.

Quiver arrows $Q_1$ correspond to a bi-fundamentally charged 4d $\CN=1$ chiral multiplet also compactified to 1d $\CN=4$:
\begin{equation}
	(q_{I},\psi_{I,\alpha},F_{I})\,,
\end{equation}
so that the field corresponding to an arrow $I\in\{a\to b\}$ is charged fundamentally with respect to $U(d_b)$ and anti-fundamentally with respect to $U(d_a)$.
Furthermore, each arrow field we couple additionally to $U(1)$ flavor symmetry with $\Phi=h_I$, where $h_I\in \IC$.
We call these parameters (equivariant) weights of the arrow fields.
In general, the equivariant weights and R-charges of the chiral fields are unconstrained.

Finally, the 4d $\CN=1$ SYM theory and, respectively, $\CN=4$ SQM, admit a potential term defined by a holomorphic in fields $q_I$ function $W$ called \emph{superpotential}.
We impose certain constraints on $W$.
The superpotential may be decomposed into a sum of monomials in $q_I$.
The gauge invariance requires ordered multipliers $q_I$ in such a monomial respective arrows $I\in Q_1$ to form a closed loop.
The superpotential must be flavor invariant as well, and its R-charge is +2. This imposes loop constraints on the weights and R-charges of fields entering each superpotential monomial:
\begin{equation}\label{loop}
	\sum\lm_{I\in{\rm loop}}h_I=0,\quad \sum\lm_{I\in{\rm loop}}R_I=2\,.
\end{equation}

We are interested in the classical vacua of this theory, which is described by values of all fields minimizing the potential modulo gauge transformations.

The potential contains three contributions: D-term, F-term and equivariant vector field:
\begin{equation}
	U=\sum\lm_{a\in Q_0}\Tr|D_a|^2+\sum\lm_{I\in Q_1}\Tr|F_I|^2+\sum\lm_{I\in Q_1} |v_I|^2\,,
\end{equation}
where
\begin{equation}\label{vacuum}
	\begin{aligned}
		& D_a=\sum\lm_{b\in Q_0}\sum\lm_{I\in\{b\to a\}}q_Iq_I^{\dagger}-\sum\lm_{b\in Q_0}\sum\lm_{J\in\{a\to b\}}q_J^{\dagger}q_J-\zeta_a \bbone_{d_a\times d_a}\,,\\
		& F_I=\p_{q_I}W\,,\\
		& v_{I:a\to b}=\Tr(\Phi_b q_I-q_I\Phi_a-h_I q_I)\frac{\p}{\p q_I}\,.
	\end{aligned}
\end{equation}
We expect that in the classical vacuum only fields $\Phi_a$  from the gauge multiplets and fields $q_I$ from the chiral multiplets acquire expectation values that one could define from the $U$-minimization constraint.

\subsection{Toric quivers for toric Calabi-Yau three-folds} \label{sec:toric_quiver}

\begin{figure}[ht!]
	\centering
	\begin{tikzpicture}
		\tikzset{arr/.style={ 
				postaction={decorate},
				decoration={markings, mark= at position 0.6 with {\arrow{stealth}}},
				thick}}
		\draw[thick, burgundy, postaction={decorate},decoration={markings, 
			mark= at position 0.7 with {\arrow{stealth}}}] (0,0) to[out=60,in=0] (0,1) to[out=180,in=120] (0,0);
		\node[above] at (0,1) {$\scriptstyle B_1$};
		\begin{scope}[rotate=120]
			\draw[thick, \myblue, postaction={decorate},decoration={markings, 
				mark= at position 0.7 with {\arrow{stealth}}}] (0,0) to[out=60,in=0] (0,1) to[out=180,in=120] (0,0);
			\node[below left] at (0,1) {$\scriptstyle B_2$};
		\end{scope}
		\begin{scope}[rotate=240]
			\draw[thick, black!70!green, postaction={decorate},decoration={markings, 
				mark= at position 0.7 with {\arrow{stealth}}}] (0,0) to[out=60,in=0] (0,1) to[out=180,in=120] (0,0);
			\node[below right] at (0,1) {$\scriptstyle B_3$};
		\end{scope}
		\draw[fill=gray] (0,0) circle (0.08);
		\node[below] at (1.5,-1) {$W=(\Delta_1)-(\Delta_2)=\Tr \,B_1\left[B_2,B_3\right]$};
		\begin{scope}[shift={(5,0)}]
			\draw[fill=white!70!orange] (0,0) -- (0.866025, 0.5) -- (0.866025, -0.5) -- cycle;
			\node at (0.55,0) {$\scriptstyle \Delta_1$};
			\begin{scope}[shift={(1.73205,0)}]
				\draw[fill=white!70!green] (0,0) -- (-0.866025, 0.5) -- (-0.866025, -0.5) -- cycle;
				\node at (-0.55,0) {$\scriptstyle \Delta_2$};
			\end{scope}
			\draw[arr,burgundy] (0,0) -- (0.866025, 0.5);
			\draw[arr,black!70!green] (0,1) -- (0,0);
			\draw[arr,\myblue] (0,0) -- (-0.866025, 0.5);
			\draw[arr,burgundy] (-0.866025, -0.5) -- (0,0);
			\draw[arr,black!70!green] (0,0) -- (0,-1);
			\draw[arr,\myblue] (0.866025, -0.5) -- (0,0);
			\draw[arr,\myblue] (0.866025, 0.5) -- (0,1);
			\draw[arr,\myblue] (0,-1) -- (-0.866025, -0.5);
			\draw[arr,black!70!green] (0.866025, 0.5) -- (0.866025, -0.5);
			\draw[arr,black!70!green] (-0.866025, 0.5) -- (-0.866025, -0.5);
			\draw[arr,burgundy] (0,-1) -- (0.866025, -0.5);
			\draw[arr,burgundy] (-0.866025, 0.5) -- (0,1);
			\begin{scope}[shift={(1.73205,0)}]
				\draw[arr,burgundy] (0,0) -- (0.866025, 0.5);
				\draw[arr,black!70!green] (0,1) -- (0,0);
				\draw[arr,\myblue] (0,0) -- (-0.866025, 0.5);
				\draw[arr,burgundy] (-0.866025, -0.5) -- (0,0);
				\draw[arr,black!70!green] (0,0) -- (0,-1);
				\draw[arr,\myblue] (0.866025, -0.5) -- (0,0);
				\draw[arr,\myblue] (0.866025, 0.5) -- (0,1);
				\draw[arr,\myblue] (0,-1) -- (-0.866025, -0.5);
				\draw[arr,black!70!green] (0.866025, 0.5) -- (0.866025, -0.5);
				\draw[arr,burgundy] (0,-1) -- (0.866025, -0.5);
				\draw[arr,burgundy] (-0.866025, 0.5) -- (0,1);
			\end{scope}
			\draw[fill=gray] (0,0) circle (0.08) (0.866025, 0.5) circle (0.08) (0.866025, -0.5) circle (0.08) (-0.866025, 0.5) circle (0.08) (-0.866025, -0.5) circle (0.08) (0,1) circle (0.08) (0,-1) circle (0.08);
			\begin{scope}[shift={(1.73205,0)}]
				\draw[fill=gray] (0,0) circle (0.08) (0.866025, 0.5) circle (0.08) (0.866025, -0.5) circle (0.08) (0,1) circle (0.08) (0,-1) circle (0.08);
			\end{scope}
		\end{scope}
		\draw[<->] (1.5,0) -- (3.5,0);
		\draw[<->] (8,0) -- (9.3,0);
		\begin{scope}[shift={(11,0)}]
		\begin{scope}[scale=0.3]
			\draw[dashed,burgundy,thick] (-1.02193,-0.105365) to[out=-22.2158,in=106.049] (-0.802668,-0.370446) to[out=-73.951,in=72.143] (-0.893237,-1.62267) to[out=-107.857,in=54.0665] (-1.31348,-2.51262) to[out=-125.933,in=346.667] (-1.97807,-2.94886);
			\draw[burgundy, thick, postaction={decorate},decoration={markings, 
				mark= at position 0.75 with {\arrow{stealth}}}] (-1.97807,-2.94886) to[out=157.784,in=276.205] (-2.23672,-2.50453) to[out=96.2049,in=252.143] (-2.10676,-1.43155) to[out=72.143,in=224.046] (-1.5,-0.313585) to[out=44.0455,in=166.667] (-1.02193,-0.105365);
			\draw[dashed, thick, \myblue] (4.49901,-0.0553934) to[out=91.0207,in=299.793] (4.30794,0.764499) to[out=119.793,in=328.494] (3.34415,1.76987) to[out=148.494,in=337.329] (2.7207,2.10685) to[out=157.329,in=351.785] (1.21014,2.5476) to[out=171.785,in=2.4716] (-0.18844,2.64271) to[out=-177.528,in=11.1893] (-1.30064,2.53214) to[out=-168.811,in=29.4085] (-3.01109,1.96564) to[out=-150.592,in=52.0844] (-4.0307,1.17608) to[out=-127.916,in=84.9049] (-4.49901,0.0553934);
			\draw[\myblue, thick, postaction={decorate},decoration={markings, 
				mark= at position 0.75 with {\arrow{stealth}}}] (-4.49901,0.0553934) to[out=-88.9793,in=119.793] (-4.30794,-0.764499) to[out=-60.2067,in=148.494] (-3.34415,-1.76987) to[out=-31.5057,in=157.329] (-2.7207,-2.10685) to[out=-22.6711,in=171.785] (-1.21014,-2.5476) to[out=-8.21525,in=182.472] (0.18844,-2.64271) to[out=2.4716,in=191.189] (1.30064,-2.53214) to[out=11.1893,in=209.408] (3.01109,-1.96564) to[out=29.4085,in=232.084] (4.0307,-1.17608) to[out=52.0844,in=264.905] (4.49901,-0.0553934);
			\draw[black!70!green, dashed, thick] (-3.78098,-1.81208) to[out=126.741,in=283.437] (-3.89513,-1.62499) to[out=103.437,in=259.101] (-3.92043,-1.42511) to[out=79.1007,in=239.534] (-3.86041,-1.22025) to[out=59.5342,in=225.774] (-3.72181,-1.0176) to[out=45.7736,in=216.212] (-3.51408,-0.823297) to[out=36.2119,in=204.231] (-2.9389,-0.477807) to[out=24.2309,in=197.387] (-2.23984,-0.204643) to[out=17.3873,in=193.794] (-1.52045,-0.000738347) to[out=13.7938,in=193.299] (-0.712434,0.190969);
			\draw[black!70!green, thick, postaction={decorate},decoration={markings, 
				mark= at position 0.65 with {\arrow{stealth}}}] (-0.712434,0.190969) to[out=13.7258,in=201.029] (1.25809,1.03284) to[out=21.0289,in=165.609] (2.01524,1.13539) to[out=-14.3915,in=103.426] (2.56769,0.666454) to[out=-76.5742,in=53.3254] (2.31755,-0.35877) to[out=-126.675,in=27.0056] (0.934744,-1.51178) to[out=-152.994,in=17.1527] (-0.0903552,-1.9508) to[out=-162.847,in=5.13454] (-1.47116,-2.25577) to[out=-174.865,in=353.797] (-2.48862,-2.27785) to[out=173.797,in=316.395] (-3.78098,-1.81208);
			\draw[thick] (-4.5, 0) to[out=-89.2907,in=105.179] (-4.42408,-0.617196) to[out=-74.8209,in=117.767] (-4.21596,-1.17591) to[out=-62.2326,in=132.059] (-3.79285,-1.79865) to[out=-47.9408,in=149.235] (-2.97356,-2.4843) to[out=-30.7655,in=161.238] (-2.08097,-2.91289) to[out=-18.7621,in=173.685] (-0.782795,-3.21655) to[out=-6.31531,in=181.511] (0.147084,-3.26172) to[out=1.51116,in=192.489] (1.27968,-3.13638) to[out=12.4889,in=201.002] (2.08097,-2.91289) to[out=21.0019,in=212.778] (2.97356,-2.4843) to[out=32.7783,in=229.493] (3.79285,-1.79865) to[out=49.4928,in=244.291] (4.21596,-1.17591) to[out=64.2912,in=257.005] (4.42408,-0.617196) to[out=77.0047,in=269.291] (4.5,0);
			\draw[thick] (4.5,0) to[out=90.7126,in=285.194] (4.42389,0.617974) to[out=105.194,in=297.782] (4.21563,1.17656) to[out=117.782,in=312.072] (3.79238,1.79918) to[out=132.072,in=329.253] (2.97257,2.48492) to[out=149.253,in=341.254] (2.07952,2.91341) to[out=161.254,in=349.109] (1.27826,3.13667) to[out=169.109,in=359.522] (0.0645347,3.26304) to[out=179.522,in=12.4749] (-1.27826,3.13667) to[out=-167.525,in=20.9848] (-2.07952,2.91341) to[out=-159.015,in=32.7627] (-2.97257,2.48492) to[out=-147.237,in=49.4803] (-3.79238,1.79918) to[out=-130.52,in=64.276] (-4.21563,1.17656) to[out=-115.724,in=76.9875] (-4.42389,0.617974) to[out=-103.013,in=89.2874] (-4.5,0);
			\draw[thick] (-1.73408, 0.219328) to[out=-25.5642,in=154.436] (-1.28303,0.00356717) to[out=-25.5642,in=164.371] (-0.936696,-0.132695) to[out=-15.6286,in=168.389] (-0.722834,-0.187921) to[out=-11.6113,in=173.568] (-0.434561,-0.236794) to[out=-6.43187,in=175.879] (-0.277753,-0.252593) to[out=-4.12116,in=181.176] (0.0495026,-0.263016) to[out=1.17554,in=187.703] (0.434561,-0.236794) to[out=7.70346,in=193.377] (0.722834,-0.187921) to[out=13.3772,in=197.959] (0.936696,-0.132695) to[out=17.9589,in=205.564] (1.28303,0.00356717) to[out=25.5642,in=-154.436] (1.73408, 0.219328);
			\draw[thick] (1.28132,-0.00269801) to[out=154.501,in=344.421] (0.93446,0.133369) to[out=164.421,in=348.435] (0.720476,0.188434) to[out=168.435,in=353.618] (0.431659,0.237152) to[out=173.618,in=355.929] (0.27468,0.252831) to[out=175.929,in=1.09512] (-0.0451912,0.263072) to[out=-178.905,in=7.64423] (-0.431045,0.237228) to[out=-172.356,in=13.3201] (-0.719977,0.188542) to[out=-166.68,in=17.8946] (-0.933988,0.133511) to[out=-162.105,in=25.4851] (-1.28096,-0.00251451);
			\draw[fill=gray] (-2.25,-2.29067) circle (0.27);
		\end{scope}
		\end{scope}
	\end{tikzpicture}
	\caption{Quiver on a torus for $\IC^3$}\label{fig:quiv_torus}
\end{figure}

We refer the reader to \cite{Hanany:2005ve,Franco:2005rj,Franco:2005sm,Hanany:2005ss,Kennaway:2007tq,Yamazaki:2008bt} for an extensive review of quiver construction for the toric Calabi-Yau three-folds (CY${}_3$) via brane tiling mechanism.
Here we remind solely aspects relevant for our construction.

In this case quiver diagram $Q$ and superpotential $W$ in unframed quiver data (when the framing nodes, respective arrows and superpotential terms are eliminated) may be identified with a quiver on a torus $\hat Q$, or, equivalently, periodic quiver lattice $\CL$ in $\IR^2$ covering this torus (see Fig.~\ref{fig:quiv_torus}).

This correspondence goes as follows.
The quiver corresponds to the quiver on a torus as an abstract graph.
The superpotential is constructed as a signed sum over monomials, where the monomials correspond to loops lying on face boundaries and the sign is defined by the boundary orientation (see Fig.\ref{fig:quiv_torus}):
\begin{equation}\label{CY_sup}
	W=\sum\lm_{\Delta\in{\rm faces}}\prod\lm_{I\in\p\Delta}^{\curvearrowleft}q_I\,.
\end{equation}

In this setting the loop weight constraint \eqref{loop} could be resolved in general \cite{Li:2020rij}: all the arrow weights take values in a 2d lattice we parameterize by canonical weights $\epsilon_{1,2}$:
\begin{equation}
	h_I=x_I\epsilon_1+y_I\epsilon_2,\quad x_I,y_I\in\IZ\,,
\end{equation}
moreover vector $(a_I,b_I)$ is identified with an arrow $I$ vector in $\CL$ (see Fig.~\ref{fig:quiv_torus}).

The torus structure imposes specific relations on paths in $\CL$.
Let us consider a generic path $\wp\in\CL$.
Following this path we pass a sequence of arrows $I_1,\,I_2,\ldots,I_n$.
Let us consider a product of chiral field matrices in an order imposed by $\wp$:
\begin{equation}
	q_{\wp}=q_{I_n}\cdot\ldots\cdot q_{I_2}\cdot q_{I_1}\,.
\end{equation} 

If calculated on vacuum solution to \eqref{vacuum} $q_{\wp}=q_{\wp'}$ for $\wp$ and $\wp'$ being homotopic equivalent.
This allows us to classify uniquely all such operator expectation values by its complex weight and R-charge: 
\begin{equation}
	\begin{aligned}
	&{\rm class}(\wp)=(x_{\wp},y_{\wp},R_{\wp})\,,\\
	&x_\wp=x_{I_1}+x_{I_2}+\ldots+x_{I_n}, \quad y_\wp=y_{I_1}+y_{I_2}+\ldots+y_{I_n},\quad R_\wp=R_{I_1}+R_{I_2}+\ldots+R_{I_n}\,.
	\end{aligned}
\end{equation}
Let us note that the R-charge for path $\wp$ is twice the number of loops $\wp$ encircles in $\CL$ due to the constraint for the R-charge \eqref{loop}.

\subsection{Molten crystals and Donaldson-Thomas invariants}\label{sec:molten}

We will narrow the family of quiver data we consider further.
We assume that the framing node in our quiver $Q$ (let us denote it as $\ff$) is single and has $d_\ff=1$.
We assume that all the stability parameters $\zeta_a>0$ (this constraint corresponds to a so called \emph{cyclic} chamber in the parameter space).
One of the gauge quiver nodes is chosen; let us call it a \emph{root} node $r$, then we choose a canonical arrow $R$ flowing from $\ff$ to $r$:
\begin{equation}
	\{\ff\to a\}=\left\{\begin{array}{ll} 
	\{R\},& \mbox{ if } a=r\,,\\
	\varnothing, &\mbox{ otherwise}\,. 
	\end{array}\right.
\end{equation}
And we do not impose constraints on arrows flowing towards $\ff$ yet.

In this case, counting the classical vacua coincides with enumerating Donaldson-Thomas invariants by molten crystals.
Moreover, one is able to reconstruct explicit expectation values of fields $\Phi_a$ and $q_I$ from geometric data of a crystal.

The first molten crystal counting appeared in relation to topological strings on $\IC^3$ in \cite{Okounkov:2003sp} and has been developed intensively afterwards \cite{Ooguri:2009ijd, Szendroi:2007nu,Jafferis:2008uf,Dimofte:2009bv,Nagao:2009rq,Aganagic:2009cg,Nishinaka:2010fh,Ooguri:2010yk} to describe respective Donaldson-Thomas invariants.
We are using this model for counting \emph{fixed points} on quiver representation moduli spaces, which is a mathematical code name for vacuum solutions to $D=F=v=0$ in \eqref{vacuum}.
However, in practice, it turns out to be much more convenient to work with complex algebraic varieties, in particular, in view of exploiting equivariant integration techniques.
To perform this translation, we use  Narasimhan–Seshadri-Kobayashi–Hitchin (NSKH) correspondence (see \cite{DUY,king1994moduli,NakajimaALE}), allowing one to trade the real moment map and the unitary gauge group to a stability condition and a complexified gauge group:
\begin{equation}\label{NSKH}
	\left\{\begin{array}{c}
		D_a=0\,;\\
		F_I=0\,;\\
		v_I=0\,;
	\end{array}\right\}/\prod\lm_{a\in Q_0}U(d_a)\quad\Longleftrightarrow\quad\left\{\begin{array}{c}
	\mbox{{stable}};\\
	F_I=0\,;\\
	v_I=0\,;
	\end{array}\right\}/\prod\lm_{a\in Q_0}GL(d_a,\IC)\,.
\end{equation}

We will not unwrap the definition of stability here.
Instead, we simply switch to the construction of crystals that turn out to count stable representations in the cyclic chamber we use in what follows.

Since under our assumptions $d_\ff=1$ map $R$ is a morphism from complex numbers to $\IC^{d_r}$, in other words, it is a vector.
Chamber $\zeta_a>0$ is called \emph{cyclic}.
In this case, vector $R$ is \emph{cyclic} as well, implying that all the vectors of spaces $V_a:=\IC^{d_a}$, such that chiral fields $q_I$ are matrices of linear transformations between those $q_{I:a\to b}\in{\rm Hom}(V_a,V_b)$, can be constructed as a sequence of morphisms $q_I$ acting on $R$:
\begin{equation}\label{cyc}
	V_a={\rm Span}\left\{q_{I_n}\cdot\ldots\cdot q_{I_2}\cdot q_{I_1}\cdot R \right\}\,.
\end{equation}

As we have already mentioned in Sec.~\ref{sec:toric_quiver}, all the monomial words in $q_I$ are characterized uniquely by a triplet of numbers and may be associated with paths in $\CL$.
Let us assign to vector $R$ position $(0,0,0)$ in the 3d space, then all the other vectors in $\bigoplus_a V_a$ are characterized by points $(x_{\wp},y_{\wp},R_{\wp})$ in the same space.
We call these points \emph{atoms}, and a union of atoms representing vectors of a stable quiver representation is called a \emph{crystal}\footnote{We should emphasize that traditionally, one calls a molten crystal a \emph{complement} to our crystal in a lattice formed by all the possible words in $q_I$.
We believe our definition is more suitable for our properties.}.

F-term constraints translated into equality $q_{\wp}=q_{\wp'}$ for homotopic paths in $\CL$ is transformed further in the crystal language to a melting rule:
\begin{tcolorbox}
	If for crystal $\Lambda$ there is any atom at position $(x,y,R)$ and an atom at position $(x',y',R')\in\Lambda$ such that $(x',y',R')=(x+x_I,y+y_I,R+R_I)$ then $(x,y,R)\in \Lambda$.
\end{tcolorbox}
However, in practice, we are planning to check F-term relations directly.

As well, we introduce a \emph{color} characteristic for atoms.
If atom $\Box\in\Lambda$ corresponds to a sequence of arrows $R, I_1,\, I_2,\ldots,I_n$ then color $a\in Q_0$ of $\Box$ is defined as a node corresponding to the head of arrow $I_n$.
If we need to stress an atom color, we use the following notation $\sqbox{$a$}$.

Finally, having a crystal $\Lambda$, we describe how one constructs a fixed point.
First of all, we enumerate vectors of $V_a$:
\begin{equation}
	V_a=\bigoplus_{\ssqbox{$a$}\in \Lambda}\IC|\sqbox{$a$}\rangle, \quad a\in Q_0\,.
\end{equation} 
Respectively, we define numbers $d_a$ as numbers of atoms of color $a$.

Then we construct expectation values of fields $\Phi_a$:
\begin{equation}\label{Phi}
	\Phi_a={\rm diag}\left\{\phi_{\ssqbox{$a$}_1},\phi_{\ssqbox{$a$}_2},\ldots,\phi_{\ssqbox{$a$}_{d_a}}\right\},\quad \phi_{\ssqbox{$a$}_i}=x_{\ssqbox{$a$}_i}\epsilon_1+y_{\ssqbox{$a$}_i}\epsilon_2\,.
\end{equation}
Physically, the eigen expectation value $\phi_{\Box}$ corresponds to the weight of the operator atom $\Box$ represents.

Finally, we define explicitly the matrices of $q_I$ in fixed points ($i=1,\ldots,d_b$, $j=1,\ldots,d_a$):
\begin{equation}\label{morphisms}
	\left(q_{I:a\to b}\right)_{ij}=\left\{\begin{array}{ll}
		1, & \mbox{ if } (x_{\ssqbox{$a$}_j}+x_I,y_{\ssqbox{$a$}_j}+y_I,R_{\ssqbox{$a$}_j}+R_I)=(x_{\ssqbox{$b$}_i},y_{\ssqbox{$b$}_i},R_{\ssqbox{$b$}_i})\,;\\
		0, &\mbox{ otherwise}\,.
	\end{array}\right.
\end{equation}
Let us stress here that we have constructed a point on the complex variety -- an element in an orbit in the r.h.s. of \eqref{NSKH}.
Matrix elements $(q_I)_{ij}$ \emph{do not} solve equations $D_a=0$.
However, as the NSKH correspondence states, they represent a single point in a complexified gauge group orbit containing the orbit in the l.h.s. of \eqref{NSKH}.
So we acquire only counting of vacuum solutions.
To acquire an explicit solution to vacuum equations \eqref{vacuum} one has to allow the ``imaginary'' part of the complexified gauge group to \emph{rescale} absolute values of non-zero matrix elements $|(q_I)_{ij}|$.

\subsection{Quiver Yangians and its crystal representations} \label{sec:cry_rep}

The quiver Yangian algebra is defined in the following way \cite{Li:2020rij}.
Consider quiver data for the unframed quiver only.
To each node $a\in Q_0$, we associate a triplet of generator series joined in generating functions:
\begin{equation}
	e^{(a)}(z)=\sum\lm_{n=0}^{\infty}\frac{e_n^{(a)}}{z^{n+1}},\quad f^{(a)}(z)=\sum\lm_{n=0}^{\infty}\frac{f_n^{(a)}}{z^{n+1}},\quad \psi^{(a)}(z)=\sum\lm_{n=-\fs_a}^{\infty}\frac{\psi_n^{(a)}}{z^{n+1}}\,.
\end{equation}
Quantities $\fs_a$ are called shifts \cite{Galakhov:2021xum,Noshita:2021dgj,Kodera:2016faj}.
This is a super algebra.
Generators $e^{(a)}(z)$ and $f^{(a)}(z)$ have parity:
\begin{equation}
	|a|=(|a\to a|+1)\;{\rm mod}\;2\,.
\end{equation}
Generators $\psi^{(a)}(z)$ always have parity 0.

Quiver Yangian is defined with the help of a function called a bond factor constructed according to the weights of arrows of the unframed quiver:
\begin{equation}\label{bonding}
	\varphi_{a,b}(z)=\frac{\prod\lm_{I\in\{a\to b\}}(z+h_I)}{\prod\lm_{J\in\{b\to a\}}(z-h_J)}\,.
\end{equation}
In a compact form, one can rewrite the system of generator relations for the respective generating functions in the following way:
\begin{equation}\label{Yangian}
	\begin{aligned}
		&e^{(a)}(z)e^{(b)}(w)\simeq (-1)^{|a||b|}\varphi_{a,b}(z-w)\,e^{(b)}(w)e^{(a)}(z)\,,\\
		&\psi^{(a)}(z)e^{(b)}(w)\simeq \varphi_{a,b}(z-w)e^{(b)}(w)\,\psi^{(a)}(z)\,,\\
		&f^{(a)}(z)f^{(b)}(w)\simeq (-1)^{|a||b|}\varphi_{a,b}(z-w)^{-1}f^{(b)}(w)f^{(a)}(z)\,,\\
		&\psi^{(a)}(z)f^{(b)}(w)\simeq \varphi_{a,b}(z-w)^{-1}f^{(b)}(w)\psi^{(a)}(z)\,,\\
		&\psi^{(a)}(z)\psi^{(b)}(w)=\psi^{(b)}(w)\psi^{(a)}(z)\,,\\
		&\left[e^{(a)}(z),f^{(b)}(w)\right\}\simeq-\delta_{ab}\frac{\psi^{(a)}(z)-\psi^{(b)}(w)}{z-w}\,,
	\end{aligned}
\end{equation}
where $[*,*\}$ denotes a super-commutator, and $\simeq$ denotes an equivalence of series up to $z^{n}w^{k\geq 0}$ and $z^{n\geq 0}w^{k}$.
An unfolded form of these relations in terms of modes could be found in \cite{Li:2020rij}.

In addition to these quadratic relations, one considers a set of higher order relations \cite{Negut:2022pka}, sometimes called Serre relations.

This definition incorporates a natural question of whether the choice of the superpotential $W$ in the quiver data affects the quiver Yangian at all.
Indeed, there is no direct dependence of \eqref{Yangian} on $W$.
However, we should stress that a choice of concrete $W$ imposes loop constraints \eqref{loop} on weights $h_I$ explicitly appearing in \eqref{bonding}.
This forces, for example, all the quiver Yangians constructed from toric Calabi-Yau three-fold quivers (with fixed superpotentials \eqref{CY_sup}) to be two-parametric families of algebras depending on $\epsilon_{1,2}$ parameterizing the equivariant toric action.

Quiver Yangians admit crystal representations.
Vectors in these modules are labeled by the crystals we described in Sec.~\ref{sec:molten}.
The action of the quiver Yangian generators reads in this case:
\begin{equation}\label{rep}
	\begin{aligned}
	&e^{(a)}(z)|\Lambda\rangle=\sum\lm_{\ssqbox{$a$}\in \Lambda^+}\frac{{\bf E}_{\Lambda,\Lambda+\ssqbox{$a$}}}{z-\phi_{\ssqbox{$a$}}}|\Lambda+\sqbox{$a$}\rangle\,,\\
	&f^{(a)}(z)|\Lambda\rangle=\sum\lm_{\ssqbox{$a$}\in \Lambda^-}\frac{{\bf F}_{\Lambda,\Lambda-\ssqbox{$a$}}}{z-\phi_{\ssqbox{$a$}}}|\Lambda-\sqbox{$a$}\rangle\,,\\
	&\psi^{(a)}(z)|\Lambda\rangle=\bPsi_{\Lambda}^{(a)}(z)|\Lambda\rangle\,,
	\end{aligned}
\end{equation}
where $\Lambda^{\pm}$ are sets of atoms that can be added (removed) to a crystal $\Lambda$ so that a new set of atoms $\Lambda\pm\Box$ is also a crystal.

It is easy to show that for \eqref{rep} to be a representation of quiver Yangian \eqref{Yangian} matrix elements have to satisfy a set of \emph{hysteresis} relations:
\begin{equation}\label{general hysteresis}
\begin{aligned}
	&\bPsi^{(a)}_{\Lambda}(z)=\prod\lm_{I\in\{a\to a\}}\left(-\frac{1}{h_I}\right)\times\frac{\prod\lm_{K\in\{a\to\ff\}}(-z-h_K)}{\prod\lm_{J\in\{\ff\to a\}}(z-h_I)}\times \prod\lm_{b\in Q_0}\prod\lm_{\ssqbox{$b$}\in \Lambda}\varphi_{a,b}(z-\phi_{\ssqbox{$b$}})\,,\\
	&{\bf E}_{\Lambda+\ssqbox{$a$},\Lambda+\ssqbox{$a$}+\ssqbox{$b$}}{\bf F}_{\Lambda+\ssqbox{$a$}+\ssqbox{$b$},\Lambda+\ssqbox{$b$}}=(-1)^{|a||b|}{\bf F}_{\Lambda+\ssqbox{$a$},\Lambda}{\bf E}_{\Lambda,\Lambda+\ssqbox{$b$}}\,,\\
	&\frac{{\bf E}_{\Lambda,\Lambda+\ssqbox{$a$}}{\bf E}_{\Lambda+\ssqbox{$a$},\Lambda+\ssqbox{$a$}+\ssqbox{$b$}}}{{\bf E}_{\Lambda,\Lambda+\ssqbox{$b$}}{\bf E}_{\Lambda+\ssqbox{$b$},\Lambda+\ssqbox{$a$}+\ssqbox{$b$}}}\varphi_{a,b}(\phi_{\ssqbox{$a$}}-\phi_{\ssqbox{$b$}})=(-1)^{|a||b|}\,,\\
	&\frac{{\bf F}_{\Lambda+\ssqbox{$a$}+\ssqbox{$b$},\Lambda+\ssqbox{$a$}}{\bf F}_{\Lambda+\ssqbox{$a$},\Lambda}}{{\bf F}_{\Lambda+\ssqbox{$a$}+\ssqbox{$b$},\Lambda+\ssqbox{$b$}}{\bf F}_{\Lambda+\ssqbox{$b$},\Lambda}}\varphi_{a,b}(\phi_{\ssqbox{$a$}}-\phi_{\ssqbox{$b$}})=(-1)^{|a||b|}\,,\\
	&{\bf E}_{\Lambda,\Lambda+\ssqbox{$a$}}{\bf F}_{\Lambda+\ssqbox{$a$},\Lambda}=\mathop{\rm res}\lm_{z=\phi_{\ssqbox{$a$}}}\Psi^{(a)}_{\Lambda}(z)\,.
\end{aligned}
\end{equation}

\subsection{Matrix elements from equivariant integrals} \label{sec:equiv}

To construct a crystal representation, one has to prepare a system of matrix elements ${\bf E}_{\Lambda,\Lambda+\ssqbox{$a$}}$, ${\bf F}_{\Lambda+\ssqbox{$a$},\Lambda}$.
A proposition for this system was established in \cite{Prochazka:2015deb} for $\IC^3$, and then extended to generic quiver Yangians in \cite{Li:2020rij}.
We call this representation a \emph{square-root} representation:
\begin{equation}\label{sq_rt}
	{\bf E}_{\Lambda,\Lambda+\ssqbox{$a$}}={\bf F}_{\Lambda+\ssqbox{$a$},\Lambda}\sim\sqrt{\mathop{\rm res}\lm_{z=\phi_{\ssqbox{$a$}}}\Psi^{(a)}_{\Lambda}(z)}\,.
\end{equation}
A strong disadvantage of using this representation is that there is no proposed canonical way to define which branch of the square root function in \eqref{sq_rt} one should use.

Another approach to constructing ${\bf E}_{\Lambda,\Lambda+\ssqbox{$a$}}$ and ${\bf F}_{\Lambda+\ssqbox{$a$},\Lambda}$ is geometric and implements an equivariant integration over quiver representation moduli spaces \cite{Nakajima_lect,Rapcak:2018nsl,Rapcak:2020ueh}.
Alternative physical motivations for this construction may be found in \cite{Galakhov:2020vyb}.
Now we review this construction briefly.

The word ``equivariant'' allows one to apply powerful localization techniques \cite{berline1982classes,Witten:1982im,Cordes:1994fc,guillemin2013supersymmetry}: equivariant forms are exact outside neighborhoods of fixed loci and can be easily integrated, and in the vicinity of the fixed points, a perturbative analysis for tangent spaces is applicable.
In this framework, a natural characteristic of a fixed point is the corresponding Euler class that is constructed from equivariant weights $w_i$ of tangent directions $z_i$, where the equivariant vector field is transformed to a locally diagonal form:
\begin{equation}
	v=\sum\lm_i w_i \, z_i\frac{\p}{\p z_i}\,.
\end{equation}

The only problem with a direct application of these algorithms to our situation is that the varieties in question might turn out to be singular.
These singularities are reflected in jumps in the dimensions of the tangent spaces.
In \cite{Galakhov:2020vyb}, a regularization procedure was proposed.
To establish this procedure, we use a modified version of the Euler class.
Suppose the tangent space  $\CN$ is spanned by vectors $z_i$ with weights $w_i$, some of those weights may be zeroes.
Then we define:
\begin{equation}\label{reg_Eul}
	{\rm Eul}\;\CN\,=\,(-1)^{\left\lfloor\frac{1}{2}\#\{i:\, w_i=0\}\right\rfloor}\prod\lm_{i:\,w_i\neq 0}w_i\,.
\end{equation}

We have identified fixed points with crystals $\Lambda$ in Sec.~\ref{sec:molten}.
Depending on whether the quiver variety is smooth or singular, we define two types of the tangent spaces to the corresponding fixed points:
\begin{enumerate}
	\item $\mathsf{T}_{\Lambda}\mathscr{M}^{({\rm smth})}$ is defined as all the directions $q_I$ in the chiral fields parallel to the surface $F_I=0$ modulo gauge d.o.f.
	\item $\mathsf{T}_{\Lambda}\mathscr{M}^{({\rm sing})}$ is defined as all the directions $q_I$  modulo gauge d.o.f. In this case, we \emph{do not require} these d.o.f. to be parallel to the F-term locus.
\end{enumerate}
Then we define the corresponding Euler class:
\begin{equation}
	{\rm Eul}_{\Lambda}={\rm Eul}\; \mathsf{T}_{\Lambda}\mathscr{M}^{({\rm smth/sing})}\,.
\end{equation}

The matrix elements ${\bf E}_{\Lambda,\Lambda'}$, ${\bf F}_{\Lambda',\Lambda}$ are defined for two neighboring crystals $\Lambda'=\Lambda+\Box$.
For this construction, we will need a homomorphism of quiver representations $q_I$ and $q_I'$: a set of maps $\tau_a$, $a\in Q_0$ making the following diagrams commutative:
\begin{equation}\label{incidence}
	\begin{array}{c}
		\begin{tikzpicture}
			\node(A) at (0,0) {$V_a$};
			\node(B) at (2,0) {$V_b$};
			\node(C) at (0,-1.2) {$V_a'$};
			\node(D) at (2,-1.2) {$V_b'$};
			\path (A) edge[->] node[above] {$\scriptstyle q_{I:a\to b}$} (B) (C) edge[->] node[above] {$\scriptstyle q_{I:a\to b}'$} (D) (A) edge[->] node[left] {$\scriptstyle \tau_a$} (C) (B) edge[->] node[right] {$\scriptstyle \tau_b$} (D);
		\end{tikzpicture}
	\end{array},\quad q_{I:a\to b}'\cdot \tau_a=\tau_b\cdot q_{I:a\to b},\quad \forall I\in Q_1\,.
\end{equation}
We say that two representations $q_I$ and $q_I'$ are homomorphic if such a homomorphism exists.

Having two fixed points (neighboring crystals) $\Lambda$ and $\Lambda'=\Lambda+\Box$ we construct an \emph{incidence locus} $\CI$ as a surface in the Cartesian product of two representations $q_I$ and $q_I'$ where representations are homomorphic.
The tangent space to the incidence locus $\mathsf{T}_{\Lambda,\Lambda'}\CI\subset \mathsf{T}_{\Lambda}\mathscr{M}\oplus \mathsf{T}_{\Lambda'}\mathscr{M}$ is an equivariantly weighted space.
Therefore, we are able to define the corresponding Euler class:
\begin{equation}
	{\rm Eul}_{\Lambda,\Lambda'}={\rm Eul}\;\mathsf{T}_{\Lambda,\Lambda'}\CI\,.
\end{equation}

Geometrically matrix elements are defined as Fourier-Mukai transforms from $\mathsf{T}_{\Lambda}\mathscr{M}$ to $\mathsf{T}_{\Lambda'}\mathscr{M}$ or in the inverse direction with a kernel given by the structure sheaf of $\CI$:
\begin{equation}
	\begin{array}{c}
		\begin{tikzpicture}
			\node (A) at (-2.5,0) {$\mathsf{T}_{\Lambda}\mathscr{M}$};
			\node (B) at (0,0) {$\mathsf{T}_{\Lambda}\mathscr{M}\oplus\mathsf{T}_{\Lambda'}\mathscr{M}$};
			\node (C) at (2.5,0) {$\mathsf{T}_{\Lambda'}\mathscr{M}$};
			\path (B) edge[->] node[above] {$\scriptstyle p$} (A) (B) edge[->] node[above] {$\scriptstyle p'$} (C);
			\draw[->] (A.north) to[out=30,in=180] (0,0.8) to[out=0,in=150] (C.north);
			\node[above] at (0,0.8) {$\scriptstyle {\bf E}_{\Lambda,\Lambda'}$};
			\draw[<-] (A.south) to[out=330,in=180] (0,-0.8) to[out=0,in=210] (C.south);
			\node[below] at (0,-0.8) {$\scriptstyle {\bf F}_{\Lambda',\Lambda}$};
		\end{tikzpicture}
	\end{array}\,.
\end{equation}
This mathematical construction boils down to a standard pullback-pushforward integration \cite{Nakajima_lect,Rapcak:2020ueh,huybrechts2006fourier} for respective cohomological classes, or, in plain words, we should perform a Fourier transform with a kernel given by a delta-function of the locus $\CI$.
Fortunately, all the integrals are equivariant, so the canonical localization formulae are applicable.

Then, applying the canonical Atiyah-Bott localization formula:
\begin{equation}
	\int\lm_X\alpha=\sum\lm_{f\in F}\int\lm_f\frac{i^*\alpha}{{\rm Eul}_f}\,,
\end{equation}
we derive:
\begin{equation}
	\begin{aligned}
	&{\bf E}_{\Lambda,\Lambda+\Box}=\int\lm_{\CI\,{\rm near}\,(\Lambda,\Lambda+\Box)}p^*{\rm Eul}_{\Lambda}=\frac{{\rm Eul}_{\Lambda}}{{\rm Eul}_{\Lambda,\Lambda+\Box}}\,,\\
	&{\bf F}_{\Lambda+\Box,\Lambda}=\int\lm_{\CI\,{\rm near}\,(\Lambda,\Lambda+\Box)}p'^*{\rm Eul}_{\Lambda+\Box}=\frac{{\rm Eul}_{\Lambda+\Box}}{{\rm Eul}_{\Lambda,\Lambda+\Box}}\,.
	\end{aligned}
\end{equation}


\section{Algorithm for crystals}\label{sec:Algo_Cry}

We propose a recursive procedure to construct crystals in the cyclic chamber for quiver data of a toric Calabi-Yau three-fold.
We will call a recursion level, or simply a level, the total number of atoms in a crystal.
So having constructed all crystals at level $k$, our procedure offers a construction of all crystals at level $k+1$.

For computational purposes, it is simpler to keep track of all information about crystal atoms.
In addition to a position in the 3d space we discussed in Sec.~\ref{sec:molten}, we assigned to an atom the color characteristic.
We would like to present this information to a machine as a 4d vector where the fourth component is the color:
\begin{equation}
	\sqbox{$a$} \mbox{ at } (x,y,R)\quad\rightsquigarrow\quad (x,y,R,a)\,.
\end{equation}

We will discuss the recursion algorithm momentarily; however, to apply it successfully, one has to determine all the crystals at level $k=1$.
Fortunately, this level contains a single canonical crystal consisting of a single atom $(0,0,0,r)$.
This sets an explicit solution to the r.h.s. of \eqref{NSKH} with $d_{a}=\delta_{a,r}$ and $R=(1)$, a 1-by-1 matrix, whereas all the other $q_I=0$, or, equivalently, to the l.h.s. of  \eqref{NSKH} with $R=(\sqrt{\zeta_r})$.

Now let us discuss the recursion step.
It consists of three stages:
\begin{enumerate}
	\item Determining vacant atom positions
	\item Constructing new crystals, checking F-term consistency
	\item Deleting duplicates
\end{enumerate}

\subsection{Determining vacant atom positions}

First, having a crystal $\Lambda$ at level $k$, we determine where one might try to add an atom in principle.
For this purpose, we construct a new set of atoms by adding all the possible vectors $(x_I,y_I,R_I)$ to all atoms such that the tail of arrow $I$ corresponds to the color of the respective atom, and we switch colors accordingly:
\begin{equation}
	S(\Lambda):=\bigcup\lm_{(x,y,R,a)\in\Lambda, I\in Q_1, b\in Q_0}\left\{\begin{array}{ll}
		(x+x_I, y+y_I, R+R_I, b), & \mbox{ if }I\in\{a\to b\}\,,\\
		\varnothing, &\mbox{ otherwise}\,.
	\end{array}\right.
\end{equation}
In this way, we determine all the possible cyclic vectors \eqref{cyc} by simply multiplying those that are already present by all the operators, so that the matrix multiplication rule is not compromised.

In what follows, we need only new operators; therefore, we subtract the old ones to acquire a vacant set:
\begin{equation}
	V(\Lambda):=S(\Lambda)\setminus \Lambda\,.
\end{equation}

\subsection{Constructing new crystals, checking F-term consistency} \label{sec:alg_new_cry}

The next step in our algorithm is rather straightforward.
We have defined a set of new vacant atom positions in the previous stage.
So now, for each crystal $\Lambda$ at level $k$, we try to make new crystals by adding a single atom from $V(\Lambda)$.
However, not all the resulting unions of atoms are crystals.
We form a set of crystals at level $k+1$ by checking all the combinations $\Lambda+\Box$, $\Box\in V(\Lambda)$ from crystals $\Lambda$ at level $k$.
We will turn to the check condition momentarily; however, let us stress here that in this approach, the same crystals may appear in this set multiple times.
We will deal with this issue in what follows.

To check if a union of atoms $\Lambda$ \emph{is a crystal or not}, it is sufficient to check if the F-term relations for potential $q_I$ constructed on $\Lambda$ as a crystal according to rule \eqref{morphisms} are satisfied or not.
Let us repeat this construction in terms adopted in this section.
We assume that the expectation values for all the arrows flowing towards $\ff$ are zeros.
$R$ is a $d_r$-dimensional vector with elements:
\begin{equation}
	R_i=\left\{\begin{array}{ll}
		1,&\mbox{ if }\sqbox{$r$}_i=(0,0,0,r)\,,\\
		0,&\mbox{ otherwise}\,.
	\end{array}\right.
\end{equation}

To arrows in the unframed quiver, we associate the following matrix elements ($1\leq i\leq d_b$, $1\leq j\leq d_a$):\footnote{Here and in what follows, we denote a vacuum expectation of field $X$ as $\bar X$.
It should not be confused with the complex conjugation we denote as $X^{\dagger}$.
}
\begin{equation}\label{algo_vac}
	\bar q_{I:a\to b}=\left\{\begin{array}{ll}
		1, & \mbox{ if }\sqbox{$a$}_j=(x,y,R)\mbox{ and }\sqbox{$b$}_i=(x+x_I,y+y_I,R+R_I)\,,\\
		0, & \mbox{ otherwise}\,.
	\end{array}\right.
\end{equation}

It is easy to illustrate this process in the example with Young diagrams from Sec.~\ref{sec: Y(gl(1)) fixed points}:
\begin{equation}
	\begin{array}{c}
		\begin{tikzpicture}
			\node (A) at (0,0) {$\Lambda = \begin{array}{c}
					\begin{tikzpicture}[scale = 0.3]
						\draw[-stealth] (0.5,0.5) -- (2.5,0.5);
						\draw[-stealth] (0.5,0.5) -- (0.5,1.5);
						\draw[fill=white] (0,0) -- (2,0) -- (2,1) -- (0,1) -- cycle;
						\draw (1,0) -- (1,1);
						\node[right] at (2.5,0.5) {$\scriptstyle x$};
						\node[above] at (0.5,1.5) {$\scriptstyle y$};
					\end{tikzpicture}
				\end{array}$};
			\node (B) at (5,0) {$V(\Lambda) = \begin{array}{c}
					\begin{tikzpicture}[scale = 0.3]
						\draw[-stealth] (0.5,0.5) -- (3.5,0.5);
						\draw[-stealth] (0.5,0.5) -- (0.5,2.5);
						\draw[fill=gray] (0,0) -- (2,0) -- (2,1) -- (0,1) -- cycle;
						\draw (1,0) -- (1,1);
						\draw[fill=white] (0,1) -- (0,2) -- (2,2) -- (2,1) -- cycle;
						\draw (1,1) -- (1,2);
						\draw[fill=white] (2,0) -- (3,0) -- (3,1) -- (2,1) -- cycle;
						\node[right] at (3.5,0.5) {$\scriptstyle x$};
						\node[above] at (0.5,2.5) {$\scriptstyle y$};
					\end{tikzpicture}
				\end{array}$};
			\node (C) at (0,-3) {$\begin{array}{c}
					\begin{tikzpicture}[scale = 0.3]
						\draw[-stealth] (0.5,0.5) -- (3.5,0.5);
						\draw[-stealth] (0.5,0.5) -- (0.5,2.5);
						\node[right] at (3.5,0.5) {$\scriptstyle x$};
						\node[above] at (0.5,2.5) {$\scriptstyle y$};
						\begin{scope}
							\draw[fill=white] (0,0) -- (0,1) -- (1,1) -- (1,0) -- cycle;
							\node at (0.5,0.5) {$\scriptstyle 1$};
						\end{scope}
						\begin{scope}[shift = {(1,0)}]
							\draw[fill=white] (0,0) -- (0,1) -- (1,1) -- (1,0) -- cycle;
							\node at (0.5,0.5) {$\scriptstyle 2$};
						\end{scope}
						\begin{scope}[shift = {(0,1)}]
							\draw[fill=white] (0,0) -- (0,1) -- (1,1) -- (1,0) -- cycle;
							\node at (0.5,0.5) {$\scriptstyle 3$};
						\end{scope}
					\end{tikzpicture}\\
					\scalebox{0.7}{$\bar B_1=\left(\begin{array}{ccc}
						0 & 0 & 0\\
						1 & 0 & 0\\
						0 & 0 & 0\\
					\end{array}\right),\, \bar B_2=\left(\begin{array}{ccc}
					0 & 0 & 0\\
					0 & 0 & 0\\
					1 & 0 & 0\\
				\end{array}\right)$}\\
				\scalebox{0.7}{$F=[\bar B_1,\bar B_2]=\left(\begin{array}{ccc}
						0 & 0 & 0\\
						0 & 0 & 0\\
						0 & 0 & 0\\
					\end{array}\right)$}\\
				\mbox{\bf\color{black!40!green}a crystal}
				\end{array}$};
			\node (D) at (5,-3) {$\begin{array}{c}
					\begin{tikzpicture}[scale = 0.3]
						\draw[-stealth] (0.5,0.5) -- (3.5,0.5);
						\draw[-stealth] (0.5,0.5) -- (0.5,2.5);
						\node[right] at (3.5,0.5) {$\scriptstyle x$};
						\node[above] at (0.5,2.5) {$\scriptstyle y$};
						\begin{scope}
							\draw[fill=white] (0,0) -- (0,1) -- (1,1) -- (1,0) -- cycle;
							\node at (0.5,0.5) {$\scriptstyle 1$};
						\end{scope}
						\begin{scope}[shift = {(1,0)}]
							\draw[fill=white] (0,0) -- (0,1) -- (1,1) -- (1,0) -- cycle;
							\node at (0.5,0.5) {$\scriptstyle 2$};
						\end{scope}
						\begin{scope}[shift = {(1,1)}]
							\draw[fill=white] (0,0) -- (0,1) -- (1,1) -- (1,0) -- cycle;
							\node at (0.5,0.5) {$\scriptstyle 3$};
						\end{scope}
					\end{tikzpicture}\\
					\scalebox{0.7}{$\bar B_1=\left(\begin{array}{ccc}
							0 & 0 & 0\\
							1 & 0 & 0\\
							0 & 0 & 0\\
						\end{array}\right),\, \bar B_2=\left(\begin{array}{ccc}
							0 & 0 & 0\\
							0 & 0 & 0\\
							0 & 1 & 0\\
						\end{array}\right)$}\\
					\scalebox{0.7}{$F=[\bar B_1,\bar B_2]=\left(\begin{array}{ccc}
							0 & 0 & 0\\
							0 & 0 & 0\\
							{\bf -1} & 0 & 0\\
						\end{array}\right)$}\\
					\mbox{\bf\color{burgundy} not a crystal}
				\end{array}$};
			\node (E) at (10,-3) {$\begin{array}{c}
					\begin{tikzpicture}[scale = 0.3]
						\draw[-stealth] (0.5,0.5) -- (3.5,0.5);
						\draw[-stealth] (0.5,0.5) -- (0.5,2.5);
						\node[right] at (3.5,0.5) {$\scriptstyle x$};
						\node[above] at (0.5,2.5) {$\scriptstyle y$};
						\begin{scope}
							\draw[fill=white] (0,0) -- (0,1) -- (1,1) -- (1,0) -- cycle;
							\node at (0.5,0.5) {$\scriptstyle 1$};
						\end{scope}
						\begin{scope}[shift = {(1,0)}]
							\draw[fill=white] (0,0) -- (0,1) -- (1,1) -- (1,0) -- cycle;
							\node at (0.5,0.5) {$\scriptstyle 2$};
						\end{scope}
						\begin{scope}[shift = {(2,0)}]
							\draw[fill=white] (0,0) -- (0,1) -- (1,1) -- (1,0) -- cycle;
							\node at (0.5,0.5) {$\scriptstyle 3$};
						\end{scope}
					\end{tikzpicture}\\
					\scalebox{0.7}{$\bar B_1=\left(\begin{array}{ccc}
							0 & 0 & 0\\
							1 & 0 & 0\\
							0 & 1 & 0\\
						\end{array}\right),\, \bar B_2=\left(\begin{array}{ccc}
							0 & 0 & 0\\
							0 & 0 & 0\\
							0 & 0 & 0\\
						\end{array}\right)$}\\
					\scalebox{0.7}{$F=[\bar B_1,\bar B_2]=\left(\begin{array}{ccc}
							0 & 0 & 0\\
							0 & 0 & 0\\
							0 & 0 & 0\\
						\end{array}\right)$}\\
					\mbox{\bf\color{black!40!green}a crystal}
				\end{array}$};
			\path (A) edge[->] (B);
			\draw[->] ([shift={(-1,0)}]B.south) to ([shift={(2,0)}]C.north);
			\draw[->] (B.south) to (D.north);
			\draw[->] ([shift={(1,0)}]B.south) to ([shift={(-2,0)}]E.north);
			\begin{scope}[shift={(5,-3)}]
				\draw[thick, burgundy] (-1.5,-1.5) -- (1.5,1.5) (-1.5,1.5) -- (1.5,-1.5);
			\end{scope}
		\end{tikzpicture}
	\end{array}
\end{equation}

In practical applications, we recommend to try to parallelize this computation.
Each of the concurrent processes may acquire as incoming data a crystal $\Lambda$ on level $k$, construct $V(\Lambda)$, and further all the possible crystals derived from $\Lambda$ by adding an atom.
Then the result may be written to common storage memory, like a file.

\subsection{Deleting duplicates}

One issue in this algorithm inherited from the previous stage is that, as a result, we will acquire the same crystals multiple times.
This observation becomes transparent from the following diagram, where we depict links between crystals differing by a single atom:
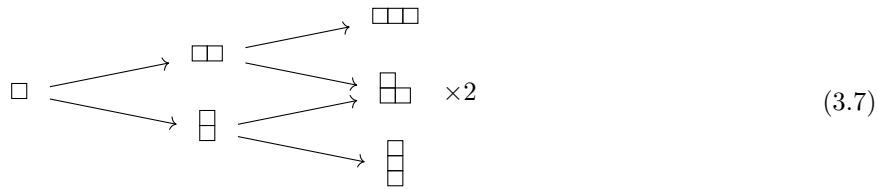
\begin{equation}
	\begin{array}{c}
		\begin{tikzpicture}
			\node(A) at (0,0) {$\begin{array}{c}
					\begin{tikzpicture}[scale=0.2]
						\foreach \i/\j in {0/0}
						{
							\draw (\i,\j) -- (\i,\j+1) -- (\i+1,\j+1) -- (\i+1,\j) -- cycle;
						}
					\end{tikzpicture}
				\end{array}$};
			\node(B) at (2.5,0.5) {$\begin{array}{c}
					\begin{tikzpicture}[scale=0.2]
						\foreach \i/\j in {0/0, 1/0}
						{
							\draw (\i,\j) -- (\i,\j+1) -- (\i+1,\j+1) -- (\i+1,\j) -- cycle;
						}
					\end{tikzpicture}
				\end{array}$};
			\node(C) at (2.5,-0.5) {$\begin{array}{c}
					\begin{tikzpicture}[scale=0.2]
						\foreach \i/\j in {0/0, 0/1}
						{
							\draw (\i,\j) -- (\i,\j+1) -- (\i+1,\j+1) -- (\i+1,\j) -- cycle;
						}
					\end{tikzpicture}
				\end{array}$};
			\node(D) at (5,1) {$\begin{array}{c}
					\begin{tikzpicture}[scale=0.2]
						\foreach \i/\j in {0/0, 1/0, 2/0}
						{
							\draw (\i,\j) -- (\i,\j+1) -- (\i+1,\j+1) -- (\i+1,\j) -- cycle;
						}
					\end{tikzpicture}
				\end{array}$};
			\node(E) at (5,0) {$\begin{array}{c}
					\begin{tikzpicture}[scale=0.2]
						\foreach \i/\j in {0/0, 1/0, 0/1}
						{
							\draw (\i,\j) -- (\i,\j+1) -- (\i+1,\j+1) -- (\i+1,\j) -- cycle;
						}
					\end{tikzpicture}
				\end{array}$};
			\node(F) at (5,-1) {$\begin{array}{c}
					\begin{tikzpicture}[scale=0.2]
						\foreach \i/\j in {0/0, 0/1, 0/2}
						{
							\draw (\i,\j) -- (\i,\j+1) -- (\i+1,\j+1) -- (\i+1,\j) -- cycle;
						}
					\end{tikzpicture}
				\end{array}$};
			\path (A) edge[->] (B) (A) edge[->] (C) (B) edge[->] (D) (B) edge[->] (E) (C) edge[->] (E) (C) edge[->] (F);
			\node[right] at (E.east) {$\times 2$};
		\end{tikzpicture}
	\end{array}
\end{equation}

Apparently, our algorithm will contribute partition $[2,1]$ twice from two parents $[2]$ and $[1,1]$.
To avoid this repetition, we have to delete duplicates.
This could be achieved by any deleting duplicates algorithm if all crystals are stored in any canonical form.
So far, we have represented a crystal as an array of 4d vectors -- atom labels.
A canonical form for an array would be an ordered array with respect to any ordering function, since there are no equivalent elements (no atoms at the same position) in that array.
In practice, for simplicity, we choose a lexicographic order.


\section{Algorithm for matrix elements}\label{sec:Algo_ME}

\subsection{Digression: useful function \texorpdfstring{$\mbox{\texttt{\small ReduceSolve()}}$}{} for linear systems}\label{ssec:RedSlv}

Probably all the modern symbolic computational systems have an implementation for solving linear systems:
\begin{equation}\label{linear}
	A\vec x =\vec y\,.
\end{equation}
However, in many cases, this implementation is rather simplistic; it solves the most common type of this problem.
An algorithm checks if $A$ is invertible, and if the check succeeds, it returns the answer $\vec x = A^{-1}\vec y$.

In this subsection, we would like to discuss a canonical realization of a function that extends the analysis of \eqref{linear} to the cases when $A$ is allowed to have a non-trivial kernel or/and a co-kernel.
We call this function $\mbox{\texttt{\small ReduceSolve()}}$.
To do so, let us assume that both $\vec x$ and $\vec y$ are (linear combinations of) elements of a symbolic alphabet $\CA$.
Then we call a set of equations $\mathscr{E}$ a set of linear combinations of $\CA$.
Furthermore, a part of $\CA$ we call variables $\CV$, for which we would like $\mathscr{E}$ to be solved.
Procedure $\mbox{\texttt{\small ReduceSolve($\mathscr{E}$,$\CV$)}}$ is expected to return a triplet of results:
\begin{enumerate}
	\item {\bf Ker:} an explicit parametrization of ${\rm ker}\;A$ in terms of $\CV$.
	\item {\bf Sol:} an explicit solution for variables in $\CV\setminus {\rm ker}\;A$.
	\item {\bf Obs:} an obstruction in terms of $\CA\setminus \CV$ for $\mathscr{E}$ to have a solution.
\end{enumerate}
Let us illustrate how this function works in the following primitive example.
Let us assume that we have five symbols $\{x_1,\ldots,x_5\}$, and choose to be variables only $x_1$ and $x_2$.
Furthermore, we try to solve the following system of equations:
\begin{equation}
	\left\{\begin{array}{l}
		x_1+x_2=x_3\,;\\
		x_4+x_5=0\,.
	\end{array}\right.
\end{equation}
It is easy to solve this system for $x_1$ in terms of $x_2$; however, the solution is valid if and only if $x_4+x_5=0$.
Therefore, in this case:
\begin{equation}
	{\bf Ker}=\{x_2\}, \quad {\bf Sol}=\{x_1=-x_2+x_3\}, \quad {\bf Obs}=\{x_4+x_5\}\,.
\end{equation}

In this setting, a recursive implementation of this function becomes apparent:
\begin{enumerate}
	\item At each recursion $n^{\rm th}$ step, we choose a variable $x_n\in\CV$.
	\item We search for an equation $e_k\in\mathscr{E}$ such that $\p_{x_n}e_k\neq 0$:
	\begin{enumerate}
		\item If there is no such $e_k$, we increase ${\bf Ker}$ by this variable ${\bf Ker}:={\bf Ker}\cup \{x_n\}$. 
		The next step is skipped.
		\item Otherwise, we solve $e_k=0$ for $x_n$ and acquire a solution $x_n^*$. 
		This solution is to be added to ${\bf Sol}$.
	\end{enumerate}
	\item We redefine the set of equations by substituting the solution found at the previous step $\mathscr{E}:=\mathscr{E}\big|_{x_n= x_n^*}\setminus \{0\}$.
\end{enumerate}
Eventually, after all the variables in $\CV$ are used, ${\bf Obs}=\mathscr{E}$.

\subsection{Tangent spaces and weights}

Having a crystal $\Lambda$, we can describe a tangent space to a fixed point on a quiver representation.
First, we describe the respective vector spaces $\IC^{d_a}$ as atoms of color $a$ and construct expectation values \eqref{algo_vac} of field $\bar q_I$ as it was discussed in Sec.~\ref{sec:Algo_Cry}.

In the basis where vectors are labeled by atoms, fields $\Phi_a$ acquire diagonal expectation values as well (cf. \eqref{Phi}):
\begin{equation}
	\bar\Phi_a|(x,y,R,a)\rangle=(x\epsilon_1+y\epsilon_2)\times |(x,y,R,a)\rangle\,.
\end{equation}

We parameterize the tangent space to fixed point $\Lambda$ by small chiral matrix deformations:
\begin{equation}
	\delta q_{I:a\to b} \in {\rm Mat}_{d_b\times d_a}(\IC),\quad \forall I\in Q_1 \,,
\end{equation}
so that $q_I=\bar q_I+\delta q_I$.

Equivariant weights for the equivariant vector field in \eqref{vacuum} are defined by expectation values of $\Phi_a$:
\begin{equation}\label{weights}
	w\left[(\delta q_{I:a\to b})_{ij}\right]=(\bar\Phi_b)_{ii}-(\bar\Phi_a)_{jj}-h_I\,.
\end{equation}

\subsection{Gauge fixing and F-terms}\label{sec: gauge and F-terms}

Let us parameterize the action of the gauge algebra $\fg_{\IC}=\bigoplus\lm_{a\in Q_0}\fg\fl(\IC,d_a)$ by variables $g_{a,ij}$, $i,j=1,\ldots,d_a$.
Then the gauge action on the tangent degrees of freedom is linearized:
\begin{equation}
	\fg_{\IC}:\quad \delta q_{I:a\to b}\,\mapsto\, \delta q_{I:a\to b}+g_b\cdot \bar q_{I:a\to b}-\bar q_{I:a\to b}\cdot g_a\,.
\end{equation}
For the framing node $g_{\ff}=(0)$.
To fix the gauge, we implement the following algorithm.
Consider the action $\fg_{\IC}(\delta q_I)$ for all arrows as a set of equations with zeroes in the r.h.s., then we apply the solving procedure from Section \ref{ssec:RedSlv}:
\begin{equation}
	\mbox{\texttt{\small ReduceSolve($\fg_{\IC}(\delta q_I)$,$\{g_a\}$)}}\,.
\end{equation}
It turns out that in this case ${\bf Ker}=\varnothing$ and we acquire a solution $g_a^*$ for all the gauge variables.
Now we are in a position to define which matrix elements of $\delta q_I$ parameterize gauge directions in the tangent space (those are Higgsed away in the IR description) and which are gauge-invariant ``mesonic'' directions.
If after substituting solution $g_a^*$ into the action of $\fg_{\IC}$ the respective degree of freedom is annihilated it implies this d.o.f. is not gauge-invariant:
\begin{equation}
	\begin{array}{ccl}
		\fg_{\IC}\left[(\delta q_I)_{ij}\right]\big|_{g_a= g_a^*} = 0 & \Leftrightarrow & (\delta q_I)_{ij}\mbox{ is Higgsed}\,,\\
		\fg_{\IC}\left[(\delta q_I)_{ij}\right]\big|_{g_a= g_a^*} \neq 0 & \Leftrightarrow & (\delta q_I)_{ij}\mbox{ is gauge-invariant}\,.
	\end{array}
\end{equation}
Using this criterion, we simply eliminate (substitute by zeros) those degrees of freedom that are not gauge-invariant.

In some cases we will also need to define those tangent directions that are parallel to the F-term equations.
Since superpotential $W$ is gauge-invariant, the F-term equations are automatically covariant.
This implies that the gauge degrees of freedom are parallel to the F-term surface in the space of chiral fields $q_I$.
Therefore the procedures of selecting parallel and gauge-invariant d.o.f. commute.

To define parallel to the F-term surface d.o.f. we linearize those equations first:
\begin{equation}
	\mathscr{E}=\left\{\p_tF_J(\bar q_I+t\,\delta q_I)\big|_{t=0}\right\}\,.
\end{equation}
Then the respective $\bf Ker$ of $\mbox{\texttt{\small ReduceSolve($\mathscr{E}$,$\{\delta q_I\}$)}}$ describes a parametrization of the tangent plane to the F-term surface, and the respective $\bf Sol$ describes an embedding of this tangent space in the tangent space of chiral fields $q_I$.

Both of these procedures allow one to select among matrix elements $(\delta q_I)_{ij}$ those that satisfy the gauge-invariant or F-term criteria, or simultaneously both:
\begin{equation}
	\begin{aligned}
		&\mathsf{T}_{\Lambda}\mathscr{M}=\{\delta q_I\}/\{{\rm gauge},\mbox{F-term}\},\quad \mbox{ when the quiver variety is smooth}\,,\\
		&\mathsf{T}_{\Lambda}\mathscr{M}= \{\delta q_I\}/\{{\rm gauge}\},\quad \mbox{ when the quiver variety is singular}\,.
	\end{aligned}
\end{equation}
Then the Euler classes are defined according to \eqref{reg_Eul} with weights \eqref{weights}.

\subsection{Incidence loci}

We construct tangent spaces to the incidence loci for pairs of crystals $\Lambda$ and $\Lambda'$ that differ by an atom $\Lambda'=\Lambda+\Box$.
For each crystal in this pair, the respective expectation values $\bar q_I$ and $\bar q_I'$ and tangent spaces $\mathsf{T}_{\Lambda}\mathscr{M}$ and $\mathsf{T}_{\Lambda'}\mathscr{M}$ could be constructed with the help of the algorithms discussed above.

Expectation values for the singular gauge transform are constructed accordingly:
\begin{equation}
	\bar\tau_a|\sqbox{$a$}_i\in\Lambda'\rangle=\left\{\begin{array}{ll}
		|\sqbox{$a$}_i\rangle,&\mbox{ if }\sqbox{$a$}_i\in\Lambda\,,\\
		0, &\mbox{ otherwise}\,.
	\end{array}\right.
\end{equation}
Similarly, we parameterize the tangent directions for these singular gauge d.o.f.:
\begin{equation}
	\delta\tau_a\in{\rm Mat}_{d_a'\times d_a}(\IC),\quad a\in Q_0\,,
\end{equation}
For the framing node, we set $\delta\tau_f=(0)$.

The linearized incidence locus equation \eqref{incidence} reads in this case:
\begin{equation}
	\mathscr{E}=\left\{\delta\tau_b\cdot \bar q_{I:a\to b}'+\bar\tau_b\cdot \delta q_{I:a\to b}'-\bar q_{I:a\to b}\cdot \delta\tau_a-\delta q_{I:a\to b}\cdot \bar\tau_a \right\}_{I\in Q_1}\,.
\end{equation}
A set of equations $\Sigma$ describing the incidence locus is described as $\bf Obs$ in $\mbox{\texttt{\small ReduceSolve($\mathscr{E}$,$\{\delta\tau_a\}$)}}$.

Therefore, the tangent space in $\mathsf{T}_{\Lambda}\mathscr{M}\oplus \mathsf{T}_{\Lambda'}\mathscr{M}$ to the incidence locus is constructed as:
\begin{equation}
	\mathsf{T}_{\Lambda,\Lambda'}\CI=\mbox{\texttt{\small ReduceSolve($\Sigma$,$\{\delta q_I\}\cup \{\delta q_I'\}$)}}\big|_{\bf Ker}\,.
\end{equation}
The respective Euler class is defined according to \eqref{reg_Eul} with weights \eqref{weights}.


\section{Simplest classical Yangian \texorpdfstring{$\mathsf{Y}(\mathfrak{sl}_2)$}{}}\label{sec: Y(sl(2))}

In this section we apply the above prescription to explore properties of the spin representations of the quiver Yangian algebras. 
We start with the simplest quiver Yangian algebra $\mathsf{Y}(\mathfrak{sl}_{2})$.

\subsection{Commutation relations}

Yangian algebra $\mathsf{Y}(\mathfrak{sl}_{2})$ can be defined via an infinite family of Chevalley-like generators $e_n$, $f_n$, $\psi_n$, $n\in\IN_0$.
The algebra is formulated in terms of relations:
\begin{equation}\label{Y(sl(2)) relations in modes}
	\begin{aligned}
		&[\psi_{n}, \psi_{m}] = 0, \\
		&[e_{n}, f_{m}] - \psi_{n+m} = 0,\\
		&[e_{n+1}, e_{m}] - [e_{n}, e_{m+1}] - \epsilon \{e_{n}, e_{m}\} = 0,\\
		&[\psi_{n+1}, e_{m}] - [\psi_{n}, e_{m + 1}] - \epsilon \{\psi_{n}, e_{m}\} = 0,\\
		&[f_{n+1}, f_{m}] - [f_{n}, f_{m+1}] + \epsilon\{f_{n}, f_{m}\} = 0, \\
		&[\psi_{n+1}, f_{m}] - [\psi_{n}, f_{m+1}] + \epsilon\{\psi_{n}, f_{m}\} = 0.
	\end{aligned}
\end{equation}
There are additional relations that can be interpreted as ``boundary conditions'':
\begin{align}
	\begin{aligned}
		\label{sl(2) boundary conditions}
		&[\psi_{0}, e_{m}] = 2 e_{m}, \\
		&[\psi_{0}, f_{m}] = -2 f_{m}. \\
	\end{aligned}
\end{align}

Assembling the generators \eqref{Y(sl(2)) relations} in terms of generating functions:
\begin{equation}\label{generating functions}
	e(z) = \sum_{n \in \mathbb{N}_{0}}\dfrac{e_{n}}{z^{n+1}}, \quad \psi(z) = 1 + \epsilon\sum_{n\in \mathbb{N}_{0}}\dfrac{\psi_{n}}{z^{n+1}}, \quad f(z) = \sum_{n\in \mathbb{N}_{0}}\dfrac{f_{n}}{z^{n+1}}
\end{equation}
we are able to rewrite relations \eqref{Y(sl(2)) relations in modes} in a more compact form:
\begin{equation}\label{Y(sl(2)) relations}
	\begin{aligned}
		&e(z)e(w) \sim \varphi(z - w)e(w)e(z)\,,\\
		&f(z)f(w) \sim [\varphi(z - w)]^{-1}f(w)f(z)\,,\\
		&\psi(z)e(w) \simeq \varphi(z - w)e(w)\psi(z)\,,\\
		&\psi(z)f(w) \simeq [\varphi(z - w)]^{-1}f(w)\psi(z)\,,\\
		&[e(z), f(w)] \sim -\dfrac{1}{\epsilon}\frac{\psi(z) - \psi(w)}{z - w}\,,
	\end{aligned}
\end{equation}
where the bond factor (cf. \cite{Li:2020rij}) reads:
\begin{equation}
	\varphi(z) = \frac{z + \epsilon}{z - \epsilon}\,.
\end{equation}

As a consequence of the ``boundary conditions'', a triplet of zero modes $e_0, \psi_0, f_0$ forms an $\mathfrak{sl}_2$ subalgebra of the Yangian. 
In contrast, there is a special set of operators in the Yangian algebra:
\begin{equation}\label{small set sl(2)}
	\boxed{e_0, \hspace{5mm} \psi_1, \hspace{5mm} f_0} ,
\end{equation}
that allows one to construct  generators for   the whole algebra by recursion relations:
\begin{equation}
	\begin{aligned}
		&e_{n+1} = \frac{1}{2} \Big[ \psi_1, e_{n} \Big] - \frac{\epsilon}{2} \Big\{ \psi_0, e_n \Big\}, \\
		&f_{n+1} = -\frac{1}{2} \Big[ \psi_1, f_{n} \Big] - \frac{\epsilon}{2} \Big\{ \psi_0, f_n \Big\}
	\end{aligned}
\end{equation}
These relations follow directly from \eqref{Y(sl(2)) relations in modes} and \eqref{sl(2) boundary conditions}.

\subsection{Quiver equations}

\begin{figure}[h]
	\begin{center}
		\begin{tikzpicture}
			\draw[postaction={decorate},
			decoration={markings, mark= at position 0.6 with {\arrow{stealth}}}] (-2,0) to[out=20,in=160] node[pos=0.5,above] {$\scriptstyle I$} (0,0);
			\draw[postaction={decorate},
			decoration={markings, mark= at position 0.6 with {\arrow{stealth}}}] (0,0) to[out=-160,in=-20] node[pos=0.5,below] {$\scriptstyle J$} (-2,0);
			\draw[postaction={decorate},
			decoration={markings, mark= at position 0.8 with {\arrow{stealth}}}] (0,0) to[out=-30,in=270] (1,0) to[out=90,in=30] (0,0);
			\draw[fill=\myblue] (0,0) circle (0.1);
			\node[right] at (1,0) {$\scriptstyle B$};
			\begin{scope}[shift={(-2,0)}]
				\draw[fill=burgundy] (-0.08,-0.08) -- (-0.08,0.08) -- (0.08,0.08) -- (0.08,-0.08) -- cycle;
			\end{scope}
			\node[above] at (0,0.1) {$\scriptstyle n$};
		\end{tikzpicture}
	\end{center}
	\caption{The Jordan quiver with framing corresponding to $\mathsf{Y}(\mathfrak{sl}_{2})$}\label{Y(sl(2)) quiver}
\end{figure}
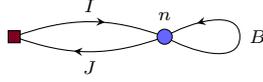

The quiver that we are going to use to describe $\mathsf{Y}(\mathfrak{sl}_{2})$ representations is depicted in Fig. \ref{Y(sl(2)) quiver}.
This quiver is also known as a Jordan quiver \cite{Ginzburg_lect}.
A similar description appeared in \cite{Bykov:2019cst,Yang:2024ubh}.

Following \cite{Bykov:2019cst}, we consider a spin-$\frac{\lambda}{2}$ superpotential:
\begin{equation}\label{Y(sl(2)) superpotential}
	W = \Tr(B^{\lambda}IJ)\,.
\end{equation}
The above definition implies that $\lambda$ is expected to be an integer.
In principle, we are also able to retrieve representations with a 
non-integral spin that corresponds to $\lambda$ being generic.
In the latter case $W=0$, however, we should keep the weight of field $J$ imposed by the loop constraint \eqref{loop}.
The weights of fields read:
\begin{equation}
	\begin{array}{c|c|c|c}
		\mbox{Fields} & B & I & J\\
		\hline
		\mbox{Weights} & \epsilon & 0 & -\lambda\epsilon
	\end{array}\,.
\end{equation}

In this case, we have the following D-term and F-term equations:
\begin{equation}\label{Y(sl(2)) D-F -terms}
	\begin{aligned}
		&[B, B^{\dagger}] + II^{\dagger} - J^{\dagger}J =\zeta \bbone_{n\times n}, &\text{ D-term}\\
		&\d_{J} \mathcal{W} = B^{\lambda}I = 0,  &\text{ F-terms} \\
		&\d_{I} \mathcal{W} = JB^{\lambda} = 0,\\
		&\d_{B}\mathcal{W} = B^{\lambda - 1}IJ + B^{\lambda - 2}IJB + \dots + IJB^{\lambda - 1} = 0
	\end{aligned}
\end{equation}
Fixed points on the above variety are defined by the following constraint:
\begin{equation}\label{Y(sl(2)) Fixed point equations}
	\begin{aligned}
		&[\Phi, B] = \epsilon B,\\
		&\Phi I = 0, \\
		&J\Phi - \epsilon\lambda = 0
	\end{aligned}
\end{equation}

\subsection{Fixed points}

As we discussed in Sec.~\ref{sec:alg_new_cry}, fixed points that we have defined by equations \eqref{Y(sl(2)) Fixed point equations} can be associated with sets of paths in the quiver in Fig.~\ref{Y(sl(2)) quiver}.  
In this case, all the paths have the simple form $B^{a}I$ and can be depicted on a line:
\begin{equation}
\begin{array}{c}
	\begin{tikzpicture}[rotate = 0, scale=0.8]
		\draw[postaction={decorate},
		decoration={markings, mark= at position 0.6 with {\arrow{stealth}}}] (0, 0) -- (2,0);
		\draw[postaction={decorate},
		decoration={markings, mark= at position 0.6 with {\arrow{stealth}}}] (2, 0) -- (4,0);
		\draw[postaction={decorate},
		decoration={markings, mark= at position 0.6 with {\arrow{stealth}}}] (4, 0) -- (6,0);
		\draw[postaction={decorate},
		decoration={markings, mark= at position 0.6 with {\arrow{stealth}}}] (6, 0) -- (8,0);
		\draw[postaction={decorate},
		decoration={markings, mark= at position 0.6 with {\arrow{stealth}}}] (8, 0) -- (9,0);
		\draw[fill=white, draw=burgundy] circle (0.4);
		\draw[fill=\myblue] (0, 0) circle (0.2);
		\draw[fill=\myblue] (2, 0) circle (0.2);
		\draw[fill=\myblue] (4, 0) circle (0.2);
		\draw[fill = \myblue] (6, 0) circle (0.2);
		\draw[fill = \myblue] (8, 0) circle (0.2);
		\draw (0, 0.6) node {$\scriptstyle I$};
		\draw (2, 0.6) node {$\scriptstyle BI$};
		\draw (4, 0.6) node {$\scriptstyle B^{2}I$};
		\draw (6, 0.6) node {$\scriptstyle B^{3}I$};
		\draw (8, 0.6) node {$\scriptstyle \dots$};
	\end{tikzpicture}
\end{array}
\end{equation}

The F-term constraint $B^{\lambda}I = 0$ plays the role of a cutoff. 
Therefore, one can classify the fixed points \eqref{Y(sl(2)) Fixed point equations} with one-dimensional Young diagrams (like [5]= \begin{tikzpicture}[scale=0.25]
	\foreach \i/\j in {0/-1, 0/0, 1/-1, 1/0, 2/-1, 2/0, 3/-1, 3/0, 4/-1, 4/0}
	{
		\draw (\i,\j) -- (\i+1,\j);
	}
	\foreach \i/\j in {0/0, 1/0, 2/0, 3/0, 4/0, 5/0}
	{
		\draw (\i,\j) -- (\i,\j-1);
	}
\end{tikzpicture}) that we denote as $[n]$.

This representation is isomorphic as a vector space to a spin-$\frac{\lambda}{2}$ representation of ordinary Lie algebra $\fs\fl_2$.
The following crystal representation is expected (see Sec.~\ref{sec:cry_rep}):
\begin{equation}\label{Y(sl(2)) Ansatz}
	\begin{aligned}
		&e^{[\lambda]}(z)|n\rangle = \dfrac{{\bf E}_{n, n+1}^{[\lambda]}}{z - n\epsilon}|n+1\rangle\,,\\
		&f^{[\lambda]}(z)|n\rangle = \dfrac{{\bf F}_{n,n-1}^{[\lambda]}}{z - (n-1)\epsilon}|n-1\rangle\,,\\
		&\psi^{[\lambda]}(z)|n\rangle = \bPsi_{n}^{[\lambda]}(z)|n\rangle\,.
	\end{aligned}
\end{equation}

The eigenvalues of $\psi^{[\lambda]}(z)$ are defined as follows:
\begin{equation}\label{eigenvalues}
	\bPsi_{n}^{[\lambda]}(z) = \frac{z - \lambda \epsilon}{z}\prod_{k = 1}^{n}\varphi\left(z - (k-1)\epsilon\right)=\frac{(z-\lambda\epsilon)(z+\epsilon)}{(z-n\epsilon)(z-(n-1)\epsilon)}\,.
\end{equation}

As we will show in what follows, matrix elements ${\bf E}_{n, n+1}^{[\lambda]}$ and ${\bf F}_{n,n-1}^{[\lambda]}$ coincide with matrix elements of spin-$\frac{\lambda}{2}$ $\fs\fl_2$ representation.
Therefore, we could conclude that our crystal representations and spin-$\frac{\lambda}{2}$ representations of $\fs\fl_2$ are isomorphic.

\subsection{Euler classes}

Following the procedure described in Sec.~\ref{sec:Algo_ME} we construct vacuum expectation values for chiral fields:
\begin{equation}
	\bar B = \begin{pmatrix}
		0 & 0 & 0 & \dots & 0\\
		1 & 0 & 0 & \dots & 0\\
		0 & 1 & 0 & \dots & 0\\
		\vdots & \vdots & \ddots & \ddots & \vdots\\
		0 & 0 & \dots & 1 & 0
	\end{pmatrix}_{n\times n}, \quad 
	\bar I = \begin{pmatrix}
		1 \\ 0 \\ 0 \\ \vdots \\0
	\end{pmatrix}_{n\times 1}, \quad 
	\bar J = \begin{pmatrix}
		0
	\end{pmatrix}_{1\times n}, \quad 
	\bar \Phi = \begin{pmatrix}
		0 & 0 & 0 & \dots &0\\
		0 & \epsilon & 0 & \dots & 0\\
		0 & 0 & 2\epsilon & \dots & 0\\
		\vdots & \vdots & \vdots & \ddots & \vdots\\
		0 & 0 & 0 & \dots & (n - 1)\epsilon
	\end{pmatrix}_{n\times n}\,.
\end{equation}

After gauge fixing the gauge invariant degrees of freedom acquire the following form:
\begin{equation}
\delta B = \begin{pmatrix}
	0 & 0 & \dots & 0 & b_{1n}\\
	0 & 0 & \dots & 0 & b_{2n}\\
	0 & 0 & \dots & 0 & b_{3n}\\
	\vdots & \vdots & \ddots & \vdots & \vdots\\
	0 & 0 & \dots & 0 & b_{nn}
\end{pmatrix}_{n\times n}, \quad \delta I = \begin{pmatrix}
	0 \\ 0 \\ 0 \\ \vdots \\ 0
\end{pmatrix}_{n \times 1}, \quad \delta J = \begin{pmatrix}
	j_{1} & j_{2} & \dots & j_{n}
\end{pmatrix}_{1\times n}\,.
\end{equation}

In cases other than $\lambda=1$, when the quiver is of Nakajima type \cite{Ginzburg_lect}, the quiver variety is not smooth.
Therefore, we apply here the scheme for the singular varieties from Sec.~\ref{sec:equiv}.
The tangent space to the fixed point is described as:
\begin{equation}
	\mathsf{T}_n\mathscr{M}^{[\lambda]}={\rm Span}\left\{b_{1n},b_{2n},\ldots,b_{nn},j_1,\ldots,j_n\right\}\,.
\end{equation}

We easily calculate respective weights:
\begin{equation}
	\begin{aligned}
		&w(b_{kn}) = w[(B)_{kn}] = \phi_{k} - \phi_{n} - \epsilon = (k - 1)\epsilon - (n - 1)\epsilon - \epsilon = (k - n - 1)\epsilon\,,\\
		&w(j_{k}) = w[(J)_{k}] = \phi_{k} - \lambda\epsilon = (k - 1)\epsilon - \lambda\epsilon = (k - \lambda - 1)\epsilon\,.
	\end{aligned}
\end{equation}

Therefore, the Euler class expression reads in this case:
\begin{equation}
	\Eul_n^{[\lambda]} = \prod_{k = 1}^{n}(\lambda + 1 - k)\epsilon (n + 1 - k)\epsilon = \dfrac{n!\lambda!}{(\lambda - n)!}\epsilon^{2n}\,.
\end{equation}

\subsection{Incidence locus}

Now we construct the tangent space to the incidence locus for two neighboring crystals $[n]$ and $[m]=[n+1]$.
Quiver morphism in the fixed point reads:
\begin{equation}
	\bar\tau  = \begin{pmatrix}
		1 & 0 & \dots & 0 & 0\\
		0 & 1 & \dots & 0 & 0\\
		\vdots & \vdots & \ddots & \vdots & \vdots\\
		0 & 0 & \dots & 1 & 0
	\end{pmatrix}_{n\times m}\,.
\end{equation}

Then we consider two tangent spaces $\mathsf{T}_n\mathscr{M}^{[\lambda]}$ and $\mathsf{T}_{m}\mathscr{M}^{[\lambda]}$:
\begin{equation}
	\begin{aligned}
		&\delta B^{(1)} = \begin{pmatrix}
			0 & 0 & \dots & 0 & b_{1n}^{(1)}\\
			0 & 0 & \dots & 0 & b_{2n}^{(1)}\\
			0 & 0 & \dots & 0 & b_{3n}^{(1)}\\
			\vdots & \vdots & \ddots & \vdots & \vdots\\
			0 & 0 & \dots & 0 & b_{nn}^{(1)}
		\end{pmatrix}_{n\times n}, \quad \delta I^{(1)} = \begin{pmatrix}
			0 \\ 0 \\ 0 \\ \vdots \\ 0
		\end{pmatrix}_{n \times 1}, \quad \delta J^{(1)} = \begin{pmatrix}
			j_{1}^{(1)} & j_{2}^{(1)} & \dots & j_{n}^{(1)}
		\end{pmatrix}_{1\times n}\,,\\
		&\delta B^{(2)} = \begin{pmatrix}
			0 & 0 & \dots & 0 & b_{1m}^{(2)}\\
			0 & 0 & \dots & 0 & b_{2m}^{(2)}\\
			0 & 0 & \dots & 0 & b_{3m}^{(2)}\\
			\vdots & \vdots & \ddots & \vdots & \vdots\\
			0 & 0 & \dots & 0 & b_{mm}^{(2)}
		\end{pmatrix}_{m\times m}, \quad \delta I^{(2)} = \begin{pmatrix}
			0 \\ 0 \\ 0 \\ \vdots \\ 0
		\end{pmatrix}_{m \times 1}, \quad \delta J^{(2)} = \begin{pmatrix}
			j_{1}^{(2)} & j_{2}^{(2)} & \dots & j_{m}^{(2)}
		\end{pmatrix}_{1\times m}\,.
	\end{aligned}
\end{equation}

The incidence locus cuts out constraints $b_{km}^{(2)}=0$, $k\leq n$, $j_{m}^{(2)}=0$, $j_k^{(2)}=j_k^{(1)}$, $k\leq n$:
\begin{equation}
	\begin{aligned}
		&\delta B^{(1)} = \begin{pmatrix}
			0 & 0 & \dots & 0 & b_{1n}^{(1)}\\
			0 & 0 & \dots & 0 & b_{2n}^{(1)}\\
			0 & 0 & \dots & 0 & a_{3n}^{(1)}\\
			\vdots & \vdots & \ddots & \vdots & \vdots\\
			0 & 0 & \dots & 0 & b_{nn}^{(1)}
		\end{pmatrix}_{n\times n}, \quad \delta I^{(1)} = \begin{pmatrix}
			0 \\ 0 \\ 0 \\ \vdots \\ 0
		\end{pmatrix}_{n \times 1}, \quad \delta J^{(1)} = \begin{pmatrix}
			j_{1}^{(1)} & j_{2}^{(1)} & \dots & j_{n}^{(1)}
		\end{pmatrix}_{1\times n}\,,\\
		&\delta B^{(2)} = \begin{pmatrix}
			0 & 0 & \dots & 0 & 0\\
			0 & 0 & \dots & 0 & 0\\
			0 & 0 & \dots & 0 & 0\\
			\vdots & \vdots & \ddots & \vdots & \vdots\\
			0 & 0 & \dots & 0 & 0\\
			0 & 0 & \dots & 0 & b_{mm}^{(2)}
		\end{pmatrix}_{m\times m}, \quad \delta I^{(2)} = \begin{pmatrix}
			0 \\ 0 \\ 0 \\ \vdots \\ 0
		\end{pmatrix}_{m \times 1}, \quad \delta J^{(2)} = \begin{pmatrix}
			j_{1}^{(1)} & j_{2}^{(1)} & \dots & j_{n}^{(1)} & 0
		\end{pmatrix}_{1\times m}\,.
	\end{aligned}
\end{equation}

Respective Euler class reads:
\begin{equation}
	\Eul_{n,n+1}^{[\lambda]} =-\epsilon \prod_{k = 1}^{n}(k - \lambda - 1)\epsilon (k - n - 1)\epsilon = \dfrac{n!\lambda!}{(\lambda - n)!}(-\epsilon)^{2n + 1}\,.
\end{equation}

\subsection{Amplitudes}

We derive the following matrix coefficients:
\begin{equation}\label{coefficients}
	{\bf E}_{n, n+1}^{[\lambda]} = \dfrac{\Eul_n^{[\lambda]}}{\Eul_{n, n+1}^{[\lambda]}} = -\dfrac{1}{\epsilon}, \quad {\bf F}_{n, n - 1}^{[\lambda]} = \dfrac{\Eul_{n}^{[\lambda]}}{\Eul_{ n-1, n}^{[\lambda]}} = -n(\lambda - n +1)\epsilon\,.
\end{equation}

Let us show that these matrix coefficients are isomorphic to spin-$\frac{\lambda}{2}$ representations of $\mathfrak{sl}_{2}$.
We use the standard notations $|j, m\rangle$ for the vectors of $\mathfrak{sl}_{2}$. 
In this case, $\lambda = 2j$, $n = m + j$. 
As well, we rescale the vectors so that they are all normalized $|\lambda, n\rangle = \sqrt{\Eul_{(\lambda, n)}}|\lambda, \tilde{n}\rangle$.
Applying these notations we derive:
\begin{align}
	&e_{0}|\lambda, \tilde{n}\rangle  = \sqrt{(n + 1)(\lambda - n)}|\lambda, \widetilde{n + 1}\rangle, \\
	&f_{0}|\lambda, \tilde{n}\rangle  = \sqrt{n(\lambda - n + 1)}|\lambda, \widetilde{n - 1}\rangle; \\
	&e_{0}|j, m\rangle = \sqrt{(m + j + 1)(2j - m - j)}|j, m + 1\rangle = \sqrt{(m + j + 1)(j - m)}|j, m + 1\rangle, \\
	&f_{0}|j, m\rangle = \sqrt{(m + j)(2j - m - j + 1)}|j, m - 1\rangle = \sqrt{(m + j)(j - m + 1)}|j, m - 1\rangle
\end{align}
This is the canonical spin-$j$ representation for $\fs\fl_2$ familiar to the reader from any textbook on quantizing angular momenta in quantum mechanics, e.g. \cite{landau2013quantum}.

Finally, we check a single hysteresis relation  surviving in this case:
\begin{equation}\label{Y(sl(2)) EF relations}
	{\bf E}_{n,n+1}^{[\lambda]}{\bf F}_{n+1,n}^{[\lambda]} = -\dfrac{1}{\epsilon}\mathop{\rm res}\lm_{z = n\epsilon}\bPsi_{n}^{[\lambda]}(z)=(\lambda-n)(n+1)\,.
\end{equation}

\subsection{Representation on polynomials}

The description of the algebra representation in terms of differential operators has proven to be useful. 
We treat generator $e_{0}$ as a generator that adds a box $[n] \to [n + 1]$ to diagram $[n]$ and generator $f_{0}$ as an operator that removes a box $[n] \to [n - 1]$. 
We define the generators of the algebra as some differential operators of one variable $x$, and states (vectors these operators act upon) are defined as monomials $[n] \to \# x^{n} = P_{[n]}$.

Let us fix generator $f_{0}$ as:
\begin{equation}
	f_{0} = \d_{x}.
\end{equation}

Generator $e_{0}$ is searched for with the help of the following ansatz:
\begin{equation}
	e_{0} = a \lambda x + b x^{2}\d_{x} + c_{3} x^{3}\d^{2}_{x} + c_{4}x^{4}\d^{3}_{x}\dots
\end{equation}
since it adds one box $x^{n} \to x^{n + 1}$.
\begin{equation}
	\begin{aligned}
		&e_{0}P_{[n]} = {\bf E}_{n, n + 1}^{[\lambda]}P_{[n + 1]} = P_{[n + 1]},\\
		&f_{0}P_{[n]} = {\bf F}_{n, n - 1}^{[\lambda]}P_{[n - 1]} = n(\lambda - n + 1) P_{[n - 1]}
	\end{aligned}
\end{equation}

Let us consider the fundamental representation where $\lambda = 1$:
\begin{equation}
	[0] \to 1, \quad [1] \to x
\end{equation}

The generator action reads:
\begin{equation}
	\begin{aligned}
		&f_{0}x = 1, \quad e_{0}1 = x = a x, \\
		&e_{0}x = 0 = ax^{2} + bx^{2}, \quad e_{0}x^{2} = 0 = c_{3}x^{3}, \dots ; \\
		&\text{Therefore : } a = 1, \quad b = -a = -1, \quad c_{i} = 0.
	\end{aligned}
\end{equation}

Finally, we derive the following representation for operators:
\begin{equation}
	e_0 = \lambda x - x^2 \frac{\partial}{\partial x}, \hspace{10mm} \psi_1 = 2 (1 - \lambda) \epsilon \cdot  x \frac{\partial}{\partial x} + 3 \epsilon \cdot x^2 \frac{\partial^2}{\partial x^2}, \hspace{10mm} f_0 = \frac{\partial}{\partial x}
\end{equation}
where we have chosen $\psi_{1}$ so that it satisfies relations \eqref{Y(sl(2)) relations in modes}.

\section{Towards representations of \texorpdfstring{$\mathsf{Y}(\fs\fl_n)$}{}}\label{sec: sl(n)}

In this section we would like to use the crystal construction to enumerate vectors in representations of $\mathsf{Y}(\fs\fl_n)$.
We are not planning to go into details and describe the representations.
Instead, we count crystals for these representations and compare their counting with specific representations of $\fs\fl_n$ corresponding to rectangular Young diagrams.

Consider a family of \emph{framed} quivers with superpotentials:
\begin{equation}
	\CQ_{n,p,\lambda}=\left\{\begin{array}{c}
		\begin{tikzpicture}
			\draw[postaction=decorate, decoration={markings, mark= at position 0.7 with {\arrow{stealth}}}] (0,0) to[out=20,in=160] node[pos=0.5,above] {$\scriptstyle A_1$} (1.5,0);
			\draw[postaction=decorate, decoration={markings, mark= at position 0.7 with {\arrow{stealth}}}] (1.5,0) to[out=200,in=340] node[pos=0.5,below] {$\scriptstyle B_1$} (0,0);
			\draw[postaction=decorate, decoration={markings, mark= at position 0.8 with {\arrow{stealth}}}] (0,0) to[out=60,in=0] (0,0.6) to[out=180,in=120] (0,0);
			\node[above] at (0,0.6) {$\scriptstyle C_1$};
			\begin{scope}[shift={(1.5,0)}]
				\draw[postaction=decorate, decoration={markings, mark= at position 0.8 with {\arrow{stealth}}}] (0,0) to[out=60,in=0] (0,0.6) to[out=180,in=120] (0,0);
				\node[above] at (0,0.6) {$\scriptstyle C_2$};
			\end{scope}
			\draw[fill=\myblue] (0,0) circle (0.08) (1.5,0) circle (0.08);
			\begin{scope}[shift = {(4,0)}]
				\draw[postaction=decorate, decoration={markings, mark= at position 0.7 with {\arrow{stealth}}}] (0,-1.2) to[out=100,in=260] node[pos=0.3, left] {$\scriptstyle R$} (0,0);
				\draw[postaction=decorate, decoration={markings, mark= at position 0.7 with {\arrow{stealth}}}] (0,0) to[out=280,in=80] node[pos=0.7, right] {$\scriptstyle S$} (0,-1.2);
				\draw[postaction=decorate, decoration={markings, mark= at position 0.7 with {\arrow{stealth}}}] (0,0) to[out=20,in=160] node[pos=0.5,above] {$\scriptstyle A_p$} (1.5,0);
				\draw[postaction=decorate, decoration={markings, mark= at position 0.7 with {\arrow{stealth}}}] (1.5,0) to[out=200,in=340] node[pos=0.5,below] {$\scriptstyle B_p$} (0,0);
				\begin{scope}[shift={(-1.5,0)}]
					\draw[postaction=decorate, decoration={markings, mark= at position 0.7 with {\arrow{stealth}}}] (0,0) to[out=20,in=160] node[pos=0.5,above] {$\scriptstyle A_{p-1}$} (1.5,0);
					\draw[postaction=decorate, decoration={markings, mark= at position 0.7 with {\arrow{stealth}}}] (1.5,0) to[out=200,in=340] node[pos=0.5,below] {$\scriptstyle B_{p-1}$} (0,0);
				\end{scope}
				\draw[postaction=decorate, decoration={markings, mark= at position 0.8 with {\arrow{stealth}}}] (0,0) to[out=60,in=0] (0,0.6) to[out=180,in=120] (0,0);
				\node[above] at (0,0.6) {$\scriptstyle C_p$};
				\begin{scope}[shift={(1.5,0)}]
					\draw[postaction=decorate, decoration={markings, mark= at position 0.8 with {\arrow{stealth}}}] (0,0) to[out=60,in=0] (0,0.6) to[out=180,in=120] (0,0);
					\node[above] at (0,0.6) {$\scriptstyle C_{p+1}$};
				\end{scope}
				\begin{scope}[shift={(-1.5,0)}]
					\draw[postaction=decorate, decoration={markings, mark= at position 0.8 with {\arrow{stealth}}}] (0,0) to[out=60,in=0] (0,0.6) to[out=180,in=120] (0,0);
					\node[above] at (0,0.6) {$\scriptstyle C_{p-1}$};
				\end{scope}
				\draw[fill=\myblue] (-1.5,0) circle (0.08) (0,0) circle (0.08) (1.5,0) circle (0.08);
				\begin{scope}[shift={(0,-1.2)}]
					\draw[fill=burgundy] (-0.08,-0.08) -- (-0.08,0.08) -- (0.08,0.08) -- (0.08,-0.08) -- cycle;
				\end{scope}
			\end{scope}
			\begin{scope}[shift = {(8,0)}]
				\begin{scope}[shift={(-1.5,0)}]
					\draw[postaction=decorate, decoration={markings, mark= at position 0.7 with {\arrow{stealth}}}] (0,0) to[out=20,in=160] node[pos=0.5,above] {$\scriptstyle A_{n-2}$} (1.5,0);
					\draw[postaction=decorate, decoration={markings, mark= at position 0.7 with {\arrow{stealth}}}] (1.5,0) to[out=200,in=340] node[pos=0.5,below] {$\scriptstyle B_{n-2}$} (0,0);
				\end{scope}
				\draw[postaction=decorate, decoration={markings, mark= at position 0.8 with {\arrow{stealth}}}] (0,0) to[out=60,in=0] (0,0.6) to[out=180,in=120] (0,0);
				\node[above] at (0,0.6) {$\scriptstyle C_{n-1}$};
				\begin{scope}[shift={(-1.5,0)}]
					\draw[postaction=decorate, decoration={markings, mark= at position 0.8 with {\arrow{stealth}}}] (0,0) to[out=60,in=0] (0,0.6) to[out=180,in=120] (0,0);
					\node[above] at (0,0.6) {$\scriptstyle C_{n-2}$};
				\end{scope}
				\draw[fill=\myblue] (-1.5,0) circle (0.08) (0,0) circle (0.08);
			\end{scope}
			\draw[fill=black] (1.75,0) circle (0.03) (2,0) circle (0.03) (2.25,0) circle (0.03) (5.75,0) circle (0.03) (6,0) circle (0.03) (6.25,0) circle (0.03);
		\end{tikzpicture}\\
		W=\Tr\left[A_1C_1B_1+\sum\lm_{a=2}^{n-2}\left(A_aC_aB_a-B_{a-1}C_aA_{a-1}\right)-B_{n-2}C_{n-1}A_{n-2}+{\color{burgundy}R C_p^{\lambda}S}\right]
	\end{array}\right\}\,.
\end{equation}
Here we choose the superpotential analogous to superpotential \eqref{Y(sl(2)) superpotential}.

The unframed quiver is independent of fields $R$, $S$ and of parameters $p$, $\lambda$.
Pictorially, it represents simply a Dynkin diagram for $\fs\fl_n$.
We expect that the framing choice, depending explicitly on $p$ and $\lambda$, cuts out a specific representation of $\fs\fl_n$, classified by the following Young diagram:
\begin{equation}
	\Upsilon_{p,\lambda}=\begin{array}{c}
		\begin{tikzpicture}[scale=0.4]
			\foreach \i in {0,1,2,3}
			{
				\draw[thick] (0,\i) -- (4,\i);
			}
			\foreach \i in {0,1,2,3,4}
			{
				\draw[thick] (\i,0) -- (\i,3);
			}
			\draw (0,0) to[out=270, in=180] (0.25,-0.25) -- (1.75,-0.25) to[out=0,in=90] (2,-0.5) to[out=90,in=180] (2.25,-0.25) -- (3.75,-0.25) to[out=0,in=270] (4,0);
			\draw (4,0) to[out=0,in=270] (4.25,0.25) -- (4.25,1.25) to[out=90,in=180] (4.5,1.5) to[out=180,in=270] (4.25,1.75) -- (4.25,2.75) to[out=90,in=0] (4,3);
			\node[below] at (2,-0.5) {$\scriptstyle p$ \scriptsize columns};
			\node[right] at (4.5,1.5) {$\scriptstyle \lambda$ \scriptsize rows};
		\end{tikzpicture}
	\end{array}\,.
\end{equation}
To present an argument in favor of this correspondence, consider a generating function for DT invariants of a theory associated with quiver $Q$.
Suppose the theory has a set $\CV$ (that might be finite or infinite) of fixed points -- SUSY vacua -- on quiver varieties of different dimensions $\vec d$.
In practice, $\vec d$ are related to fractional charges of D-branes if $Q$ represents a Calabi-Yau manifold.
Therefore we introduce the following generating function for DT invariants:
\begin{equation}
	\mathsf{DT}_{Q}\left(\{q_k\}\right):=\sum\lm_{v\in\CV}\prod\lm_{a\in Q_0} q_a^{d_a(v)}\,.
\end{equation}

Applying directly the fixed point counting algorithm we presented in Sec.~\ref{sec:Algo_Cry} for various values of parameters $n$, $p$, and $\lambda$, we find that the generating function for DT invariants coincides with the  character of $\fs\fl_n$ representation $\Upsilon_{p,\lambda}$, in particular:
\begin{equation}\label{main}
	\mathsf{DT}_{\CQ_{n,p,\lambda}}\left(\{q_k\}\right)=\prod\lm_{i=1}^pq_i^{-\lambda(p-i)}\times{\rm Sch}_{\Upsilon_{p,\lambda}}\left(\{p_k\}\right)\big|_{\fs\fl_n}\,,
\end{equation}
where ${\rm Sch}_R\left(\{p_k\}\right)$ are Schur functions for Young diagrams $R$, and by $\fs\fl_n$ character substitution we imply the following:
\begin{equation}
	p_k \big|_{\fs\fl_n}:=\sum\lm_{i=1}^n\left(\prod\lm_{j=1}^{i-1}q_i^k\right)\,.
\end{equation}

Relation \eqref{main} is a strong indication that, in general, BPS algebra on BPS states of quiver $\CQ_{n,p,\lambda}$ theory forms a $\Upsilon_{p,\lambda}$ representation of Yangian $\mathsf{Y}(\fs\fl_n)$.

\section{Simplest affine Yangian \texorpdfstring{$\mathsf{Y}(\widehat{\fg\fl}_1)$}{}}\label{sec: Y(gl(1))}

The next example of the quiver Yangian algebra that we would like to discuss is $\mathsf{Y}(\widehat{\fg\fl}_1)$.

\subsection{Commutation relations}

Algebra $\mathsf{Y}(\widehat{\fg\fl}_1)$ is constructed in terms of a triplet family of generators $e_n$, $f_n$ and $\psi_n$, $n\in\IN_0$ satisfying the following set of relations \cite{Prochazka:2015deb}:
\begin{equation}\label{gl(1) relations in modes}
	\begin{aligned}
		&[\psi_{i}, \psi_{k}] = 0, \\
		&[e_{j+3}, e_{k}] - 3[e_{j+2}, e_{k+1}] + [e_{j+1}, e_{k+2}] - [e_{j}, e_{k+3}] + \sigma_{2}[e_{j+1}, e_{k}] - \sigma_{2}[e_{j}, e_{k+1}] - \sigma_{3}\{e_{j}, e_{k}\} = 0, \\
		&[f_{j+3}, f_{k}] - 3[f_{j+2}, f_{k+1}] + [f_{j+1}, f_{k+2}] - [f_{j}, f_{k+3}] + \sigma_{2}[f_{j+1}, f_{k}] - \sigma_{2}[f_{j}, f_{k+1}] - \sigma_{3}\{f_{j}, f_{k}\} = 0, \\
		&[e_{j}, f_{k}] - \psi_{j+k} = 0,\\
		&[\psi_{j+3}, e_{k}] - 3[\psi_{j+2}, e_{k+1}] + [\psi_{j+1}, e_{k+2}] - [\psi_{j}, e_{k+3}] + \sigma_{2}[\psi_{j+1}, e_{k}] - \sigma_{2}[\psi_{j}, e_{k+1}] - \sigma_{3}\{\psi_{j}, e_{k}\} = 0, \\
		&[\psi_{j+3}, f_{k}] - 3[\psi_{j+2}, f_{k+1}] + [\psi_{j+1}, f_{k+2}] - [\psi_{j}, f_{k+3}] + \sigma_{2}[\psi_{j+1}, f_{k}] - \sigma_{2}[\psi_{j}, f_{k+1}] - \sigma_{3}\{\psi_{j}, f_{k}\} = 0, \\
		&\text{Sym}_{(j_{1}, j_{2}, j_{3})}[e_{j_{1}}, [e_{j_{2}}, e_{j_{3}+1}]] = 0,\\
		&\text{Sym}_{(j_{1}, j_{2}, j_{3})}[f_{j_{1}}, [f_{j_{2}}, f_{j_{3}+1}]] = 0,
	\end{aligned}
\end{equation}
where $\sigma_{2} = \epsilon_{1}\epsilon_{2} + \epsilon_{2}\epsilon_{3} + \epsilon_{3}\epsilon_{1}$, $\sigma_{3} = \epsilon_{1}\epsilon_{2}\epsilon_{3}$. 
This algebra has two free parameters $\epsilon_1$ and $\epsilon_2$.
Parameter $\epsilon_3$ is constrained by a Calabi-Yau condition $\epsilon_1+\epsilon_2+\epsilon_3=0$.
There are additional ``boundary conditions'' for zero modes:
\begin{equation}
	\begin{aligned}
		&[ \psi_0, e_k] = 0 &\hspace{10mm} &[ \psi_1, e_k] = 0  &\hspace{10mm} &[ \psi_2, e_k] = 2 e_k\\
		&[ \psi_0, f_k] = 0 &\hspace{10mm} &[ \psi_1, f_k] = 0 &\hspace{10mm} &[ \psi_2, f_k] = -2 f_k\\
	\end{aligned}
\end{equation}

As before the generator families could be assembled in generating functions with the help of a spectral parameter $z$:
\begin{equation}\label{generating functions of gl(1)}
	e(z) = \sum_{n\in \mathbb{N}_{0}}\dfrac{e_{n}}{z^{n+1}}, \quad \psi(z) = 1 + \sigma_{3}\sum_{n\in \mathbb{N}_{0}}\dfrac{\psi_{n}}{z^{n+1}}, \quad f(z) = \sum_{n\in \mathbb{N}_{0}}\dfrac{f_{n}}{z^{n+1}}.
\end{equation}

In terms of these generating functions relations \eqref{gl(1) relations in modes} could be rewritten in the following form:
\begin{equation}\label{gl(1) relations}
	\begin{aligned}
		&e(z)e(w) \sim \varphi(z - w)e(w)e(z);\\
		&f(z)f(w) \sim [\varphi(z - w)]^{-1}f(w)f(z);\\
		&\psi(z)e(w) \simeq \varphi(z - w)e(w)\psi(z);\\
		&\psi(z)f(w) \simeq [\varphi(z - w)]^{-1}f(w)\psi(z);\\
		&\sigma_{3}[e(z), f(w)] \sim -\frac{\psi(z) - \psi(w)}{z - w};\\
		&\text{Sym}_{z_{1}, z_{2}, z_{3}}(z_{2} - z_{3})[e(z_{1})[e(z_{2}), e(z_{3})]] = 0;\\
		&\text{Sym}_{z_{1}, z_{2}, z_{3}}(z_{2} - z_{3})[f(z_{1})[f(z_{2}), f(z_{3})]] = 0,
	\end{aligned}
\end{equation}
where notations $\sim$ ($\simeq$) imply that both sides coincide in expansion in $x$ and $y$ up to monomials $x^{j}y^{k\geq 0}$, $x^{j\geq 0}y^{k}$ ($x^{j\geq 0}y^{k\geq 0}$).
Here the bond factor reads:
\begin{equation}\label{gl(1) bonding factor}
	\varphi(z) = \dfrac{(z + \epsilon_{1})(z + \epsilon_{2})(z + \epsilon_{3})}{(z - \epsilon_{1})(z - \epsilon_{2})(z - \epsilon_{3})}\,.
\end{equation}

A triplet of the following operators:
\begin{equation}\label{small set gl(1)}
	\boxed{e_0, \hspace{5mm} \psi_3, \hspace{5mm} f_0} ,
\end{equation}
allows one to generate all the remaining generators of the algebra \cite{morozov20233schurs} via the following relations:
\begin{equation}\label{resursion for gl(1)}
	\begin{aligned}
		&e_{k + 1} = \dfrac{1}{6}[\psi_{3}, e_{k}] - \dfrac{1}{3}\psi_{0}\sigma_{3}e_{k}, \\
		&f_{k + 1} = - \dfrac{1}{6}[\psi_{3}, f_{k}] - \dfrac{1}{3}\psi_{0}\sigma_{3}f_{k}.
	\end{aligned}
\end{equation}

\subsection{Quiver equations}

\begin{figure}[ht!]
	\begin{center}
		\begin{tikzpicture}[rotate=-90]
			\draw[postaction={decorate},decoration={markings,
				mark= at position 0.75 with {\arrow{stealth}}}] (0,0) to[out=60,in=0] (0,1) to[out=180,in=120] (0,0);
			\node[right] at (0,1) {$\scriptstyle B_2$};
			\begin{scope}[rotate=90]
				\draw[postaction={decorate},decoration={markings,
					mark= at position 0.75 with {\arrow{stealth}}}] (0,0) to[out=60,in=0] (0,1) to[out=180,in=120] (0,0);
				\node[above] at (0,1) {$\scriptstyle B_1$};
			\end{scope}
			\begin{scope}[rotate=270]
				\draw[postaction={decorate},decoration={markings,
					mark= at position 0.75 with {\arrow{stealth}}}] (0,0) to[out=60,in=0] (0,1) to[out=180,in=120] (0,0);
				\node[below] at (0,1) {$\scriptstyle B_3$};
			\end{scope}
			\draw[postaction={decorate},decoration={markings,
				mark= at position 0.65 with {\arrow{stealth}}}] (0,-1.5) to[out=120,in=240] node[pos=0.5,above] {$\scriptstyle I$} (0,0);
			\draw[postaction={decorate},decoration={markings,
				mark= at position 0.65 with {\arrow{stealth}}}] (0,0) to[out=300,in=60] node[pos=0.5,below] {$\scriptstyle J$} (0,-1.5);
			\begin{scope}[shift={(0,-1.5)}]
				\draw[fill=burgundy] (-0.1,-0.1) -- (-0.1,0.1) -- (0.1,0.1) -- (0.1,-0.1) -- cycle;
			\end{scope}
			\draw[fill=\myblue] (0,0) circle (0.1);
		\end{tikzpicture}
	\end{center}
	\caption{The quiver corresponding to $\mathbb{C}^{3}$}\label{fig:C3quiver}
\end{figure}
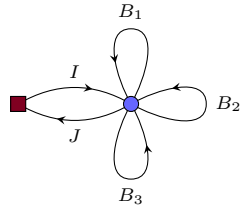

The quiver used to describe the Fock module of $\mathsf{Y}(\widehat{\fg\fl}_1)$ is depicted in Fig.~\ref{fig:C3quiver}.

We choose the following superpotential (see also \cite{Rapcak:2018nsl,Galakhov:2023mak}):
\begin{equation}\label{sup_C3}
	W = \Tr B_{3}\left([B_{1}, B_{2}] + IJ\right)\,.
\end{equation}

This situation describes the canonical representation of a Hilbert scheme of points on $\IC^2$ in terms of a quiver variety \cite{Nakajima_lect}.
To see this, one should note that field $B_3$ in the superpotential \eqref{sup_C3} plays the role of a ``Lagrange multiplier'', and it is set to zero.
Then vacuum equations \eqref{vacuum} transform in this case to the following set of equations:
\begin{equation}\label{ADHM equations for gl(1)}
	\begin{aligned}
		&[B_{1}, B_{1}^{\dagger}] + [B_{2}, B_{2}^{\dagger}] + II^{\dagger} - J^{\dagger}J = \zeta \bbone_{n\times n}\\
		&[B_{1}, B_{2}] + IJ = 0.
	\end{aligned}
\end{equation}
This is a canonical description of moduli spaces ${\rm Hilb}^n(\IC^2)$ with the help of ADHM construction \cite{Nakajima_lect,NakajimaALE}, where $n$ denotes the dimension in the single gauge node of the quiver depicted in Fig.~\ref{fig:C3quiver}.
This quiver variety is well-known to be smooth.

\subsection{Fixed points}\label{sec: Y(gl(1)) fixed points}

\paragraph{Equivariant action.}

Equations \eqref{ADHM equations for gl(1)} also permit an Abelian action of $U(1)\times U(1)$ with two complex parameters, $\epsilon_{1}, \epsilon_{2}$.
For the fixed points in this case, one obtains the following constraints:
\begin{equation}\label{fixed points gl(1)}
	\begin{aligned}
		&[\Phi, B_{i}] = \epsilon_{i}B_{i}, \\
		&\Phi I = 0, \\
\		&-J\Phi = (\epsilon_{1} + \epsilon_{2})J\,.
	\end{aligned}
\end{equation}

\paragraph{Path algebra.}

The fixed points defined by \eqref{fixed points gl(1)} are related to sets of paths on the quiver depicted in Fig.~\ref{fig:C3quiver}. 
The paths may be represented as words of the letters $B_{1}, B_{2}, B_{3}, I, J$. 
Equations \eqref{ADHM equations for gl(1)} and \eqref{fixed points gl(1)} allow us to set $J = 0$.

In our settings, $J = 0$ and $B_{3} = 0$, so the possible paths take the form $B_{1}^{a}B_{2}^{b}I$. 
The cardinality of the set of paths up to $F$-term equivalence is defined by dimension $n$.
The set of paths can be represented as points on a lattice according to Sec.~\ref{sec:Algo_Cry}, where each point corresponds to the path starting from $I$.
The equivalence relation $[B_{1}, B_{2}] = 0$ implies that we do not track the order of $B_{1}, B_{2}$, therefore the following paths are equivalent:
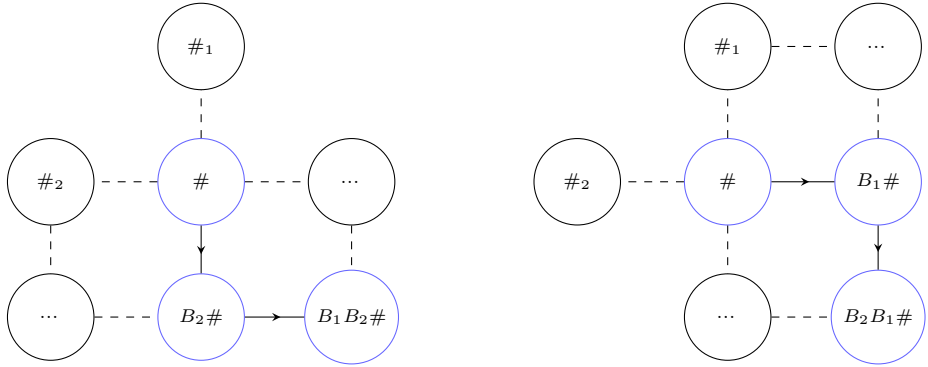
\begin{center}
	\begin{tikzpicture}[yscale=-1, every node/.style={circle, draw = black, minimum size=11.5mm}]
		\draw
		(0, 0) node[draw = \myblue] (1) {$\scriptstyle\#$}
		(0, 1.8) node[draw = \myblue] (2) {$\scriptstyle B_{2}\#$}
		(2, 1.8) node[draw = \myblue] (3) {$\scriptstyle B_{1}B_{2}\#$}
		(7, 0) node[draw = \myblue] (4) {$\scriptstyle \#$}
		(9, 0) node[draw = \myblue] (5) {$\scriptstyle B_{1}\#$}
		(9, 1.8) node[draw = \myblue] (6) {$\scriptstyle B_{2}B_{1}\#$}
		(2, 0) node (1gh1) {$\scriptstyle \dots$}
		(-2, 0) node (1gh2) {$\scriptstyle \#_{2}$}
		(0, -1.8) node (1gh3) {$\scriptstyle \#_{1}$}
		(-2, 1.8) node (1gh4) {$\scriptstyle \dots$}
		(5, 0) node (2gh1) {$\scriptstyle \#_{2}$}
		(7, 1.8) node (2gh2) {$\scriptstyle \dots$}
		(7, -1.8) node (2gh3) {$\scriptstyle \#_{1}$}
		(9, -1.8) node (2gh4) {$\scriptstyle \dots$}
		;
		\draw[postaction={decorate}, decoration={markings, mark= at position 0.6 with {\arrow{stealth}}}] (1) to (2);
		\draw[postaction={decorate}, decoration={markings, mark= at position 0.6 with {\arrow{stealth}}}] (2) to (3);
		\draw[postaction={decorate}, decoration={markings, mark= at position 0.6 with {\arrow{stealth}}}] (4) to (5);
		\draw[postaction={decorate}, decoration={markings, mark= at position 0.6 with {\arrow{stealth}}}] (5) to (6);
		\draw[-, dashed] (1) to (1gh2);
		\draw[-, dashed] (1) to (1gh3);
		\draw[-, dashed] (1) to (1gh1);
		\draw[-, dashed] (1gh4) to (2);
		\draw[-, dashed] (1gh2) to (1gh4);
		\draw[-, dashed] (1gh1) to (3);
		\draw[-, dashed] (4) to (2gh1);
		\draw[-, dashed] (4) to (2gh2);
		\draw[-, dashed] (4) to (2gh3);
		\draw[-, dashed] (5) to (2gh4);
		\draw[-, dashed] (2gh2) to (6);
		\draw[-, dashed] (2gh3) to (2gh4);
	\end{tikzpicture}
\end{center}

As we explained in Sec.~\ref{sec:toric_quiver}, the equivalence of such paths implies that an equivalence class of paths is described by the destination point for the path -- the atom.
And, moreover, collections of all paths are classified by crystals that, in our case, coincide with Young diagrams.
For example, if we have paths corresponding to monomials $\{I, B_{2}I, B_{1}B_2I\}$ then $B_{1}(B_2I)=B_2 (B_1 I)$, and a monomial in the brackets in the r.h.s. $B_1 I$ should acquire a non-zero expectation value as well.
Therefore set $\{I, B_{1}I, B_{2}I, B_{1}B_{2}I\}$ corresponds to a fixed point and can be identified with the following diagram:\footnote{See also an enumeration of these fixed points in terms of A-brane special Lagrangians on a mirror dual curve in \cite{Banerjee:2024smk}.}
\begin{equation}
	\begin{array}{c}
	\begin{tikzpicture}[yscale=-1, every node/.style={circle, draw = black, minimum size=11.5mm}]
		\draw
		(0, 0) node (1) {$\scriptstyle I$}
		(0, 1.8) node (2) {$\scriptstyle B_{2}I$}
		(2, 1.8) node (4) {$\scriptstyle B_{1}B_{2}I$}
		(2, 0) node (3) {$\scriptstyle B_{1}I$}
		;
		\draw[postaction={decorate}, decoration={markings, mark= at position 0.5 with {\arrow{stealth}}}] (1) to (2);
		\draw[postaction={decorate}, decoration={markings, mark= at position 0.5 with {\arrow{stealth}}}] (1) to (3);
		\draw[postaction={decorate}, decoration={markings, mark= at position 0.5 with {\arrow{stealth}}}] (2) to (4);
		\draw[postaction={decorate}, decoration={markings, mark= at position 0.5 with {\arrow{stealth}}}] (3) to (4);
	\end{tikzpicture}
	\end{array}\quad\longleftrightarrow\quad \begin{array}{c}
	\begin{tikzpicture}[scale=0.25, baseline={(0, -0.3)}]
		\foreach \i/\j in {0/-2, 0/-1, 0/0, 1/-2, 1/-1, 1/0}
		{
			\draw (\i,\j) -- (\i+1,\j);
		}
		\foreach \i/\j in {0/-1, 0/0, 1/-1, 1/0, 2/-1, 2/0}
		{
			\draw (\i,\j) -- (\i,\j-1);
		}
	\end{tikzpicture}
	\end{array}\,.
\end{equation}

Generators in crystal representations could be described in the following way (see Sec.~\ref{sec:cry_rep}):
\begin{equation}\label{ansatz for gl(1)}
	\begin{aligned}
		&e(z)|\lambda\rangle = \sum_{\Box\in \text{Add}(\lambda)}\dfrac{1}{z - h(\Box)}{\bf E}_{\lambda, \lambda+\Box}|\lambda + \Box\rangle\,,\\
		&\psi(z)|\lambda\rangle ={\bf \Psi}_{\lambda}(z)|\lambda\rangle\,,\\
		&f(z)|\lambda\rangle = \sum_{\Box\in \text{Rem}(\lambda)}\dfrac{1}{z - h(\Box)}{\bf F}_{\lambda, \lambda-\Box}|\lambda - \Box\rangle\,,
	\end{aligned}
\end{equation}
where the eigenvalues are defined by the following charge function:
\begin{equation}\label{eigenvalues gl(1)}
	{\bf \Psi}_{\lambda}(z) = \psi_{0}(z)\prod_{\Box\in \lambda}\varphi(z - h(\Box)), \quad \psi_{0}(z) = \dfrac{z + \sigma_{3}\psi_{0}}{z}.
\end{equation}

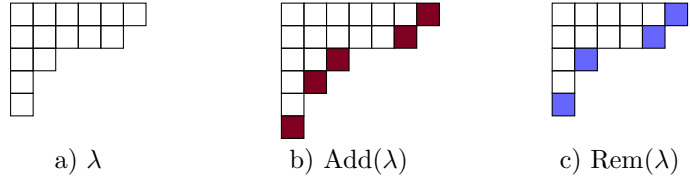
\begin{figure}[ht!]
	\begin{center}
		\begin{tikzpicture}[scale = 0.3]
			\begin{scope}
				\foreach \x/\y in {0/0, 0/1, 0/2, 0/3, 0/4, 1/0, 1/1, 1/2, 2/0, 2/1, 3/0, 3/1, 4/0, 4/1, 5/0}
				{
					\draw (\x,-\y) -- (\x+1,-\y) -- (\x+1,-\y-1) -- (\x,-\y-1) -- cycle;
				}
				\node at (3,-7) {a) $\lambda$};
			\end{scope}
			\begin{scope}[shift={(12,0)}]
				\foreach \x/\y in {0/0, 0/1, 0/2, 0/3, 0/4, 1/0, 1/1, 1/2, 2/0, 2/1, 3/0, 3/1, 4/0, 4/1, 5/0}
				{
					\draw (\x,-\y) -- (\x+1,-\y) -- (\x+1,-\y-1) -- (\x,-\y-1) -- cycle;
				}
				\foreach \x/\y in {0/5,1/3,2/2,5/1,6/0}
				{
					\draw[fill=burgundy] (\x,-\y) -- (\x+1,-\y) -- (\x+1,-\y-1) -- (\x,-\y-1) -- cycle;
				}
				\node at (3,-7) {b) ${\rm Add}(\lambda)$};
			\end{scope}
			\begin{scope}[shift={(24,0)}]
				\foreach \x/\y in {0/0, 0/1, 0/2, 0/3, 1/0, 1/1, 2/0, 2/1, 3/0, 3/1, 4/0}
				{
					\draw (\x,-\y) -- (\x+1,-\y) -- (\x+1,-\y-1) -- (\x,-\y-1) -- cycle;
				}
				\foreach \x/\y in {0/4, 1/2, 4/1, 5/0}
				{
					\draw[fill=\myblue] (\x,-\y) -- (\x+1,-\y) -- (\x+1,-\y-1) -- (\x,-\y-1) -- cycle;
				}
				\node at (3,-7) {c) ${\rm Rem}(\lambda)$};
			\end{scope}
		\end{tikzpicture}
		\caption{An example of $\text{Add}(\lambda)$ and $\text{Rem}(\lambda)$ sets for $\{5,3,2,2,2,1\}$.}\label{fig:AddRem}
	\end{center}
\end{figure}

The sets $\text{Add}(\lambda)$ and $\text{Rem}(\lambda)$ list the boxes in the plane that could be added to (resp. removed from) partition $\lambda$ so that a new set of boxes $\lambda\pm\Box$ is again a Young diagram.
See an example in Fig.~\ref{fig:AddRem}.

The weights of boxes are defined as:
\begin{equation}
	h(\Box) = x_{\Box}\epsilon_{1} + y_{\Box}\epsilon_{2}\,,
\end{equation}
where $(x_{\Box}, y_{\Box})$ are coordinates in the equivariant plane.

\subsection{Euler classes}

As an example, we consider transition $\resizebox{6mm}{!}{\ydiagram{3}} \to \resizebox{8mm}{!}{\ydiagram{4}}$. 

For $\resizebox{6mm}{!}{\ydiagram{3}}$ we construct the following solution to \eqref{ADHM equations for gl(1)}:
\begin{equation}\label{vacuum solution for example in gl(1)}
	\bar \Phi^{(1)} = \begin{pmatrix}
		0 & 0& 0\\
		0 & \epsilon_{1} & 0\\
		0 & 0 & 2\epsilon_{1}
	\end{pmatrix}, \quad
	\bar B_1^{(1)} = \begin{pmatrix}
		0 & 0 & 0\\
		\sqrt{2\zeta} & 0 & 0\\
		0 & \sqrt{\zeta} & 0
	\end{pmatrix}, \quad
	\bar B_2^{(1)} = 0_{3\times 3}, \quad
	\bar I^{(1)} = \begin{pmatrix}
		\sqrt{3\zeta} \\ 0 \\ 0
	\end{pmatrix},\quad \bar J^{(1)}=0_{1\times 3}.
\end{equation}

Similarly, the vacuum solution for $\resizebox{8mm}{!}{\ydiagram{4}}$ has the following form:
\begin{equation}\label{solution [4] for gl(1)}
	\bar \Phi^{(2)} = \begin{pmatrix}
		0 & 0& 0 & 0\\
		0 & \epsilon_{1} & 0& 0\\
		0 & 0 & 2\epsilon_{1} & 0\\
		0 & 0 & 0 & 3\epsilon_1
	\end{pmatrix},\,
	\bar B_{1}^{(2)} = \begin{pmatrix}
		0 & 0 & 0& 0\\
		\sqrt{3\zeta} & 0 & 0 & 0\\
		0 & \sqrt{2\zeta} & 0 & 0\\
		0 & 0 & \sqrt{\zeta} & 0
	\end{pmatrix}, \,
	\bar B_{2}^{(2)} = 0_{4\times 4}, \,
	\bar I^{(2)} = \begin{pmatrix}
		\sqrt{4\zeta} \\ 0\\0\\0
	\end{pmatrix},\, 
	 \bar J^{(2)}=0_{1\times 4}.
\end{equation}

We have constructed these solutions as orbit representatives in the l.h.s. in \eqref{NSKH}, so that matrix elements of the chiral fields depend on $\zeta$ explicitly.
This solution reflects the validity of the construction in the cyclic chamber $\zeta>0$ only.
If one decided to leave the cyclic chamber, the set of fixed points would experience wall-crossing phenomena \cite{Galakhov:2024foa}.
In this particular case, when $\zeta<0$ the roles of fields $I$ and $J$ are interchanged, and $J$ acquires a vacuum expectation value whereas $I=0$.
However, to calculate the Euler classes, we use an element of the r.h.s. orbit in \eqref{NSKH}, so we scale back all $\sqrt{N\zeta}\to 1$.

Thus we consider the tangent space to the fixed points in the following form:
\begin{equation}
	\begin{aligned}
		&B_{1}^{(1)} = \bar B_{1}^{(1)} + \delta B_{1}^{(1)} = \begin{pmatrix}
			a_{11} & a_{12} & a_{13}\\
			a_{21} + 1 & a_{22} & a_{23}\\
			a_{31} & 1 + a_{32} & a_{33}
		\end{pmatrix}, \\
		&B_{2}^{(1)} = \bar B_{2}^{(1)} +
		\delta B_{2}^{(1)} = \begin{pmatrix}
			b_{11} & b_{12} & b_{13}\\
			b_{21} & b_{22} & b_{23}\\
			b_{31} & b_{32} & b_{33}
		\end{pmatrix}, \\
		&I^{(1)} = \bar I^{(1)} + \delta I^{(1)} = \begin{pmatrix}
			i^{(1)}_{1} + 1\\  i^{(1)}_{2}\\i^{(1)}_{3}
		\end{pmatrix},\\
		&B_{1}^{(2)} = \bar B_{1}^{(2)} + \delta B_{1}^{(2)} = \begin{pmatrix}
			c_{11} & c_{12} & c_{13} & c_{14}\\
			c_{21} + 1 & c_{22} & c_{23} & c_{24}\\
			c_{31} & 1 + c_{32} & c_{33} & c_{34} \\
			c_{41} & c_{42} & c_{43} + 1 & c_{44}
		\end{pmatrix}, \\
		&B_{2}^{(2)} = \bar B_{2}^{(2)} +
		\delta B_{2}^{(2)} = \begin{pmatrix}
			d_{11} & d_{12} & d_{13} & d_{14}\\
			d_{21} & d_{22} & d_{23} & d_{24}\\
			d_{31} & d_{32} & d_{33} & d_{34}\\
			d_{41} & d_{42} & d_{43} & d_{44}
		\end{pmatrix}, \\
		&I^{(2)} = \bar I^{(2)} + \delta I^{(2)} = \begin{pmatrix}
			i^{(2)}_{1} + 1\\ i^{(2)}_{2}\\ i^{(2)}_{3} \\ i^{(2)}_{4}
		\end{pmatrix}.
	\end{aligned}
\end{equation}

After excluding the gauge degrees of freedom and those that are non-tangent to the F-term surface, we acquire the following parametrization of the chiral field matrices.
The rest of matrices are simply zeroes:
\begin{equation}
	\begin{aligned}
		&B_{1}^{(1)} = \begin{pmatrix}
			0 & 0 & a_{13}\\
			1 & 0 & a_{23}\\
			0 & 1 & a_{33}
		\end{pmatrix},
		&B_{2}^{(1)} = \begin{pmatrix}
			b_{11} & 0 & 0\\
			b_{21} & b_{11} & 0\\
			b_{31} & b_{21} & b_{11}
		\end{pmatrix};\\
		&B_{1}^{(2)} = \begin{pmatrix}
			0 & 0 & 0 & c_{14}\\
			1 & 0 & 0 & c_{24}\\
			0 & 1 & 0& c_{34}\\
			0 & 0 & 1 & c_{44}
		\end{pmatrix},
		&B_{2}^{(2)} = \begin{pmatrix}
			d_{11} & 0 & 0 & 0\\
			d_{21} & d_{11} & 0 & 0\\
			d_{31} & d_{21} & d_{11} & 0\\
			d_{41} & d_{31} & d_{21} & d_{11}
		\end{pmatrix}\,.
	\end{aligned}
\end{equation}

Weights of the remaining tangent directions are defined by the following expressions:
\begin{equation}
	w[(B_{a}^{(s)})_{ij}] = \bar\Phi^{(s)}_{ii} - \bar\Phi^{(s)}_{jj} - \epsilon_{a}, \quad w[I^{(s)}_{i}] = \bar\Phi^{(s)}_{ii}\,.
\end{equation}

Summarizing, we arrive at the following expression for the Euler class:
\begin{equation}\label{Eul[3] for gl(1)}
	\Eul_{\resizebox{6mm}{!}{\ydiagram{3}}} = 6\epsilon_{1}^{3}\epsilon_{2}(\epsilon_{2} - \epsilon_{1})(\epsilon_{2} - 2\epsilon_{1})\,.
\end{equation}

And the expression for the second Euler class reads:
\begin{equation}\label{Eul[4] for gl(1)}
	\Eul_{\resizebox{8mm}{!}{\ydiagram{4}}} = 24\epsilon_{1}^{4}\epsilon_{2}(\epsilon_{2} - \epsilon_{1})(\epsilon_{2} - 2\epsilon_{1})(\epsilon_{2} - 3\epsilon_{1})\,.
\end{equation}

The next step is to calculate $\Eul_{\resizebox{6mm}{!}{\ydiagram{3}}, \resizebox{8mm}{!}{\ydiagram{4}}}$.

\subsection{Incidence locus}

To define the incidence locus, we have to construct a homomorphism $\tau$ between fixed points that acts as follows:
\begin{equation}\label{homomorphism example for gl(1)}
	\begin{aligned}
		&\tau B_{i}^{(2)} = B_{i}^{(1)}\tau, \quad \tau I^{(2)} = I^{(1)}, \\
		&\tau = \bar\tau + \delta \tau, \\
		&\tau = \begin{pmatrix}
			1 & 0 & 0 & 0\\
			0 & 1 & 0 & 0\\
			0 & 0 & 1 & 0
		\end{pmatrix} +
		\begin{pmatrix}
			t_{11} & t_{12} & t_{13} & t_{14}\\
			t_{21} & t_{22} & t_{23} & t_{24}\\
			t_{31} & t_{32} & t_{33} & t_{34}
		\end{pmatrix}.
	\end{aligned}
\end{equation}

Homomorphism equations impose constraints on $t_{ij}$:

\begin{equation}
	\begin{Bmatrix}
		t_{11} = t_{12} = t_{13} = 0, & t_{21} = t_{31} = 0 & t_{22} = t_{11} & t_{23} = t_{12}\\
		t_{24} - t_{13} - a_{23} = 0 & t_{14} = a_{13} & t_{14} = c_{24} & t_{24} = c_{34}\\
		t_{34} - t_{23} - a_{33} = 0 & t_{33} = t_{22} & t_{21} = t_{32} & \\
		d_{11} = b_{11} & d_{21} = b_{21} & d_{31} = b_{31} & c_{14} = 0
	\end{Bmatrix}
\end{equation}

Solving constraints on $t_{ij}$ we get the incidence locus:
\begin{equation}
	\begin{Bmatrix}
		&c_{14} = 0 & c_{24} = 0 & c_{34} = 0\\
		&d_{11} = b_{11} & d_{21} = b_{21} & d_{31} = b_{31}
	\end{Bmatrix}.
\end{equation}

Tangent directions now read:
\begin{equation}
	\begin{aligned}
		&\delta B_{1}^{(1)} = \begin{pmatrix}
			0 & 0 & a_{13}\\
			0 & 0 & a_{23}\\
			0 & 0 & a_{33}
		\end{pmatrix},
		&\delta B_{2}^{(1)} = \begin{pmatrix}
			b_{11} & 0 & 0\\
			b_{21} & b_{11} & 0\\
			b_{31} & b_{21} & b_{11}
		\end{pmatrix};\\
		&\delta B_{1}^{(2)} = \begin{pmatrix}
			0 & 0 & 0 & 0\\
			0 & 0 & 0 & 0\\
			0 & 0 & 0& 0\\
			0 & 0 & 0 & c_{44}
		\end{pmatrix},
		&\delta B_{2}^{(2)} = \begin{pmatrix}
			b_{11} & 0 & 0 & 0\\
			b_{21} & b_{11} & 0 & 0\\
			b_{31} & b_{21} & b_{11} & 0\\
			d_{41} & b_{31} & b_{21} & b_{11}
		\end{pmatrix}\,.
	\end{aligned}
\end{equation}

Finally, the character is $\Eul_{\resizebox{6mm}{!}{\ydiagram{3}}, \resizebox{8mm}{!}{\ydiagram{4}}} = 6\epsilon_{1}^{4}\epsilon_{2}(\epsilon_{1} - \epsilon_{2})(\epsilon_{2} - 2\epsilon_{1})(\epsilon_{2} - 3\epsilon_{1})$, and matrix element has the form:
\begin{equation}
	{\bf E}_{\resizebox{6mm}{!}{\ydiagram{3}}, \resizebox{8mm}{!}{\ydiagram{4}}} = \dfrac{\Eul_{\resizebox{6mm}{!}{\ydiagram{3}}}}{\Eul_{\resizebox{6mm}{!}{\ydiagram{3}}, \resizebox{8mm}{!}{\ydiagram{4}}}} = \dfrac{1}{\epsilon_{1}(\epsilon_{2} - 3\epsilon_{1})}.
\end{equation}

One can derive in a similar manner other coefficients:
\begin{equation}
	\Eul_{\resizebox{6mm}{!}{\ydiagram{3}}, \resizebox{6mm}{!}{\ydiagram{3, 1}}} = 2\epsilon_{1}^{3}(\epsilon_{1} - \epsilon_{2})(\epsilon_{2} - 2\epsilon_{1})(\epsilon_{2} - 3\epsilon_{1})\epsilon_{2}^{2}, \quad {\bf E}_{\resizebox{6mm}{!}{\ydiagram{3}}, \resizebox{6mm}{!}{\ydiagram{3, 1}}} = \dfrac{3}{\epsilon_{2}(3\epsilon_{1} - \epsilon_{2})}.
\end{equation}

\subsection{Amplitudes and hook formulas}

One could use the algorithm described above to derive ``hook'' formulas for Euler characters (cf. \cite{Galakhov:2023mak}):
\begin{equation}\label{gl(1) hook formulas}
	\begin{aligned}
		&\Eul_{\lambda} = \prod_{\Box \in \lambda}\left[\epsilon_{2}\mathbf{l}_{\lambda}(\Box) - \epsilon_{1}\bigl(\mathbf{a}_{\lambda}(\Box) + 1\bigr)\right]\cdot \left[\epsilon_{1}\mathbf{a}_{\lambda}(\Box) - \epsilon_{2}\bigl(\mathbf{l}_{\lambda}(\Box) + 1\bigr)\right]\,,\\
		&\Eul_{\lambda, \lambda + \Box} = \epsilon_{1}\epsilon_{2}\prod_{\Box'\in \lambda}\left[\epsilon_{2}\mathbf{l}_{\lambda'}(\Box') - \epsilon_{1}\bigl(\mathbf{a}_{\lambda'}(\Box') + 1 - \delta_{y_{\Box}, y_{\Box'})}\bigr)\right]\left[\epsilon_{1}\mathbf{a}_{\lambda'}(\Box') - \epsilon_{2}\bigl(\mathbf{l}_{\lambda'}(\Box') + 1 - \delta_{x_{\Box}, x_{\Box'}}\bigr)\right]\,.
	\end{aligned}
\end{equation}

\begin{figure}[h]
	\begin{center}
		\begin{tikzpicture}
		\begin{scope}[scale=0.4]
			\foreach \x/\y/\z/\w in {0/-5/1/-5, 0/-4/0/-5, 0/-4/1/-4, 0/-3/0/-4, 0/-3/1/-3, 0/-2/0/-3, 0/-2/1/-2, 0/-1/0/-2, 0/-1/1/-1, 0/0/0/-1, 0/0/1/0, 1/-5/2/-5, 1/-4/1/-5, 1/-4/2/-4, 1/-3/1/-4, 1/-3/2/-3, 1/-2/1/-3, 1/-2/2/-2, 1/-1/1/-2, 1/-1/2/-1, 1/0/1/-1, 1/0/2/0, 2/-4/2/-5, 2/-4/3/-4, 2/-3/2/-4, 2/-3/3/-3, 2/-2/2/-3, 2/-2/3/-2, 2/-1/2/-2, 2/-1/3/-1, 2/0/2/-1, 2/0/3/0, 3/-3/3/-4, 3/-3/4/-3, 3/-2/3/-3, 3/-2/4/-2, 3/-1/3/-2, 3/-1/4/-1, 3/0/3/-1, 3/0/4/0, 4/-3/5/-3, 4/-2/4/-3, 4/-2/5/-2, 4/-1/4/-2, 4/-1/5/-1, 4/0/4/-1, 4/0/5/0, 5/-3/6/-3, 5/-2/5/-3, 5/-2/6/-2, 5/-1/5/-2, 5/-1/6/-1, 5/0/5/-1, 5/0/6/0, 6/-2/6/-3, 6/-2/7/-2, 6/-1/6/-2, 6/-1/7/-1, 6/0/6/-1, 6/0/7/0, 7/-1/7/-2, 7/-1/8/-1, 7/0/7/-1, 7/0/8/0, 8/-1/9/-1, 8/0/8/-1, 8/0/9/0, 9/0/9/-1}
			{
				\draw (\x,\y) -- (\z,\w);
			}
			\draw[fill=burgundy] (1,-1) -- (2,-1) -- (2,-2) -- (1,-2) -- cycle;
			\draw[fill=\myblue] (1,-2) -- (2,-2) -- (2,-5) -- (1,-5) -- cycle;
			\node[rotate=90,white] at (1.5,-3.5) {\small leg};
			\draw[fill=black!70!green] (2,-1) -- (7,-1) -- (7,-2) -- (2,-2) -- cycle;
			\node[white] at (4.5,-1.5) {\small arm};
		\end{scope}
		\end{tikzpicture}
		\caption{The arm and the leg.}\label{fig:armleg}
	\end{center}
\end{figure}
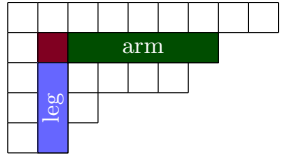

Funstions $\mathbf{a}_{\lambda}(\Box)$ and $\mathbf{l}_{\lambda}(\Box)$ are an ``arm'' and a ``leg'' of box $\Box$ in diagram $\lambda$. 
The ``leg'' is the length of a string of all the boxes below $\Box$. 
An ``arm'' is the length of all the boxes located to the right of box $\Box$.
See an example in Fig.~\ref{fig:armleg}.

\subsection{Representation on polynomials}

The states of the Fock representation of the Yangian algebra $\mathsf{Y}(\widehat{\fg\fl}_1)$ can be labeled by 2-dimensional Young diagrams. Young diagrams have natural correspondence with so-called time variables.
\begin{equation}
	\begin{tblr}{|c|c|c|}
		\hline
		n & \mbox{time-variables} & \mbox{Young diagram} \\ \hline
		1 & p_{1} & \ydiagram{1} \\ \hline
		2 & p_{1}^{2}, p_{2} & \ydiagram{2}, \ydiagram{1, 1}\\ \hline
		3 & p_{1}^{3}, p_{2}p_{1}, p_{3} & \ydiagram{3}, \ydiagram{2, 1}, \ydiagram{1, 1, 1} \\ \hline
		4 & p_{1}^{4}, p_{3}p_{1}, p_{2}^{2}, p_{2}p_{1}^{2}, p_{4} & \ydiagram{4}, \ydiagram{3, 1}, \ydiagram{2, 2}, \ydiagram{2, 1, 1}, \ydiagram{1, 1, 1, 1} \\ \hline
		\vdots & \dots & \dots \\ \hline
	\end{tblr}
\end{equation}

One can associate the states of the representation with some polynomial of time variables and the generators of the algebra with some differential generators. The representation of $\mathsf{Y}(\widehat{\fg\fl}_1)$ in these terms is well-known in the literature \cite{Prochazka:2015deb}.

We consider generator $e_{0}$ as an operator that adds a box to a diagram, and generator $f_{0}$ as an operator that removes a box from the diagram. In terms of time-variables, these generators take the form:
\begin{equation}
	e_{0} = p_{1}, \quad f_{0} = -\dfrac{\d}{\d p_{1}} = -\d_{1}.
\end{equation}

To define all the generators of the algebra we have to describe generator $\psi_{3}$ in terms of time variables.
The algebra has parameters $\epsilon_{1}, \epsilon_{2}, \epsilon_{3}$ that are constrained by $\epsilon_{1} + \epsilon_{2} + \epsilon_{3} = 0$. 
We set generator $\psi_{3}$ in the time variable description as follows:
\begin{equation}\label{cut-and-join gl(1)}
	\psi_{3} = 3\sum_{a, b = 1}^{\infty} \bigl(ab\cdot p_{a + b}\d_{a}\d_{b} + (a + b)p_{a}p_{b}\d_{a + b}\bigr) + \sigma_{3}\sum_{a = 1}^{\infty}a(3a - 1)p_{a}\d_{a}\,,
\end{equation}
where $\d_{a} = \frac{\d}{\d p_{a}}$. 
Actually, the time variable description gives a one-parametric Yangian representation since expression \eqref{cut-and-join gl(1)} satisfies relations \eqref{gl(1) relations in modes} only if
\begin{equation}
	\sigma_{3}^{2} + \sigma_{2} + 1 = 0.
\end{equation}

This constraint can be solved as $\epsilon_{1} = -h,~\epsilon_{2} = \frac{1}{h},~\epsilon_{3} = h - \frac{1}{h}$. Picking $h = 1$ results in $\sigma_{3} = 0$, and the Yangian algebra $\mathsf{Y}(\widehat{\fg\fl}_1)$ with these parameters becomes the $\mathcal{W}_{1 + \infty}$ algebra \cite{Prochazka:2015deb}.

Using equations \eqref{resursion for gl(1)} or \eqref{gl(1) relations in modes}, one can derive the full set of generators.

The states are defined as eigenvalues of \eqref{cut-and-join gl(1)} and coincide with the famous Jack polynomials. If $h = 1$, Jack polynomials become Schur functions. For example:
\begin{equation}\label{Jacks}
	\begin{aligned}
		&J_{\ydiagram{1}} = p_{1}, \quad J_{\ydiagram{2}} = p_{1}^{2} - hp_{2}, \quad J_{\ydiagram{1, 1}} = p_{1}^{2} + \dfrac{1}{h}p_{2}, \\
		&J_{\resizebox{6mm}{!}{\ydiagram{3}}} = p_{1}^{3} -3hp_{1}p_{2} + 2h^{2}p_{3},\quad
		J_{\resizebox{!}{2mm}{\ydiagram{2, 1}}} = p_{1}^{3} +\left(\dfrac{1}{h} - h\right)p_{2}p_{1} - p_{3},\\
		&J_{\resizebox{2mm}{!}{\ydiagram{1, 1, 1}}} = p_{1}^{2} + \dfrac{3}{h}p_{2}p_{1} +\dfrac{2}{h^{2}}p_{3}.
	\end{aligned}
\end{equation}

Since we treat states as $J_{\lambda}$ one can associate coefficients $E_{\lambda, \lambda + \Box}$ to Littlewood-Richardson coefficients:
\begin{equation}\label{Lit-Rich coeffs gl(1)}
	e_{0}J_{\lambda} = \sum_{\Box \in \text{Add}(\lambda)}{\bf E}_{\lambda, \lambda+\Box}J_{\lambda + \Box}.
\end{equation}

This leads to another way to define polynomials $J_{\lambda}$ if matrix coefficients ${\bf E}_{\lambda, \lambda + \Box}$ are known.

The prescription defined above has far-reaching applications and generalizations. 
For example, the generalization of 2d-Young diagrams to 3d-Young diagrams \cite{morozov20233schurs} should provide a definition of 3-Schur functions. 
Generalization of the algebra to $\mathsf{Y}(\widehat{\fg\fl}_r)$ gives rise to Uglov polynomials \cite{galakhov2024simple}.

\section{Simplest affine super Yangian \texorpdfstring{$\mathsf{Y}(\widehat{\fg\fl}_{1|1})$}{}}\label{sec: Y(gl(1|1))}

\subsection{Commutation relations}

In this case, the algebra corresponds to a $\widehat{\fg\fl}_{1|1}$ affine Dynkin diagram that has two odd nodes we denote as $+$ and $-$:
\begin{equation}
	\begin{array}{c}
	\begin{tikzpicture}
		\tikzstyle{col1} = [fill=gray]
		\tikzstyle{col2} = [fill=\myblue]
		\draw (0,0) to[out=30,in=150] (2,0);
		\draw (2,0) to[out=210,in=330] (0,0);
		\draw[col1] (0,0) circle (0.1);
		\draw[col2] (2,0) circle (0.1);
		\node[left] at (-0.1,0) {$+$};
		\node[right] at (2.1,0) {$-$};
	\end{tikzpicture}
	\end{array}\,.
\end{equation}

We have two respective families of odd (fermionic) raising and lowering generators $e^{\pm}_k$, $f^{\pm}_k$, $k\in\IN_0$, and of even (bosonic) Cartan generators $\psi_n^{\pm}$, $n\in \IZ$.

The algebra is defined by the following (super-)commutation relations among the generators:
\begin{equation}\label{gl(1|1) relations in modes}
	\begin{aligned}
		&\{e^{+}_{n}, e^{-}_{k}\} = \{f^{+}_{n}, f^{+}_{k}\} = \{e^{-}_{n}, e^{-}_{k}\} = \{f^{-}_{n}, f^{-}_{k}\} = 0, \\
		&[\psi^{+}_{n}, e^{+}_{k}] = [\psi^{+}_{n}, f_{k}^{+}] = [\psi_{n}^{-}, e_{k}^{-}] = [\psi^{-}_{n}, f_{k}^{-}] = 0, \\
		&\{e^{+}_{n + 2}, e^{-}_{k}\} - 2\{e_{n + 1}^{+}, e^{-}_{k + 1}\} + \{e_{n}^{+}, e_{k + 2}^{-}\} - \dfrac{h^{2}_{1} + h_{2}^{2}}{2}\{e_{n}^{+}, e_{k}^{-}\} + \dfrac{h_{1}^{2} - h_{2}^{2}}{2}[e^{+}_{n}, e_{k}^{-}] = 0, \\
		&\{f^{+}_{n + 2}, f^{-}_{k}\} - 2\{f_{n + 1}^{+}, f^{-}_{k + 1}\} + \{f_{n}^{+}, f_{k + 2}^{-}\} - \dfrac{h^{2}_{1} + h_{2}^{2}}{2}\{f_{n}^{+}, f_{k}^{-}\} + \dfrac{h_{1}^{2} - h_{2}^{2}}{2}[f^{+}_{n}, f_{k}^{-}] = 0,\\
		&[\psi^{+}_{n + 2}, e^{-}_{k}] - 2[\psi_{n + 1}^{+}, e_{k + 1}^{-}] + [\psi_{n}^{+}, e_{k + 2}^{-}] - \dfrac{h_{1}^{2} + h_{2}^{2}}{2}[\psi_{n}^{+}, e_{k}^{-}] + \dfrac{h_{1}^{2} - h_{2}^{2}}{2}\{\psi_{n}^{+}, e_{k}^{-}\} = 0,\\
		&[\psi^{+}_{n + 2}, f^{-}_{k}] - 2[\psi_{n + 1}^{+}, f_{k + 1}^{-}] + [\psi_{n}^{+}, f_{k + 2}^{-}] - \dfrac{h_{1}^{2} + h_{2}^{2}}{2}[\psi_{n}^{+}, f_{k}^{-}] - \dfrac{h_{1}^{2} - h_{2}^{2}}{2}\{\psi_{n}^{+}, f_{k}^{-}\} = 0,\\
		&[\psi^{-}_{n + 2}, e^{+}_{k}] - 2[\psi_{n + 1}^{-}, e_{k + 1}^{+}] + [\psi_{n}^{-}, e_{k + 2}^{+}] - \dfrac{h_{1}^{2} + h_{2}^{2}}{2}[\psi_{n}^{-}, e_{k}^{+}] - \dfrac{h_{1}^{2} - h_{2}^{2}}{2}\{\psi_{n}^{-}, e_{k}^{+}\} = 0,\\
		&[\psi^{-}_{n + 2}, f^{+}_{k}] - 2[\psi_{n + 1}^{-}, f_{k + 1}^{+}] + [\psi_{n}^{-}, f_{k + 2}^{+}] - \dfrac{h_{1}^{2} + h_{2}^{2}}{2}[\psi_{n}^{-}, f_{k}^{+}] + \dfrac{h_{1}^{2} - h_{2}^{2}}{2}\{\psi_{n}^{-}, f_{k}^{+}\} = 0,\\
		&[\psi_{n}^{+}, \psi_{k}^{+}] = [\psi_{n}^{-}, \psi^{-}_{k}] = [\psi_{n}^{+}, \psi_{k}^{-}] = [\psi_{n}^{-}, \psi_{k}^{+}] = 0, \\
		&\{e_{n}^{+}, f_{k}^{+}\} + \psi^{+}_{n + k} = 0, \quad \{e_{n}^{-}, f_{k}^{-}\} + \psi^{-}_{n + k} = 0, \\
		&\{e_{n}^{+}, f_{k}^{-}\} = 0, \quad \{e_{n}^{-}, f_{k}^{+}\} = 0.
	\end{aligned}
\end{equation}

Similarly to the case considered in the previous section the generators can be assembled in generating functions:
\begin{equation}
	e^{\pm}(z) = \sum_{n \in \mathbb{N}_{0}}\dfrac{e_{n}^{\pm}}{z^{n + 1}}, \quad \psi^{\pm}(z) = \sum_{n\in \mathbb{Z}}\dfrac{\psi_{n}}{z^{n + 1}}, \quad f^{\pm}(z) = \sum_{n\in\mathbb{N}_{0}}\dfrac{f^{\pm}_{n}}{z^{n + 1}}
\end{equation}

In terms of the generating functions the (super)commutation relations acquire the following form:
\begin{equation}\label{gl(1|1) relations}
	\begin{aligned}
		&e^{(v)}(z)e^{(v)}(w) \sim -e^{(v)}(w)e^{(v)}(z), \\
		&e^{+}(z)e^{-}(w) \sim -\varphi^{-+}(z - w)e^{-}(w)e^{+}(z), \\
		&e^{-}(z)e^{+}(w) \sim -\varphi^{+-}(z - w)e^{+}(w)e^{-}(z), \\
		&f^{(v)}(z)f^{(v)}(w) \sim -f^{(v)}(w)f^{(v)}(z), \\
		&f^{+}(z)f^{-}(w) \sim -\bigl[\varphi^{-+}(z - w)\bigr]^{-1}f^{-}(w)f^{+}(z), \\
		&f^{-}(z)f^{+}(w) \sim -\bigl[\varphi^{+-}(z - w)\bigr]^{-1}f^{+}(w)f^{-}(z), \\
		&\psi^{(v)}(z)e^{(v)}(w) \sim e^{(v)}(w)\psi^{(v)}(z), \\
		&\psi^{+}(z)e^{-}(w) \sim \varphi^{-+}(z - w)e^{-}(w)\psi^{+}(z), \\
		&\psi^{-}(z)e^{+}(w) \sim \varphi^{+-}(z - w)e^{+}(w)\psi^{-}(z), \\
		&\psi^{(v)}(z)f^{(v)}(w) \sim f^{(v)}(w)\psi^{(v)}(z),\\
		&\psi^{+}(z)f^{-}(w) \sim \bigl[\varphi^{-+}(z - w)\bigr]^{-1} f^{-}(w)\psi^{+}(z),\\
		&\psi^{-}(z)f^{+}(w) \sim \bigl[\varphi^{+-}(z - w)\bigr]^{-1} f^{+}(w)\psi^{-}(z), \\
		&\bigl[e^{(v)}(z), f^{(u)}(w)\bigr] \sim -\delta_{vu}\frac{\psi^{(v)}(z) - \psi^{(u)}(w)}{z - w}\,.
	\end{aligned}
\end{equation}
where sign $\sim$ implies an equivalence of Laurent polynomials up to $z^{j\geq 0}w^k$ and $z^{j}w^{k\geq 0}$, and the following bond factors are introduced:
\begin{equation}\label{gl(1|1) bonding factors}
	\varphi^{+-}(z) = \frac{(z + h_{1})(z - h_{1})}{(z - h_{2})(z + h_{2})}, \quad
	\varphi^{-+}(z) = \frac{(z + h_{2})(z - h_{2})}{(z - h_{1})(z + h_{1})}\,.
\end{equation}

\subsection{Quiver equations}

\begin{figure}[ht!]
	\begin{center}
		\begin{tikzpicture}
			\tikzstyle{col0} = [fill=burgundy]
			\tikzstyle{col1} = [fill=gray]
			\tikzstyle{col2} = [fill=\myblue]
			\draw[postaction={decorate},
			decoration={markings, mark= at position 0.55 with {\arrow{stealth}}}]
			(1,-2) -- (0,0) node[pos=0.5,below left] {$\scriptstyle I$};
			\draw[postaction={decorate},
			decoration={markings, mark= at position 0.55 with {\arrow{stealth}}}]
			(2,0) -- (1,-2) node[pos=0.5,below right] {$\scriptstyle J$};
			\draw[postaction={decorate},
			decoration={markings, mark= at position 0.65 with {\arrow{stealth}}, mark= at position 0.45 with {\arrow{stealth}}}] (0,0) to[out=30,in=150] node[pos=0.5, above] {$\scriptstyle A_1, A_2$} (2,0);
			\draw[postaction={decorate},
			decoration={markings, mark= at position 0.65 with {\arrow{stealth}}, mark= at position 0.45 with {\arrow{stealth}}}] (2,0) to[out=210,in=330] node[pos=0.5, below] {$\scriptstyle B_1, B_2$} (0,0);
			\draw[col1] (0,0) circle (0.1);
			\draw[col2] (2,0) circle (0.1);
			\node[left] at (-0.1,0) {$+$};
			\node[right] at (2.1,0) {$-$};
			\begin{scope}[shift={(1,-2)}]
				\draw[col0] (-0.08,-0.08) -- (-0.08,0.08) -- (0.08,0.08) -- (0.08,-0.08) -- cycle;
			\end{scope}
		\end{tikzpicture}
	\end{center}
	\caption{The quiver corresponding to the Fock module of $\mathsf{Y}(\widehat{\fg\fl}_{1|1})$.}
	\label{fig:Conifold_quiver}
\end{figure}
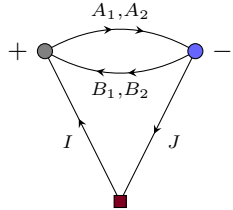

Here we consider a model corresponding to a quiver depicted in Fig.~\ref{fig:Conifold_quiver}.

The superpotential chosen for this model reads:
\begin{equation}\label{gl(1|1) superpotential}
	W = \text{Tr}\bigl(A_{1}B_{1}A_{2}B_{2} - A_{1}B_{2}A_{2}B_{1} + A_{2}IJ\bigr).
\end{equation}

A parametrization of equivariant weights for the chiral fields is the following:
\begin{equation}
	\begin{array}{c|c|c|c|c|c|c}
		\mbox{Field} & A_1 & A_2 & B_1& B_2 & I & J\\
		\hline
		\mbox{Weight} & h_1 & -h_1 & h_2& -h_2 & 0 & h_1\\
	\end{array}
\end{equation}

As in the previous section, the pair of a quiver and a superpotential carved a specific quiver variety having an algebro-geometric description of the Hilbert scheme of points on a plane; similarly, in this case our module describes stable perverse coherent sheaves on the blowup of a projective surface\footnote{We would like to thank Yu Zhao for pointing out this relation to us.} \cite{Nakajima:2008eq, Nakajima:2008ss}.
In a similar way, as before, we expect chiral field $A_2$ to play the role of a Lagrange multiplier.
So we extract from it a constraint $\p_{A_2}W=0$ and assume $A_2=0$ on the equations of motion.
Under these assumptions the quiver variety coincides with one described in the work \cite[Theorem 1.1]{Nakajima:2008eq}, where it is shown that this quiver variety is smooth.

Using the quiver data we construct D-term and F-term constraints:

\begin{equation}\label{gl(1|1) DF terms}
	\begin{aligned}
		&\text{D-terms } \colon \begin{cases}
			&-A_{1}^{\dagger}A_{1} - A_{2}^{\dagger}A_{2} + B_{1}B_{1}^{\dagger} + B_{2}B_{2}^{\dagger} + II^{\dagger} = \zeta_{1}\bbone_{n_{1}\times n_{1}};\\
			&A_{1}A_{1}^{\dagger} + A_{2}A_{2}^{\dagger} - B_{1}^{\dagger}B_{1} - B_{2}^{\dagger}B_{2} - J^{\dagger}J = \zeta_{2} \bbone_{n_{2}\times n_{2}};
		\end{cases},\\
		&\text{F-terms } \colon \begin{cases}
			&\d_{A_{1}} W = B_{1}A_{2}B_{2} - B_{2}A_{2}B_{1} = 0,\\
			&\d_{A_{2}} W = B_{2}A_{1}B_{1} - B_{1}A_{1}B_{2} + IJ =0,\\
			&\d_{B_{1}} W = A_{2}B_{2}A_{1} - A_{1}B_{2}A_{2} = 0, \\
			&\d_{B_{2}} W = A_{1}B_{1}A_{2} - A_{2}B_{1}A_{1} = 0, \\
			&\d_{I} W = JA_{2}, \\
			&\d_{J} W = A_{2}I
		\end{cases}
	\end{aligned}
\end{equation}

\subsection{Fixed points}

\paragraph{Path algebra.}

Let us consider the path algebra in the quiver depicted in Fig.~\ref{fig:Conifold_quiver}.
Again in the cyclic chamber $(\zeta_1>0,\zeta_2>0)$ we impose $J=0$.

The paths correspond to the words of the following form $m(A_1,A_2,B_1,B_2)\cdot I$, where $m$ is a monomial.
In what follows we consider possible situations:
\begin{itemize}
	\item The path contains field $A_{2}$:
	\begin{equation}
		\begin{aligned}
			&A_{2}I = 0, \\
			&A_{2}B_{1}A_{1}I = A_{1}B_{1}A_{2}I = 0, \\
			&A_{2}B_{2}A_{1}I = A_{1}B_{2}A_{2}I = 0.
		\end{aligned}
	\end{equation}
	Therefore the following monomials do not acquire expectation values in the vacuum:
	
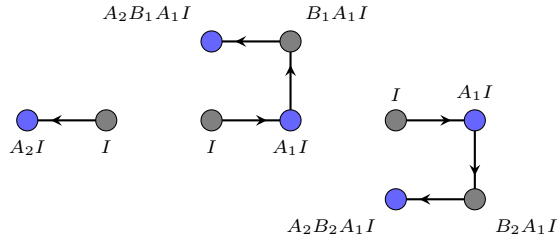
\begin{center}
		\begin{tikzpicture}[scale = 0.7]
			\draw[thick, postaction={decorate},
			decoration={markings, mark= at position 0.7 with {\arrow{stealth}}}] (0, 0) -- (1.5,0);
			\draw[thick, postaction={decorate},
			decoration={markings, mark= at position 0.7 with {\arrow{stealth}}}] (1.5, 0) -- (1.5,1.5);
			\draw[thick, postaction={decorate},
			decoration={markings, mark= at position 0.7 with {\arrow{stealth}}}] (1.5, 1.5) -- (0,1.5);
			\draw[thick, postaction={decorate},
			decoration={markings, mark= at position 0.7 with {\arrow{stealth}}}] (-2, 0) -- (-3.5,0);
			\draw[thick, postaction={decorate},
			decoration={markings, mark= at position 0.7 with {\arrow{stealth}}}] (3.5, 0) -- (5, 0);
			\draw[thick, postaction={decorate},
			decoration={markings, mark= at position 0.7 with {\arrow{stealth}}}] (5, 0) -- (5,-1.5);
			\draw[thick, postaction={decorate},
			decoration={markings, mark= at position 0.7 with {\arrow{stealth}}}] (5, -1.5) -- (3.5,-1.5);
			\draw[fill=gray] (0,0) circle (0.2) (1.5, 1.5) circle (0.2) (-2,0) circle (0.2) (3.5, 0) circle (0.2) (5, -1.5) circle (0.2);
			\draw[fill=\myblue] (1.5,0) circle (0.2) (-3.5, 0) circle (0.2) (0, 1.5) circle (0.2) (5, 0) circle (0.2) (3.5, -1.5) circle (0.2);
			\node[below] at (0, -0.2) {$\scriptstyle I$};
			\node[below] at (1.5, -0.2) {$\scriptstyle A_{1}I$};
			\node[above right] at (1.6, 1.7) {$\scriptstyle B_{1}A_{1}I$};
			\node[above left] at (-0.2, 1.7) {$\scriptstyle A_{2}B_{1}A_{1}I$};
			\node[below] at (-2, -0.2) {$\scriptstyle I$};
			\node[below] at (-3.5, -0.2) {$\scriptstyle A_{2}I$};
			\node[above] at (3.5, 0.2) {$\scriptstyle I$};
			\node[above] at (5, 0.2) {$\scriptstyle A_{1}I$};
			\node[below right] at (5.2, -1.7) {$\scriptstyle B_{2}A_{1}I$};
			\node[below left] at (3.3, -1.7) {$\scriptstyle A_{2}B_{2}A_{1}I$};
		\end{tikzpicture}
	\end{center}
	\item The path has the following form $(B_{1}A_{1})^{m}(B_{2}A_{1})^{k}I$:
	\begin{equation}
		\begin{aligned}
			&(B_{1}A_{1})^{m}(B_{2}A_{1})^{k}I = (B_{1}A_{1})^{m-1}B_{1}A_{1}B_{2}A_{1}(B_{2}A_{1})^{k-1}I = \\
			&= (B_{1}A_{1})^{m - 1}B_{2}A_{1}B_{1}A_{1}(B_{2}A_{1})^{k - 1}I = (B_{1}A_{1})^{m - 1}(B_{2}A_{1})(B_{1}A_{1})(B_{2}A_{1})^{k - 1}I = \\
			&= (B_{1}A_{1})^{m - 2}(B_{1}A_{1})(B_{2}A_{1})(B_{1}A_{1})(B_{2}A_{1})^{k - 1}I = \dots = (B_{2}A_{1})^{k}(B_{1}A_{1})^{m}I.
		\end{aligned}
	\end{equation}
	This is consequence of the property:
	\begin{equation}
		\color{\myblue} B_{1}A_{1}B_{2}\color{black}A_{1}\# = \color{\myblue} B_{2}A_{1}B_{1}\color{black}A_{1}\#\,,
	\end{equation}
	that can be depicted as an equivalence of the following paths:
	
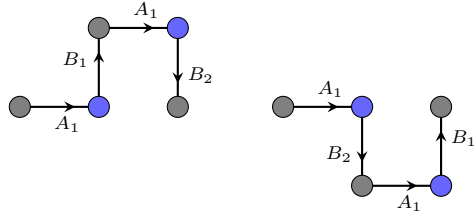
\begin{center}
		\begin{tikzpicture}[scale=0.7]
			\draw[thick, postaction={decorate},
			decoration={markings, mark= at position 0.7 with {\arrow{stealth}}}] (0, 0) -- (1.5,0) node[pos=0.6, below] {$\scriptstyle A_1$};
			\draw[thick, postaction={decorate},
			decoration={markings, mark= at position 0.7 with {\arrow{stealth}}}] (1.5, 1.5) -- (3,1.5) node[pos=0.6, above] {$\scriptstyle A_1$};
			\draw[thick, postaction={decorate},
			decoration={markings, mark= at position 0.7 with {\arrow{stealth}}}] (5, 0) -- (6.5,0) node[pos=0.6, above] {$\scriptstyle A_1$};
			\draw[thick, postaction={decorate},
			decoration={markings, mark= at position 0.7 with {\arrow{stealth}}}] (6.5, -1.5) -- (8,-1.5) node[pos=0.6, below] {$\scriptstyle A_1$};
			\draw[thick, postaction={decorate},
			decoration={markings, mark= at position 0.7 with {\arrow{stealth}}}] (1.5, 0) -- (1.5,1.5) node[pos=0.6, left] {$\scriptstyle B_{1}$};
			\draw[thick, postaction={decorate},
			decoration={markings, mark= at position 0.7 with {\arrow{stealth}}}] (8, -1.5) -- (8,0) node[pos=0.6, right] {$\scriptstyle B_{1}$};
			\draw[thick, postaction={decorate},
			decoration={markings, mark= at position 0.7 with {\arrow{stealth}}}] (3, 1.5) -- (3,0) node[pos=0.6, right] {$\scriptstyle B_{2}$};
			\draw[thick, postaction={decorate},
			decoration={markings, mark= at position 0.7 with {\arrow{stealth}}}] (6.5, 0) -- (6.5,-1.5) node[pos=0.6, left] {$\scriptstyle B_{2}$};
			\draw[fill=gray] (0,0) circle (0.2) (1.5, 1.5) circle (0.2) (3, 0) circle (0.2) (5, 0) circle (0.2) (6.5, -1.5) circle (0.2) (8, 0) circle (0.2);
			\draw[fill=\myblue] (1.5,0) circle (0.2) (3, 1.5) circle (0.2) (6.5, 0) circle (0.2) (8, -1.5) circle (0.2);
		\end{tikzpicture}
	\end{center}
	\item The path has the the following form $(B_{1}A_{1})^{m}(B_{2}A_{1})^{n}(B_{1}A_{1})^{k}I$:
	\begin{equation}
		(B_{1}A_{1})^{m}(B_{2}A_{1})^{n}(B_{1}A_{1})^{k}I = \dots = (B_{2}A_{1})^{n}(B_{1}A_{1})^{m + k}I\,.
	\end{equation}
	\item The path has the following form $(B_{2}A_{1})^{m}(B_{1}A_{1})^{n}(B_{2}A_{1})^{k}I$:
	\begin{equation}
		(B_{2}A_{1})^{m}(B_{1}A_{1})^{n}(B_{2}A_{1})^{k}I = \dots = (B_{2}A_{1})^{m + k}(B_{1}A_{1})^{n}I\,.
	\end{equation}
	\item The path has the following form $A_{1}(B_{2}A_{1})^{n}(B_{1}A_{1})^{k}I$.
\end{itemize}

Summarizing, we observe that a set of possible paths contains elements of two types:
\begin{equation}
	(B_{2}A_{1})^{n}(B_{1}A_{1})^{k}I,\quad A_{1}(B_{2}A_{1})^{n}(B_{1}A_{1})^{k}I\,,
\end{equation}
that correspond to atoms of colors $+$ and $-$ respectively.

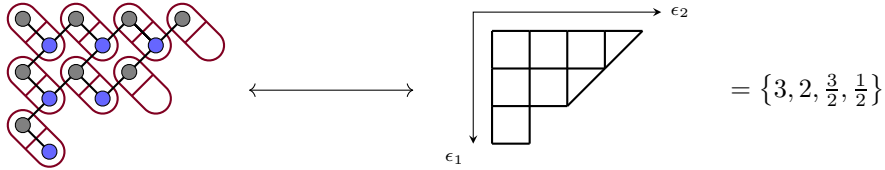
\begin{figure}[ht!]
	\centering
	\begin{tikzpicture}
		\node(A) at (0,0) {$\begin{array}{c}
				\begin{tikzpicture}[scale=0.5, rotate=-45]
					\foreach \i/\j in {0/0, 1/1, 1/-1, 2/0, 2/2, 2/-2, 3/3, 3/1}
					{
						\begin{scope}[shift={(\i,\j)}]
							\draw[burgundy, thick] (0,0.4) to[out=180,in=90] (-0.4,0) to[out=270,in=180] (0,-0.4) -- (1,-0.4) to[out=0,in=270] (1.4,0) to[out=90,in=0] (1,0.4) -- cycle;
							\draw[burgundy, thick] (0.5,-0.4) -- (0.5,0.4);
						\end{scope}
					}
					\tikzset{sty1/.style={fill=gray}}
					\tikzset{sty2/.style={fill=white!40!blue}}
					\draw[thick] (0,0) -- (1,0) (1,-1) -- (1,1) (2,-2) -- (2,2) (1,1) -- (2,1) (1,-1) -- (2,-1) (2,2) -- (3,2) (2,2) -- (3,2) (2,0) -- (3,0) (2,-2) -- (3,-2) (3,0) -- (3,3);
					\draw[sty1] (0,0) circle (0.2) (1,1) circle (0.2) (1,-1) circle (0.2) (2,0) circle (0.2) (2,2) circle (0.2) (2,-2) circle (0.2) (3,3) circle (0.2) (3,1) circle (0.2);
					\draw[sty2] (1,0) circle (0.2) (2,1) circle (0.2) (2,-1) circle (0.2) (3,2) circle (0.2) (3,0) circle (0.2) (3,-2) circle (0.2);
				\end{tikzpicture}
			\end{array}$};
		\node(B) at (6,0) {$\begin{array}{c}
				\begin{tikzpicture}[scale=0.5]
					\draw[stealth-stealth] (-0.5,-3) -- (-0.5,0.5) -- (4.5,0.5);
					\node[below left] at (-0.5,-3) {$\scriptstyle \epsilon_1$};
					\node[right] at (4.5,0.5) {$\scriptstyle \epsilon_2$};
					\foreach \i/\j in {0/-3, 0/-2, 0/-1, 0/0, 1/-2, 1/-1, 1/0, 2/-1, 2/0, 3/0}
					{
						\draw[thick] (\i,\j) -- (\i+1,\j);
					}
					\foreach \i/\j in {0/-2, 0/-1, 0/0, 1/-2, 1/-1, 1/0, 2/-1, 2/0, 3/0}
					{
						\draw[thick] (\i,\j) -- (\i,\j-1);
					}
					\foreach \i/\j in {3/-1, 4/0}
					{
						\draw[thick] (\i,\j) -- (\i-1,\j-1);
					}
				\end{tikzpicture}
			\end{array}$};
		\node[right] at (B.east) {$=\left\{3,2,\frac{3}{2},\frac{1}{2}\right\}$};
		\path (A) edge[<->] (B);
	\end{tikzpicture}
	\caption{Crystal and diagrammatic labeling of $\mathsf{Y}(\widehat{\fg\fl}_{1|1})$ semi-Fock representation vectors.}
	\label{fig:crystalYoung}
\end{figure}

As it was proposed in \cite{Galakhov:2023mak}, the resulting crystals in this case could be identified with super-Young diagrams by simply rotating the diagram clock-wise by $45^{\circ}$ and identifying pairs of atoms ($X$, $A_1X$) with complete square tiles and single atoms $X$ without a pair with triangular half-tiles (e.g. see Fig.~\ref{fig:crystalYoung}).

Sets of super-Young diagrams could be identified with sets of super-partitions.
A super-partition $\lambda$ of a non-negative semi-integer $u\in \frac{1}{2}\mathbb{N}_{0}$ is a sequence of semi-integer numbers
\begin{equation}\label{super-partition}
	\lambda_{1} \geqslant \lambda_{2} \geqslant \dots \geqslant 0,
\end{equation}
where $\sum_{i}\lambda_{i} = u$ and if $\lambda\notin\mathbb{Z}$ inequalities in \eqref{super-partition} are strict, i.e. $\lambda_{i - 1} > \lambda_{i} > \lambda_{i + 1}$. 

In a complete analogy with the generic case, we construct a crystal representation:
\begin{equation}\label{gl(1|1) crystal ansatz}
	\begin{aligned}
		&e^{\pm}(z)|\lambda\rangle  = \sum_{\ssqbox{$a$}\in \text{Add}(\lambda)^{\pm}}\frac{{\bf E}_{\lambda,  \lambda + \ssqbox{$a$}}}{z - h(\sqbox{$a$})}|\lambda + \sqbox{$a$}\rangle, \\
		&f^{\pm}(z)|\lambda\rangle = \sum_{\ssqbox{$a$}\in \text{Rem}(\lambda)^{\pm}}\frac{{\bf F}_{\lambda,  \lambda - \ssqbox{$a$}}}{z - h(\sqbox{$a$})}|\lambda - \sqbox{$a$}\rangle, \\
		&\psi^{\pm}(z)|\lambda\rangle = \bPsi^{\pm}_{\lambda}(z)|\lambda\rangle\,,
	\end{aligned}
\end{equation}
where the eigenvalues for Cartan operators read:
\begin{equation}
	\bPsi^{a}_{\lambda}(z) = \psi^{a}_{0}(z)\prod_{\ssqbox{$b$}\in \lambda}\varphi^{ab}(z - h(\sqbox{$b$})), \quad \psi^{+}_{0}(z) = \dfrac{c^{+}}{z}, \quad \psi_{0}^{-}(z) = c^{-}(z + h_{1}),
\end{equation}
where $c^{\pm}$ are some constants, and $a \in \{+, -\}$.

The weights of tiles are defined in the following way:
\begin{equation}
	\begin{aligned}
		&h(\begin{tikzpicture}[scale=0.2]
			\begin{scope}
				\draw (0, 0) -- (1, 0) -- (0, -1) -- cycle;
			\end{scope}
		\end{tikzpicture}) = h(\sqbox{$+$}) = (y_{\begin{tikzpicture}[scale=0.2]
				\begin{scope}
					\draw (0, 0) -- (1, 0) -- (0, -1) -- cycle;
				\end{scope}
		\end{tikzpicture}} + x_{\begin{tikzpicture}[scale=0.2]
				\begin{scope}
					\draw (0, 0) -- (1, 0) -- (0, -1) -- cycle;
				\end{scope}
		\end{tikzpicture}})h_{1} + (y_{\begin{tikzpicture}[scale=0.2]
				\begin{scope}
					\draw (0, 0) -- (1, 0) -- (0, -1) -- cycle;
				\end{scope}
		\end{tikzpicture}} - x_{\begin{tikzpicture}[scale=0.2]
				\begin{scope}
					\draw (0, 0) -- (1, 0) -- (0, -1) -- cycle;
				\end{scope}
		\end{tikzpicture}})h_{2} = x_{\begin{tikzpicture}[scale=0.2]
				\begin{scope}
					\draw (0, 0) -- (1, 0) -- (0, -1) -- cycle;
				\end{scope}
		\end{tikzpicture}}\epsilon_{1} + y_{\begin{tikzpicture}[scale=0.2]
				\begin{scope}
					\draw (0, 0) -- (1, 0) -- (0, -1) -- cycle;
				\end{scope}
		\end{tikzpicture}}\epsilon_{2}, \\
		&h(\begin{tikzpicture}[scale=0.2]
			\begin{scope}
				\draw (0, -1) -- (1, -1) -- (1, 0) -- cycle;
			\end{scope}
		\end{tikzpicture}) = h(\sqbox{$-$}) = (y_{\begin{tikzpicture}[scale=0.2]
				\begin{scope}
					\draw (0, -1) -- (1, -1) -- (1, 0) -- cycle;
				\end{scope}
		\end{tikzpicture}} + x_{\begin{tikzpicture}[scale=0.2]
				\begin{scope}
					\draw (0, -1) -- (1, -1) -- (1, 0) -- cycle;
				\end{scope}
		\end{tikzpicture}} + 1)h_{1} + (y_{\begin{tikzpicture}[scale=0.2]
				\begin{scope}
					\draw (0, -1) -- (1, -1) -- (1, 0) -- cycle;
				\end{scope}
		\end{tikzpicture}} - x_{\begin{tikzpicture}[scale=0.2]
				\begin{scope}
					\draw (0, -1) -- (1, -1) -- (1, 0) -- cycle;
				\end{scope}
		\end{tikzpicture}})h_{2} = x_{\begin{tikzpicture}[scale=0.2]
				\begin{scope}
					\draw (0, -1) -- (1, -1) -- (1, 0) -- cycle;
				\end{scope}
		\end{tikzpicture}}\epsilon_{1} + y_{\begin{tikzpicture}[scale=0.2]
				\begin{scope}
					\draw (0, -1) -- (1, -1) -- (1, 0) -- cycle;
				\end{scope}
		\end{tikzpicture}}\epsilon_{2} + \frac{1}{2}\epsilon_{1} + \frac{1}{2}\epsilon_{2}
	\end{aligned}
\end{equation}
where $(x_{\begin{tikzpicture}[scale=0.2]
		\begin{scope}
			\draw (0, 0) -- (1, 0) -- (0, -1) -- cycle;
		\end{scope}
\end{tikzpicture}}, y_{\begin{tikzpicture}[scale=0.2]
		\begin{scope}
			\draw (0, 0) -- (1, 0) -- (0, -1) -- cycle;
		\end{scope}
\end{tikzpicture}}, x_{\begin{tikzpicture}[scale=0.2]
		\begin{scope}
			\draw (0, -1) -- (1, -1) -- (1, 0) -- cycle;
		\end{scope}
\end{tikzpicture}}, y_{\begin{tikzpicture}[scale=0.2]
		\begin{scope}
			\draw (0, -1) -- (1, -1) -- (1, 0) -- cycle;
		\end{scope}
\end{tikzpicture}})$ are coordinates in the equivariant weight plane and $\epsilon_{1} = h_{1} - h_{2}$, $\epsilon_{2} = h_{1} + h_{2}$.

\subsection{Euler classes}
In this section, we present an explicit calculation of some matrix element ${\bf E}_{\lambda, \lambda + \ssqbox{$a$}}$.

Consider the following transition example:
\begin{equation}\label{diagrams fr gl(1,1) example}
	\lambda = \begin{tikzpicture}[scale=0.25, baseline={(0, -0.45)}]
		\foreach \i/\j in {0/-2, 0/-1, 0/0, 1/-1, 1/0, 2/0}
		{
			\draw (\i,\j) -- (\i+1,\j);
		}
		\foreach \i/\j in {0/-2, 0/-1, 0/0, 1/-1, 1/0, 2/0}
		{
			\draw (\i,\j) -- (\i,\j-1);
		}
		\foreach \i/\j in {1/-2, 3/0}
		{
			\draw (\i,\j) -- (\i-1,\j-1);
		}
	\end{tikzpicture} \quad = \quad
	\begin{array}{c}
		\begin{tikzpicture}[rotate around z=-45]
			\draw[thick, postaction={decorate},
			decoration={markings, mark= at position 0.6 with {\arrow{stealth}}}] (0,0) -- (1,0) node[pos=0.6,above] {$\scriptstyle A_1$};
			\draw[thick, postaction={decorate},
			decoration={markings, mark= at position 0.6 with {\arrow{stealth}}}] (1,1) -- (2,1) node[pos=0.6,above] {$\scriptstyle A_1$};
			\draw[thick, postaction={decorate},
			decoration={markings, mark= at position 0.6 with {\arrow{stealth}}}] (1,-1) -- (2,-1) node[pos=0.6,above] {$\scriptstyle A_1$};
			\draw[thick, postaction={decorate},
			decoration={markings, mark= at position 0.6 with {\arrow{stealth}}}] (1,0) -- (1,1) node[pos=0.4,right] {$\scriptstyle B_1$};
			\draw[thick, postaction={decorate},
			decoration={markings, mark= at position 0.6 with {\arrow{stealth}}}] (2,1) -- (2,2) node[pos=0.4,right] {$\scriptstyle B_1$};
			\draw[thick, postaction={decorate},
			decoration={markings, mark= at position 0.6 with {\arrow{stealth}}}] (1,0) -- (1,-1) node[pos=0.4,left] {$\scriptstyle B_2$};
			\draw[thick, postaction={decorate},
			decoration={markings, mark= at position 0.6 with {\arrow{stealth}}}] (2,-1) -- (2,-2) node[pos=0.4,left] {$\scriptstyle B_2$};
			\draw[fill=gray] (0,0) circle (0.1) (1,1) circle (0.1) (1,-1) circle (0.1) (2, 2) circle (0.1) (2, -2) circle (0.1);
			\draw[fill=\myblue] (1,0) circle (0.1) (2,1) circle (0.1) (2, -1) circle (0.1);
			\draw[burgundy] (0,0) circle (0.2);
			\node[below] at (0,-0.2) {$\scriptstyle I$};
			\node[above left] at (-0.1,0.1) {$\scriptstyle 1$};
			\node[left] at (0.9,1) {$\scriptstyle 2$};
			\node[left] at (0.9,-1) {$\scriptstyle 3$};
			\node[below right] at (1.05,-0.05) {$\scriptstyle 1$};
			\node[below right] at (2, 0.9) {$\scriptstyle 2$};
			\node[above right] at (2, -0.9) {$\scriptstyle 3$};
			\node[left] at (1.95, 2) {$\scriptstyle 4$};
			\node[left] at (1.95, -2) {$\scriptstyle 5$};
		\end{tikzpicture}\\
		\vec n = (5,3)
	\end{array}\quad\longrightarrow\quad \lambda' = \begin{tikzpicture}[scale=0.25, baseline = {(0, -0.45)}]
		\foreach \i/\j in {0/-2, 0/-1, 0/0, 1/-1, 1/0, 2/-1, 2/0}
		{
			\draw (\i,\j) -- (\i+1,\j);
		}
		\foreach \i/\j in {0/-2, 0/-1, 0/0, 1/-1, 1/0, 2/0, 3/0}
		{
			\draw (\i,\j) -- (\i,\j-1);
		}
		\foreach \i/\j in {1/-2}
		{
			\draw (\i,\j) -- (\i-1,\j-1);
		}
	\end{tikzpicture}\quad = \quad
	\begin{array}{c}
		\begin{tikzpicture}[rotate around z=-45]
			\draw[thick, postaction={decorate},
			decoration={markings, mark= at position 0.6 with {\arrow{stealth}}}] (0,0) -- (1,0) node[pos=0.6,above] {$\scriptstyle A_1$};
			\draw[thick, postaction={decorate},
			decoration={markings, mark= at position 0.6 with {\arrow{stealth}}}] (1,1) -- (2,1) node[pos=0.6,above] {$\scriptstyle A_1$};
			\draw[thick, postaction={decorate},
			decoration={markings, mark= at position 0.6 with {\arrow{stealth}}}] (1,-1) -- (2,-1) node[pos=0.6,above] {$\scriptstyle A_1$};
			\draw[thick, postaction={decorate},
			decoration={markings, mark= at position 0.6 with {\arrow{stealth}}}] (2,2) -- (3,2) node[pos=0.6,above] {$\scriptstyle A_1$};
			\draw[thick, postaction={decorate},
			decoration={markings, mark= at position 0.6 with {\arrow{stealth}}}] (1,0) -- (1,1) node[pos=0.4,right] {$\scriptstyle B_1$};
			\draw[thick, postaction={decorate},
			decoration={markings, mark= at position 0.6 with {\arrow{stealth}}}] (2,1) -- (2,2) node[pos=0.4,right] {$\scriptstyle B_1$};
			\draw[thick, postaction={decorate},
			decoration={markings, mark= at position 0.6 with {\arrow{stealth}}}] (1,0) -- (1,-1) node[pos=0.4,left] {$\scriptstyle B_2$};
			\draw[thick, postaction={decorate},
			decoration={markings, mark= at position 0.6 with {\arrow{stealth}}}] (2,-1) -- (2,-2) node[pos=0.4,left] {$\scriptstyle B_2$};
			\draw[fill=gray] (0,0) circle (0.1) (1,1) circle (0.1) (1,-1) circle (0.1) (2, 2) circle (0.1) (2, -2) circle (0.1);
			\draw[fill=\myblue] (1,0) circle (0.1) (2,1) circle (0.1) (2, -1) circle (0.1) (3, 2) circle (0.1);
			\draw[burgundy] (0,0) circle (0.2);
			\node[below] at (0,-0.2) {$\scriptstyle I$};
			\node[above left] at (-0.1,0.1) {$\scriptstyle 1$};
			\node[left] at (0.9,1) {$\scriptstyle 2$};
			\node[left] at (0.9,-1) {$\scriptstyle 3$};
			\node[below right] at (1.05,-0.05) {$\scriptstyle 1$};
			\node[below right] at (2, 0.9) {$\scriptstyle 2$};
			\node[above right] at (2, -0.9) {$\scriptstyle 3$};
			\node[left] at (1.95, 2) {$\scriptstyle 4$};
			\node[above right] at (3.1, 2) {$\scriptstyle 4$};
			\node[left] at (1.95, -2) {$\scriptstyle 5$};
		\end{tikzpicture}\\
		{\vec n}=(5,4)
	\end{array}\,.
\end{equation}

The crystal diagram gives us the ansatz for the vacuum expectations of the chiral fields:
\begin{equation}
	\begin{aligned}
		&A_{1}^{(1)} = \begin{pmatrix}
			A_{11} & 0 & 0 & 0 & 0\\
			0 & A_{22}& 0 & 0 & 0\\
			0 & 0 & A_{33} & 0 & 0
		\end{pmatrix},
		&B_{1}^{(1)} = \begin{pmatrix}
			0 & 0 & 0\\
			B_{21} & 0 & 0\\
			0 & 0 & 0\\
			0 & B_{42} & 0\\
			0 & 0 & 0
		\end{pmatrix},
		&B_{2}^{(1)} = \begin{pmatrix}
			0 & 0 & 0\\
			0 & 0 & 0\\
			B_{31} & 0 & 0\\
			0 & 0 & 0\\
			0 & 0 & B_{53}
		\end{pmatrix},\\
		&A_{1}^{(2)} = \begin{pmatrix}
			F_{11} & 0 & 0 & 0 & 0\\
			0 & F_{22}& 0 & 0 & 0\\
			0 & 0 & F_{33} & 0 & 0\\
			0 & 0 & 0 & F_{44} & 0\\
		\end{pmatrix},
		&B_{1}^{(2)} = \begin{pmatrix}
			0 & 0 & 0 & 0\\
			P_{21} & 0 & 0 & 0\\
			0 & 0 & 0 & 0\\
			0 & P_{42} & 0 & 0\\
			0 & 0 & 0 & 0
		\end{pmatrix},
		&B_{2}^{(2)} = \begin{pmatrix}
			0 & 0 & 0 & 0\\
			0 & 0 & 0 & 0\\
			P_{31} & 0 & 0 & 0\\
			0 & 0 & 0 & 0\\
			0 & 0 & P_{53} & 0
		\end{pmatrix},\\
		&I^{(1)} = \begin{pmatrix}
			I_{1} & 0 & 0 & 0 & 0
		\end{pmatrix}^{T},
		&I^{(2)} = \begin{pmatrix}
			I_{2} & 0 & 0 & 0 & 0
		\end{pmatrix}^{T}.
	\end{aligned}
\end{equation}

To define the coefficients, we substitute the ansatz into equations \eqref{gl(1|1) DF terms} and the solution is
\begin{equation}
	\begin{aligned}
		&A_{1}^{(1)} = \begin{pmatrix}
			\sqrt{6\zeta_{1} + 5\zeta_{2}} & 0 & 0 & 0 & 0\\
			0 & \sqrt{\zeta_{1} + \zeta_{2}}& 0 & 0 & 0\\
			0 & 0 & \sqrt{\zeta_{1} + \zeta_{2}} & 0 & 0
		\end{pmatrix},
		&B_{1}^{(1)} = \begin{pmatrix}
			0 & 0 & 0\\
			\sqrt{3\zeta_{1} + 2\zeta_{2}} & 0 & 0\\
			0 & 0 & 0\\
			0 & \sqrt{\zeta_{1}} & 0\\
			0 & 0 & 0
		\end{pmatrix}, \\
		&B_{2}^{(1)} = \begin{pmatrix}
			0 & 0 & 0\\
			0 & 0 & 0\\
			\sqrt{3\zeta_{1} + 2\zeta_{2}} & 0 & 0\\
			0 & 0 & 0\\
			0 & 0 & \sqrt{\zeta_{1}}
		\end{pmatrix},
		&B_{2}^{(2)} = \begin{pmatrix}
			0 & 0 & 0 & 0\\
			0 & 0 & 0 & 0\\
			\sqrt{3\zeta_{1} + 2\zeta_{2}} & 0 & 0 & 0\\
			0 & 0 & 0 & 0\\
			0 & 0 &
			\sqrt{\zeta_{1}} & 0
		\end{pmatrix},\\
		&A_{1}^{(2)} = \begin{pmatrix}
			\sqrt{6\zeta_{1} + 9\zeta_{2}} & 0 & 0 & 0 & 0\\
			0 & \sqrt{\zeta_{1} + 3\zeta_{2}} & 0 & 0 & 0\\
			0 & 0 & \sqrt{\zeta_{1} + \zeta_{2}} & 0 & 0\\
			0 & 0 & 0 & \sqrt{\zeta_{2}} & 0\\
		\end{pmatrix},
		&B_{1}^{(2)} = \begin{pmatrix}
			0 & 0 & 0 & 0\\
			\sqrt{3\zeta_{1} + 6\zeta_{2}} & 0 & 0 & 0\\
			0 & 0 & 0 & 0\\
			0 & \sqrt{\zeta_{1} + 2\zeta_{2}} & 0 & 0\\
			0 & 0 & 0 & 0
		\end{pmatrix}, \\
		&I^{(1)} = \begin{pmatrix}
			\sqrt{13\zeta_{1} + 10\zeta_{2}} & 0 & 0 & 0 & 0
		\end{pmatrix}^{T},
		&I^{(2)} = \begin{pmatrix}
			\sqrt{13\zeta_{1} + 18\zeta_{2}} & 0 & 0 & 0 & 0
		\end{pmatrix}^{T}.
	\end{aligned}
\end{equation}

It is sufficient to define representatives of the r.h.s. orbit in \eqref{NSKH} to define matrix coefficients of the quiver Yangian generators.
To do so, we substitute all the vacuum expectation values by simple unities $\sqrt{M\zeta_1+N\zeta_2}\to 1$.

The first step is to find $\Eul_{\begin{tikzpicture}[scale=0.19, baseline={(0, -0.45)}]
		\foreach \i/\j in {0/-2, 0/-1, 0/0, 1/-1, 1/0, 2/0}
		{
			\draw (\i,\j) -- (\i+1,\j);
		}
		\foreach \i/\j in {0/-2, 0/-1, 0/0, 1/-1, 1/0, 2/0}
		{
			\draw (\i,\j) -- (\i,\j-1);
		}
		\foreach \i/\j in {1/-2, 3/0}
		{
			\draw (\i,\j) -- (\i-1,\j-1);
		}
\end{tikzpicture}} = \Eul_{[\frac{5}{2}, 1, \frac{1}{2}]}$.

To find the Euler class, we decompose the fields around the corresponding fixed point:

\begin{equation}
	\begin{aligned}
		&\delta A_{1}^{(1)} = \begin{pmatrix}
			a_{ij}
		\end{pmatrix} \in \text{Mat}_{3\times 5},
		&\delta B_{1}^{(1)} = \begin{pmatrix}
			b_{ij}
		\end{pmatrix} \in \text{Mat}_{5\times 3},\quad
		&\delta B_{2}^{(1)} = \begin{pmatrix}
			d_{ij}
		\end{pmatrix} \in \text{Mat}_{5\times 3},\\
		&\delta A_{1}^{(2)} = \begin{pmatrix}
			c_{ij}
		\end{pmatrix}\in \text{Mat}_{4\times 5},
		&\delta B_{1}^{(2)} = \begin{pmatrix}
			f_{ij}
		\end{pmatrix} \in \text{Mat}_{5\times 4}, \quad
		&\delta B_{2}^{(2)} = \begin{pmatrix}
			h_{ij}
		\end{pmatrix}\in\text{Mat}_{5\times 4},\\
		&\delta I^{(1)} = \begin{pmatrix}
			s_{i}
		\end{pmatrix}\in\text{Mat}_{5\times 1} ,
		&\delta I^{(2)} = \begin{pmatrix}
			r_{i}
		\end{pmatrix}\in\text{Mat}_{5\times 1}.
	\end{aligned}
\end{equation}

To describe the tangent space to the fixed point following the algorithm presented in \ref{sec:equiv} we substitute the perturbed fields into equations \eqref{gl(1|1) DF terms} and exclude some degrees of freedom using gauge transformations. We also choose those that are parallel to the F-term surface:
\begin{equation}
	\begin{aligned}
		\mathsf{T}_{[\frac{5}{2}, 1, \frac{1}{2}]}\mathscr{M} = {\rm Span} \langle b_{53}, a_{12}, a_{23}, a_{35}\rangle
	\end{aligned}
\end{equation}	

To define the equivariant weights, one needs to define the expectation values of the scalar fields $\Phi_{1}^{(s)}, \Phi^{(s)}_{2}$:
\begin{equation}
	\begin{aligned}
		&\bar\Phi^{(1)}_{1} = \text{diag~}(0, h_{1} + h_{2}, h_{1} - h_{2}, 2h_{1} + 2h_{2}, 2h_{1} - 2h_{2}), &\bar\Phi^{(1)}_{2} = \text{diag~}(h_{1}, 2h_{1} + h_{2}, 2h_{1} - h_{2}), \\
		&\bar\Phi^{(2)}_{1} = \text{diag~}(0, h_{1} + h_{2}, h_{1} - h_{2}, 2h_{1} + 2h_{2}, 2h_{1} - 2h_{2}), & \bar\Phi^{(2)}_{2} = \text{diag~}(h_{1}, 2h_{1} + h_{2}, 2h_{1} - h_{2}, 3h_{1} + 2h_{2}).
	\end{aligned}
\end{equation}

The respective weights read:
\begin{equation}\label{weights gl(1|1)}
	\begin{aligned}
		&w[(A_{1}^{(s)})_{ij}] = (\bar\Phi_{2}^{(s)})_{ii} - (\bar\Phi^{(s)}_{1})_{jj} - h_{1}, \\
		&w[(B_{1}^{(s)})_{ij}] = (\bar\Phi^{(s)}_{1})_{ii} - (\bar\Phi_{2}^{(s)})_{jj} - h_{2}, \\
		&w[(B_{2}^{(s)})_{ij}] = (\bar\Phi^{(s)}_{1})_{ii} - (\bar\Phi_{2}^{(s)})_{jj} + h_{2},\\ \hline
		&w(b_{53}) = -2h_{2}, \quad w(a_{12}) = -h_{1}-h_{2}, \\
		&w(a_{23}) = 2h_{2}, \quad w(a_{35}) = h_{2} - h_{1}.
	\end{aligned}
\end{equation}

Then we calculate the corresponding Euler class:
\begin{equation}
	\Eul_{[\frac{5}{2}, 1, \frac{1}{2}]} = 4h_{2}^{2}(h_{1} + h_{2})(h_{2} - h_{1}) = \epsilon_{1}\epsilon_{2}(\epsilon_{1} - \epsilon_{2})(\epsilon_{2} - \epsilon_{1}).
\end{equation}

The second step is to define the incidence locus and its tangent space. 
In order to do that, we introduce quiver homomorphism $\tau^{a}$:
\begin{equation}\label{homomorphism gl(1|1)}
	0 = \begin{cases}
		&\tau^{-}A_{1}^{(2)} - A_{1}^{(1)}\tau^{+},\\
		&\tau^{+}B_{1}^{(2)} - B_{1}^{(1)}\tau^{-}, \\
		&\tau^{+}B_{2}^{(2)} - B_{2}^{(1)}\tau^{-}, \\
		&\tau^{+}I^{(2)} - I^{(1)}
	\end{cases}
\end{equation}

The vacuum expectation values for $\tau$ take the following form:
\begin{equation}
	\bar\tau^{+} = \begin{pmatrix}
		1 & 0 & 0 & 0 & 0\\
		0 & 1 & 0 & 0 & 0\\
		0 & 0 & 1 & 0 & 0\\
		0 & 0 & 0 & 1 & 0\\
		0 & 0 & 0 & 0 & 1
	\end{pmatrix}, \quad
	\bar\tau^{-} = \begin{pmatrix}
		1 & 0 & 0 & 0\\
		0 & 1 & 0 & 0\\
		0 & 0 & 1 & 0
	\end{pmatrix}
\end{equation}

We parameterize the tangent degrees of freedom as $\delta \tau^{+}_{ij} = (t_{ij})$, $\delta\tau^{-}_{ij} = (r_{ij})$.
The solution to the homomorphism equations for these degrees of freedom is rather bulky, so here we present just a few terms:

\begin{equation}
	\begin{Bmatrix}
		\forall i \in \overline{1, 5} \quad t_{i1} = 0 & t_{15} = t_{25} = t_{35} = t_{45} = 0 & r_{32} = b_{53} & r_{24} = 0 & t_{24} = r_{24} 
	\end{Bmatrix}
\end{equation}

The incidence locus cuts out constraints:

\begin{equation}
	\begin{Bmatrix}
		&f_{54} = 0 & f_{53} = b_{53} & f_{24} = 0 & f_{23} = 0\\
		&h_{33} = a_{35} & f_{43} = a_{23} & f_{44} = a_{12}
	\end{Bmatrix}
\end{equation}

The next step is to substitute the perturbed fields into \eqref{gl(1|1) DF terms}. 
The incidence locus becomes:
\begin{equation}
	\begin{aligned}
		&\mathsf{T}_{[\frac{5}{2}, 1, \frac{1}{2}], [\frac{5}{2}, 1, 1]}\CI = {\rm Span}\langle a_{23}, a_{12}, a_{35}, b_{53}, c_{45}\rangle\,.
	\end{aligned}
\end{equation}

The weights of the tangent directions are defined as before \eqref{weights gl(1|1)}:
\begin{equation}
	\begin{aligned}
		&w(a_{12}) = -h_{1} - h_{2}, \quad w(a_{23}) = 2h_{2}, \\
		&w(a_{35}) = h_{2} - h_{1}, \quad w(b_{53}) = -2h_{2}, \\
		&w(c_{45}) = 4h_{2}\,.
	\end{aligned}
\end{equation}

Finally, the Euler class reads in this case:
\begin{equation}
	\Eul_{[\frac{5}{2}, 1, \frac{1}{2}], [\frac{5}{2}, 1, 1]} = 16h_{2}^{3}(h_{2} - h_{1})(h_{1} + h_{2}) = \epsilon_{1}\epsilon_{2}^{2}(\epsilon_{1} - \epsilon_{2})^{2}\,.
\end{equation}

For the matrix coefficient ${\bf E}_{\lambda,\lambda + \ssqbox{$a$}}$, we acquire the following expression:
\begin{equation}
	{\bf E}_{[\frac{5}{2}, 1, \frac{1}{2}], [\frac{5}{2}, 1, 1]} = \dfrac{\Eul_{[\frac{5}{2}, 1, \frac{1}{2}]}}{\Eul_{[\frac{5}{2}, 1, \frac{1}{2}], [\frac{5}{2}, 1, 1]}} = -\dfrac{1}{\epsilon_{2}}.
\end{equation}

\subsection{Amplitudes and hook formulas}

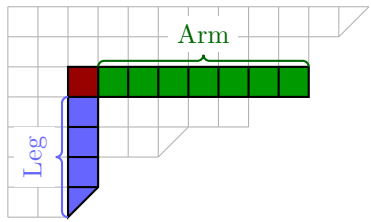
\begin{figure}[ht!]
	\begin{center}
		\begin{tikzpicture}[scale=0.4]
			\foreach \i/\j in {0/-7, 0/-6, 0/-5, 0/-4, 0/-3, 0/-2, 0/-1, 0/0, 1/-7, 1/-6, 1/-5, 1/-4, 1/-3, 1/-2, 1/-1, 1/0, 2/-6, 2/-5, 2/-4, 2/-3, 2/-2, 2/-1, 2/0, 3/-5, 3/-4, 3/-3, 3/-2, 3/-1, 3/0, 4/-5, 4/-4, 4/-3, 4/-2, 4/-1, 4/0, 5/-4, 5/-3, 5/-2, 5/-1, 5/0, 6/-4, 6/-3, 6/-2, 6/-1, 6/0, 7/-4, 7/-3, 7/-2, 7/-1, 7/0, 8/-3, 8/-2, 8/-1, 8/0, 9/-3, 9/-2, 9/-1, 9/0, 10/-1, 10/0, 11/0}
			{
				\draw[white!40!gray] (\i,\j) -- (\i+1,\j);
			}
			\foreach \i/\j in {0/-6, 0/-5, 0/-4, 0/-3, 0/-2, 0/-1, 0/0, 1/-6, 1/-5, 1/-4, 1/-3, 1/-2, 1/-1, 1/0, 2/-6, 2/-5, 2/-4, 2/-3, 2/-2, 2/-1, 2/0, 3/-5, 3/-4, 3/-3, 3/-2, 3/-1, 3/0, 4/-4, 4/-3, 4/-2, 4/-1, 4/0, 5/-4, 5/-3, 5/-2, 5/-1, 5/0, 6/-3, 6/-2, 6/-1, 6/0, 7/-3, 7/-2, 7/-1, 7/0, 8/-3, 8/-2, 8/-1, 8/0, 9/-2, 9/-1, 9/0, 10/-2, 10/-1, 10/0, 11/0}
			{
				\draw[white!40!gray] (\i,\j) -- (\i,\j-1);
			}
			\foreach \i/\j in {3/-6, 6/-4, 12/0}
			{
				\draw[white!40!gray] (\i,\j) -- (\i-1,\j-1);
			}
			\draw[thick, fill=white!40!blue] (2,-3) -- (2,-7) -- (3,-6) -- (3,-3) -- cycle;
			\foreach \i in {-4, -5, -6}
			\draw[thick] (2,\i) -- (3,\i);
			\draw[thick, fill=black!40!green] (3,-2) -- (3,-3) -- (10,-3) -- (10,-2) -- cycle;
			\foreach \i in {4, 5, 6, 7, 8, 9}
			\draw[thick] (\i,-2) -- (\i,-3);
			\draw[thick, fill=black!40!red] (2,-2) -- (2,-3) -- (3,-3) -- (3,-2) -- cycle;
			\begin{scope}[shift={(2,-3)}]
				\draw[thick, white!40!blue] (0,0) to[out=180,in=90] (-0.2,-0.2) -- (-0.2,-1.8) to[out=270,in=0] (-0.4,-2) to[out=0,in=90] (-0.2,-2.2) -- (-0.2,-3.8) to[out=270,in=180] (0,-4);
				\node[white!40!blue, rotate=90,anchor=center, fill=white] at (-1.1,-2) {Leg};
			\end{scope}
			\begin{scope}[shift={(3,-2)}]
				\draw[thick, black!60!green] (0,0) to[out=90,in=180] (0.2,0.2) -- (3.3,0.2) to[out=0,in=270] (3.5,0.4) to[out=270,in=180] (3.7,0.2) -- (6.8,0.2) to[out=0, in=90] (7,0);
				\node[black!60!green, fill=white] at (3.5,1.1) {Arm};
			\end{scope}
		\end{tikzpicture}
	\end{center}
	\caption{Hook, arm, and leg in a super-Young diagram}\label{fig:ArmLegSuper}
\end{figure}

Using the algorithm we find that expressions for Euler classes could be rewritten in terms of hook formulas (cf. \eqref{gl(1) hook formulas}):
\begin{equation}\label{gl(1|1) hook formulas}
	\begin{aligned}
		&\Eul_{\lambda} = \prod_{\Box \in \lambda}H^{\lambda}_{(ab)}(\Box),
	\end{aligned}
\end{equation}
where the product runs only over complete boxes and 
\begin{equation}
	\begin{aligned}
		&H_{(++)}^{\lambda}(\Box) = \Bigl[-\epsilon_{1}\bigl(\mathbf{a}_{\lambda}(\Box) + 1\bigr) + \epsilon_{2}\mathbf{l}_{\lambda}(\Box)\Bigr]\Bigl[\epsilon_{1}\mathbf{a}_{\lambda}(\Box) - \epsilon_{2}\bigl(\mathbf{l}_{\lambda}(\Box) + 1\bigr)\Bigr], \\
		&H_{(-+)}^{\lambda}(\Box) = \Bigl[-\epsilon_{1}\mathbf{a}_{\lambda}(\Box) + \epsilon_{2}\mathbf{l}_{\lambda}(\Box)\Bigr]\Bigl[\epsilon_{1}\mathbf{a}_{\lambda}(\Box) - \epsilon_{2}\bigl(\mathbf{l}_{\lambda}(\Box) + 1\bigr)\Bigr],\\
		&H_{(+-)}(\Box) = \Bigl[-\epsilon_{1}\bigl(\mathbf{a}_{\lambda}(\Box) + 1\bigr) + \epsilon_{2}\mathbf{l}_{\lambda}(\Box)\Bigr]\Bigl[\epsilon_{1}\mathbf{a}_{\lambda}(\Box) - \epsilon_{2}\mathbf{l}_{\lambda}(\Box)\Bigr],\\
		&H_{(--)}^{\lambda}(\Box) = 1\,.
	\end{aligned}
\end{equation}
Functions $\mathbf{a}_{\lambda}(\Box)$ and $\mathbf{l}_{\lambda}(\Box)$ denote the arm and leg functions of box $\Box$.
The arm and the leg represent tiles to the right and below box $\Box$ respectively.
Depending on whether the arm/leg is terminated by a complete tile or a half-tile, it is assumed to have even or odd parity correspondingly (see Fig.~\ref{fig:ArmLegSuper}).
Parities of the leg and of the arm contribute as subscripts $a$ and $b$ in functions $H_{(ab)}^{\lambda}(\Box)$.

\subsection{Representation on polynomials}

Using \eqref{gl(1|1) crystal ansatz} we define operators $\theta_{k}, p_{k}$:

\begin{equation}\label{super-time-variables gl(1|1)}
	\begin{aligned}
		&\theta_{1} = e_{0}^{+}, &p_{1} = \{e_{0}^{-}, e_{0}^{+}\}, \\
		&\theta_{k + 1} = \frac{1}{k}\bigl[e_{1}^{+}, \{e_{0}^{-}, \theta_{k}\}\bigr], &p_{k + 1} = \frac{1}{k}\bigl\{e_{0}^{-}, [e^{+}_{1}, p_{k}]\bigr\}, \\
		&\theta_{-1} = f_{0}^{+}, &p_{-1} = \{f_{0}^{-}, f_{0}^{+}\}, \\
		&\theta_{-k -1} = \dfrac{1}{k}\big[f_{1}^{+}, \{f_{0}^{-}, \theta_{-k}\}\big], &p_{-k -1} = \dfrac{1}{k}\bigl\{f_{0}^{-}, [f_{1}^{+}, p_{-k}]\bigr\}
	\end{aligned}
\end{equation}

We require the operators to satisfy the relations of super-Heisenberg algebra:

\begin{equation}
	\begin{aligned}
		&\{\theta_{k}, \theta_{m}\} = \delta_{m + k, 0}(-1)^{k-1}(\epsilon_{1}\epsilon_{2})^{2|k| - 2}\bbone, \\
		&[\theta_{k}, p_{m}] = 0, \\
		&[p_{k}, p_{m}] = \delta_{k + m, 0}(-1)^{k-1}k(\epsilon_{1}\epsilon_{2})^{2|k| - 1}\bbone
	\end{aligned}
\end{equation}
where we follow the convention from \cite{Galakhov:2023mak}. The expressions for zero modes generators are:

\begin{equation}\label{gl(1|1) zero modes in terms of super-time variables}
	\begin{aligned}
		&e_{0}^{+} = \theta_{1}, \quad &f_{0}^{+} = \d_{\theta_{1}}, \\
		&e_{0}^{-} = \sum_{k}p_{k}\d_{p_{k}}, \quad & f_{0}^{-} = \epsilon_{1}\epsilon_{2}\sum_{k}k\theta_{k}\d_{p_{k}}.
	\end{aligned}
\end{equation}

The variables $(p_{k}, \theta_{k})$ are analogues of time-variables $p_{k}$.

We define the states of the representation as a polynomial of $p_{k > 0}, \theta_{k > 0}$ and the generators of the algebra as differential operators.

The generating functions $\psi^{+}(z), \psi^{-}(z)$ have the following expressions:
\begin{equation}
	\begin{aligned}
		&\psi^{+}(z) = \dfrac{1}{z} - \dfrac{\epsilon_{1}\epsilon_{2}}{z^{3}}\psi_{3}^{-} - \dfrac{2\epsilon_{1}\epsilon_{2}}{z^{4}}\psi_{4}^{-} + \mathcal{O}\left(\frac{1}{z^{5}}\right), \\
		&\psi^{-}(z) = -z - \dfrac{\epsilon_{1} + \epsilon_{2}}{2} - \dfrac{\epsilon_{1}\epsilon_{2}}{z}\psi_{1}^{+} - \dfrac{2\epsilon_{1}\epsilon_{2}}{z^{2}}\left(\psi^{+}_{2} + \dfrac{\epsilon_{1} + \epsilon_{2}}{4}\psi_{1}^{+}\right) + \mathcal{O}\left(\dfrac{1}{z^{3}}\right)
	\end{aligned}
\end{equation}

Following \cite{Galakhov:2023mak}, we get the expressions:
\begin{equation}
	\begin{aligned}
		&\psi_{1}^{+} = \sum_{a = 1}^{\infty}ap_{a}\d_{p_{a}} + \sum_{a = 1}^{\infty}a\theta_{a}\d_{\theta_{a}}, \\
		&\psi^{-}_{3} = \sum_{a = 1}^{\infty}ap_{a}\d_{p_{a}} + \sum_{a = 1}^{\infty}(a - 1)\theta_{a}\d_{\theta_{a}};\\
		&\psi^{+}_{2} = \dfrac{1}{2}\sum_{a, b}\left[-\epsilon_{1}\epsilon_{2}(a + b)p_{a}p_{b}\d_{p_{a + b}} + ab p_{a + b}\d_{p_{a}}\d_{p_{b}}\right] + \\
		&+ \quad \sum_{a, b}\left[-\epsilon_{1}\epsilon_{2}bp_{a}\theta_{b}\d_{\theta_{a + b}} + ab\theta_{a + b}\d_{p_{a}}\d_{p_{b}}\right] + \sum_{a}\dfrac{\epsilon_{1} + \epsilon_{2}}{2}a(a - 1)\left(p_{a}\d_{p_{a}} + \theta_{a}\d_{\theta_{a}}\right), \\
		&\psi_{4}^{-} = \dfrac{1}{2}\sum_{a, b}\left[-\epsilon_{1}\epsilon_{2}(a + b)p_{a}p_{b}\d_{p_{a + b}} + abp_{a + b}\d_{p_{a}}\d_{p_{b}}\right] + \\
		&\quad + \sum_{a, b}\left[-\epsilon_{1}\epsilon_{2}(b - 1)p_{a}\theta_{b}\d_{\theta_{a + b}} + a(b-1)\theta_{a + b}\d_{p_{a}}\d_{\theta_{b}}\right] + \sum_{a}\dfrac{\epsilon_{1} + \epsilon_{2}}{2}\left(a^{2}p_{a}\d_{p_{a}} + (a - 1)^{2}\theta_{a}\d_{\theta_{a}}\right).
	\end{aligned}
\end{equation}

The super-Schur functions $\mathcal{S}_{\lambda}$ are defined as the eigen functions of the operators $\psi^{+}_{1}, \psi^{-}_{3}, \psi^{+}_{2}, \psi_{4}^{-}$ where the last two operators play the role of \eqref{cut-and-join gl(1)}. The super-Schur functions give rise to Littlewood-Richardson coefficients:
\begin{equation}\label{Lit-Rich gl(1|1)}
	e_{0}^{(a)}\mathcal{S}_{\lambda} =  \sum_{\ssqbox{$a$}}{\bf E}_{\lambda, \lambda + \ssqbox{$a$}}\mathcal{S}_{\lambda + \ssqbox{$a$}}.
\end{equation}

\section{Conclusion}\label{sec: Conclusion}
Simple Lie algebras in the famous Cartan classification share many common properties. 
This fact allows one to describe these algebras in a universal language. 
The main differences between the algebras in that description can be collected in simple combinatorial objects -- Dynkin diagrams. 
Almost the same classification with some peculiarities turns out to work for infinite dimensional affine Lie algebras.

In this paper, we review simple examples of quiver Yangian algebras. 
Historically, the Yangians were special infinite dimensional generalizations of classical (and affine) Lie algebras reflecting integrability properties with the help of $R$-matrix, which is a solution to Yang-Baxter equations. 
Drinfeld's ``new realization'' of Yangians, which is an analogy of Chevalley basis in simple Lie algebras, gave us hope that some classification similar to Cartan exists for Yangians as well. 
An indication that such a classification might exist comes from the recent works on algebras associated with BPS states on toric Calabi-Yau three-folds. 
Those algebras mimic various properties of Yangians and, moreover, in certain cases turn out to be isomorphic to the Yangians.
On the other hand, a classification of toric Calabi-Yau three-folds (and respective algebras) by quiver diagrams seems to be reacher than those of Cartan by Dynkin diagrams.

In addition to a sole quiver diagram, one has to specify some additional amount of data, which we call quiver data, including a gauge invariant  holomorphic function called superpotential.
If a model originates from a toric CY${}_3$, a variety of these data is restricted to a certain amount; in particular, a superpotential for an unframed quiver is uniquely fixed by the corresponding quiver on a torus.
Another piece of data known as equivariant weights is also fixed to just two free parameters of an affine Yangian.
The question what types of algebras and representations may be constructed for generic quiver data remains open \cite{Li:2023zub}.

\bigskip

In the case of Lie algebras (both classical and affine), we have the following steps to write the commutation relation in the Chevalley basis:
\begin{equation*}
	\text{Dynkin diagram} \rightarrow \text{Cartan matrix} \rightarrow \text{Commutation relation}
\end{equation*}

In a complete analogy with classical Lie algebras, in the case of Yangians, we go through the following steps:

\begin{equation*}
	\text{Quiver diagram} \rightarrow \text{Structure functions/Bond factors} \rightarrow \text{Commutation relations}
\end{equation*}

Naively, commutation relations \eqref{Yangian} are completely defined solely by the quiver diagram, and the rest of the quiver data is unimportant in this definition.
However, as we mentioned earlier, the possibility of constructing certain forms of self-consistent superpotentials imposes restrictions on the weights of quiver arrows that enter the relations \eqref{Yangian} as well.

Also there might be higher order relations, so-called generalized Serre relations; however, we did not discuss them here.

\bigskip

We should note that the Yangian algebras are not Lie algebras for generic values of parameters.
A commutator of two generators is not equal to a linear combination of generators. 
One of the implications of this fact is that relations of Yangian modes contain anti-commutators as well as usual commutators, even if the Yangian in question is not a super-algebra.
For the simplest illustration of this phenomenon, see an example of $\mathsf{Y}(\fs\fl_2)$ in Sec.~\ref{sec: Y(sl(2))}.

\bigskip

However a complete set of quiver data and a choice of quiver framing affect clearly the representations of quiver Yangians.
In this paper we considered explicit algorithms to construct so-called crystal representations and to describe respective matrix elements for quiver Yangian generators.
Our aim was to reduce maximally pre-requisites for the implementation of these algorithms so that they could be implemented within a machine framework with the help of any symbolic calculation system.

The algorithms are divided into two major parts: constructing crystals and calculating matrix elements.

The algorithm for crystals is recursive and consists of adding a $(k+1)^{\rm th}$ atom to all the vacant ``cavities'' in $k$-atom crystals.
The algorithm for matrix elements consists of just a careful analysis of tangent spaces to the fixed points labeled by crystals and implementing a geometric matrix element formula derived from the equivariant integration.

We illustrated the work of these algorithms in the following simple examples of quiver Yangians: $\mathsf{Y}(\mathfrak{sl}_{2})$, $\mathsf{Y}(\widehat{\fg\fl}_{1})$ and $\mathsf{Y}(\widehat{\fg\fl}_{1|1})$.

Also, in all the cases, we provided a realization of a representation in terms of differential operators on the spaces of polynomials. 
The simplest case $\mathsf{Y}(\mathfrak{sl}_{2})$ requires polynomials in one variable, and more complicated affine Yangians $\mathsf{Y}(\widehat{\fg\fl}_{1})$ and $\mathsf{Y}(\widehat{\fg\fl}_{1|1})$ require an infinite number of variables.

\section*{Acknowledgments}\label{sec: Acknowledgments}

This work was supported by the Russian Science Foundation (Grant No.23-41-00049).


\bibliographystyle{utphys}
\bibliography{biblio}

\end{document}